\documentclass[acmtog, nonacm]{acmart}

\usepackage{float}
\usepackage{wrapfig}
\usepackage{subcaption}
\usepackage{multirow}
\usepackage[ruled]{algorithm2e}

\newcommand{\mycc}{\cellcolor[HTML]{DAE6F1}}
\newcommand{\myrc}{\rowcolor[HTML]{DAE6F1}}

\definecolor{sperp-green}{rgb}{0.651,0.776,0.451}
\definecolor{S-blue}{rgb}{0.361,0.604,0.890}
\definecolor{cpc-purple}{rgb}{0.584,0.318,0.886}
\definecolor{S-dark-blue}{rgb}{0.061,0.304,0.590}
\definecolor{cpdiff-yellow}{rgb}{0.875,0.843,0.431}

\DeclareMathOperator*{\argmin}{arg\,min}
\newcommand{\II}{{\mathbf I}}
\newcommand{\rr}{{\mathbf r}}

\newcommand{\A}{{\mathbf A}}
\newcommand{\EE}{{\mathbf E}}
\newcommand{\LL}{{\mathbf L}}
\def\S{{\mathcal S}}
\def\C{{\mathcal C}}
\def\I{{\mathcal I}}
\def\L{{\mathcal L}}
\def\D{{\mathcal D}}
\def\G{{\mathcal G}}
\def\N{{\mathcal N}}
\def\cp{\text{cp}}
\def\t{{\mathbf t}}
\def\b{{\mathbf b}}
\def\x{{\mathbf x}}
\def\y{{\mathbf y}}
\def\z{{\mathbf z}}
\def\u{{\mathbf u}}
\def\bfv{{\mathbf v}}
\def\n{{\mathbf n}}
\def\h{{\mathbf h}}
\def\J{{\mathbf J}}

\AtBeginDocument{%
  }

\setcopyright{acmlicensed}
\copyrightyear{2024}
\acmYear{2024}
\acmDOI{XXXXXXX.XXXXXXX}

\acmISBN{978-1-4503-XXXX-X/24/06}

\begin{document}

\title{A Closest Point Method for PDEs on Manifolds with Interior Boundary Conditions for Geometry Processing}

\author{Nathan King}
\affiliation{%
  \institution{University of Waterloo}
  \city{Waterloo}
  \state{Ontario}
  \country{Canada}
}
\email{n5king@uwaterloo.ca}

\author{Haozhe Su}
\affiliation{%
  \institution{LightSpeed Studios}
  \city{Los Angeles}
  \state{California}
  \country{USA}
}
\email{haozhesu@global.tencent.com}

\author{Mridul Aanjaneya}
\affiliation{%
  \institution{Rutgers University}
  \city{Piscataway}
  \state{New Jersey}
  \country{USA}
}
\email{mridul.aanjaneya@rutgers.edu}

\author{Steven Ruuth}
\affiliation{%
  \institution{Simon Fraser University}
  \city{Burnaby}
  \state{British Columbia}
  \country{Canada}
}
\email{sruuth@sfu.ca}

\author{Christopher Batty}
\affiliation{%
  \institution{University of Waterloo}
  \city{Waterloo}
  \state{Ontario}
  \country{Canada}
}
\email{christopher.batty@uwaterloo.ca}


\begin{abstract}
Many geometry processing techniques require the solution of partial differential equations (PDEs) on manifolds embedded in $\mathbb{R}^2$ or $\mathbb{R}^3$, such as curves or surfaces. Such {\it manifold PDEs} often involve boundary conditions (e.g., Dirichlet or Neumann) prescribed at points or curves on the manifold's interior or along the geometric (exterior) boundary of an open manifold. However, input manifolds can take many forms (e.g., triangle meshes, parametrizations, point clouds, implicit functions, etc.). Typically, one must generate a mesh to apply finite element-type techniques or derive specialized discretization procedures for each distinct manifold representation.
We propose instead to address such problems in a unified manner through a novel extension of the \emph{closest point method} (CPM) to handle interior boundary conditions. CPM solves the manifold PDE by solving a volumetric PDE defined over the Cartesian embedding space containing the manifold, and requires only a closest point representation of the manifold. Hence, CPM supports objects that are open or closed, orientable or not, and of any codimension. To enable support for interior boundary conditions we derive a method that implicitly partitions the embedding space across interior boundaries. CPM's finite difference and interpolation stencils are adapted to respect this partition while preserving second-order accuracy. Additionally, we develop an efficient sparse-grid implementation and numerical solver that can scale to tens of millions of degrees of freedom, allowing PDEs to be solved on more complex manifolds. We demonstrate our method's convergence behaviour on selected model PDEs and explore several geometry processing problems: diffusion curves on surfaces, geodesic distance, tangent vector field design, harmonic map construction, and reaction-diffusion textures. Our proposed approach thus offers a powerful and flexible new tool for a range of geometry processing tasks on general manifold representations.   
\end{abstract}

\setcopyright{acmlicensed}
\acmJournal{TOG}
\acmYear{2024} \acmVolume{1} \acmNumber{1} \acmArticle{1} \acmMonth{1}\acmDOI{10.1145/3673652}

\ccsdesc[500]{Mathematics of computing~Discretization}
\ccsdesc[500]{Mathematics of computing~Partial differential equations}
\ccsdesc[500]{Computing methodologies~Shape analysis}

\keywords{manifold partial differential equations,  embedding methods, closest point method, boundary conditions, geometry processing, diffusion curves, geodesic distance, vector field design, harmonic maps, reaction-diffusion textures}


\begin{teaserfigure}
     \centering
     \begin{subfigure}[b]{\textwidth}
     \hspace{5pt}
          \begin{subfigure}[b]{0.36\textwidth}
             \centering
             \begin{subfigure}[b]{0.48\textwidth}
                \includegraphics[width=\textwidth]{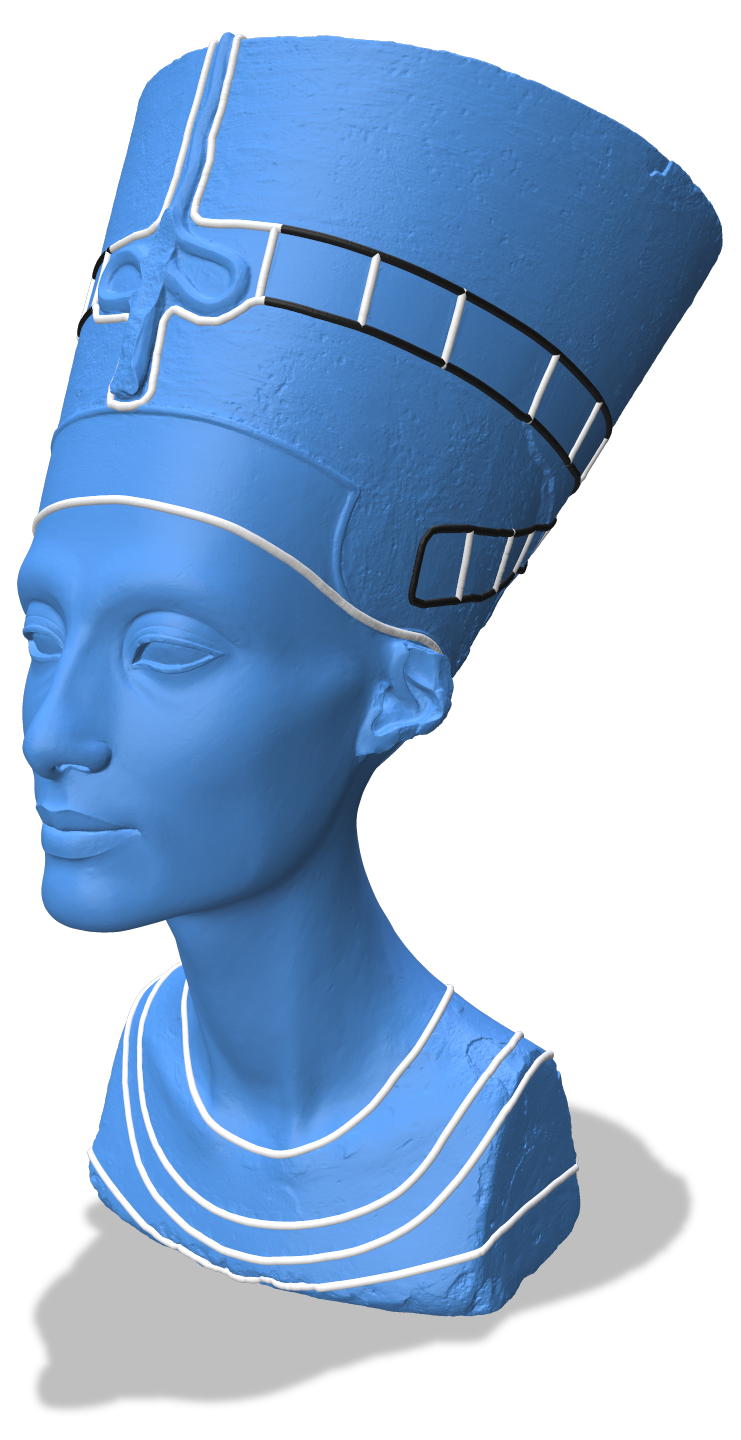}
             \end{subfigure}
             \hfill
             \begin{subfigure}[b]{0.48\textwidth}
                \includegraphics[width=\textwidth]{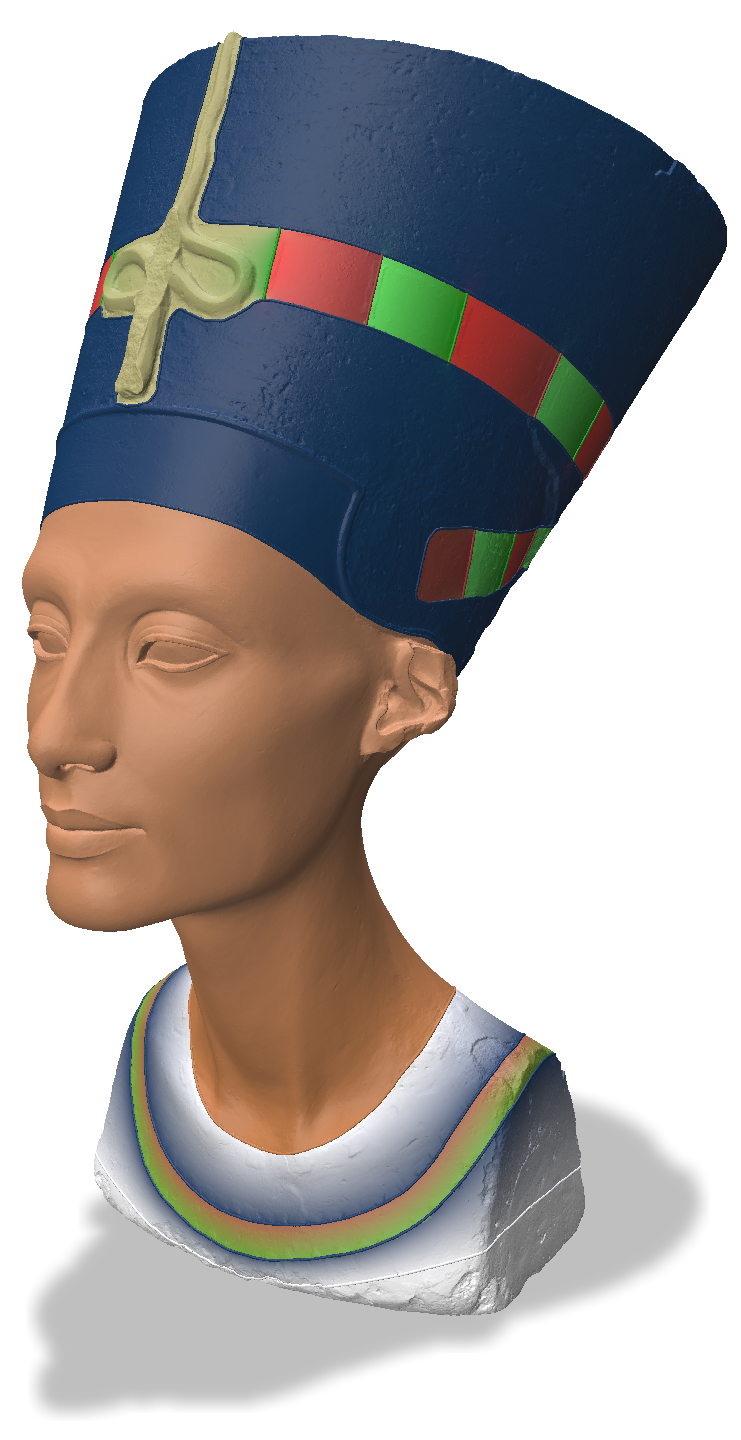}
             \end{subfigure}
             \caption{Diffusion Curves}
         \end{subfigure}
         \hspace{-0.6cm}
        \begin{subfigure}[b]{0.35\textwidth}
             \centering
             \includegraphics[width=0.8\textwidth]{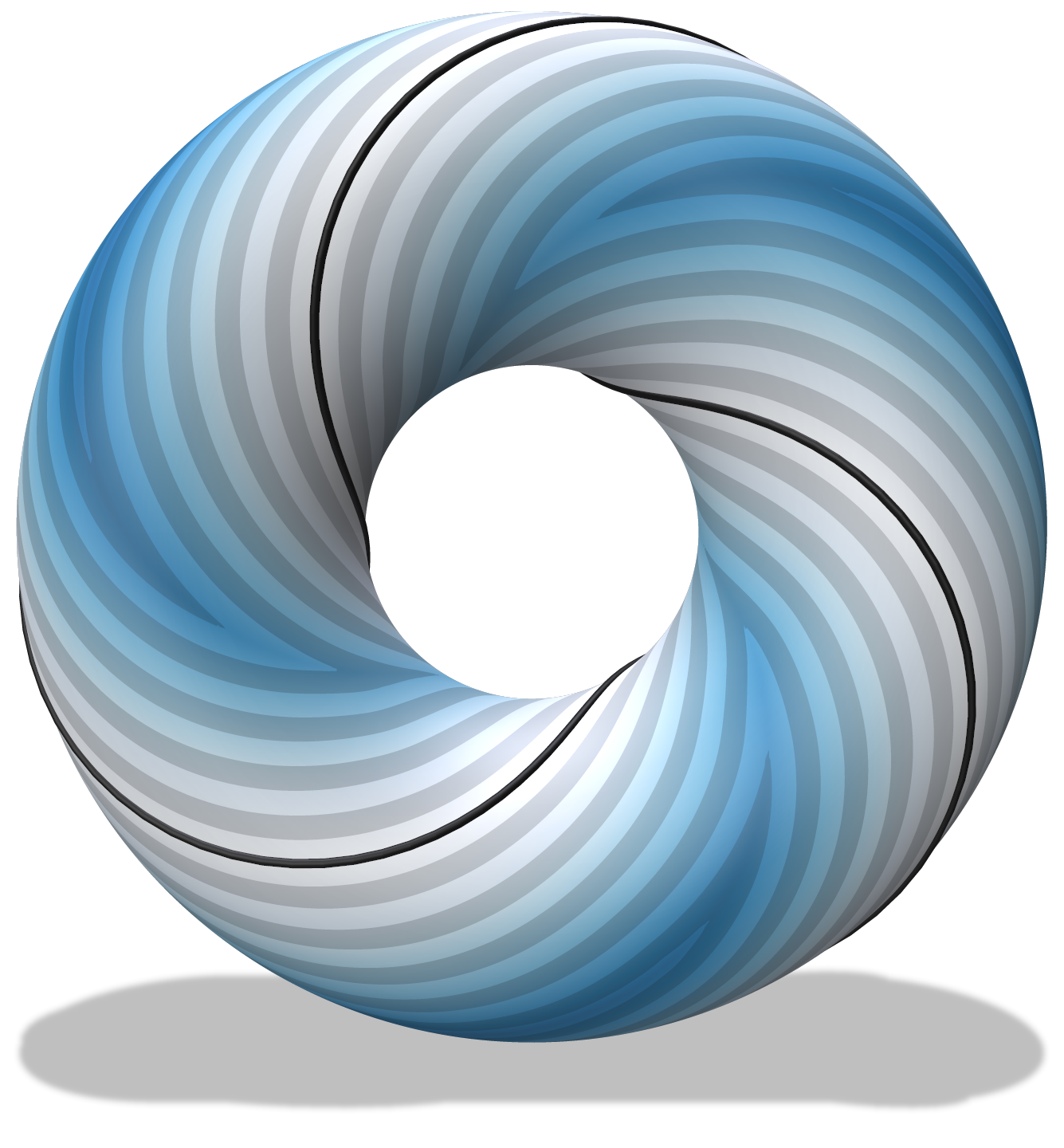}
             \caption{Geodesic Distance}
         \end{subfigure}
         \hspace{0.5cm}
         \begin{subfigure}[b]{0.21\textwidth}
         \hspace{-1cm}
         \begin{subfigure}[b]{\textwidth}
            \includegraphics{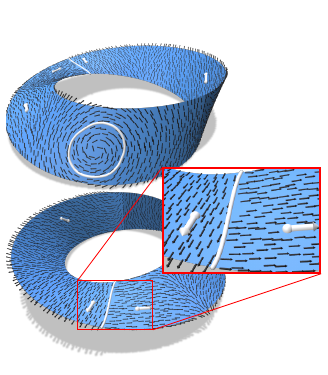}
         \end{subfigure}
         \caption{Vector Field Design}
         \end{subfigure}
     \end{subfigure}
  \caption{We extend the \emph{closest point method} to support solving PDEs on manifolds with interior boundary conditions. Our method enables the solution of various geometry processing tasks on general surfaces, given only the ability to perform closest point queries. (a) Colouring a triangulated surface using diffusion curves. (b) Geodesic distance to a parametric curve (black) on an analytical closest point surface. (c) Vector field design on a triangulation of a M\"{o}bius strip, which is an open and nonorientable surface. }
  \Description{}
  \label{fig:teaser}
\end{teaserfigure}

\maketitle

\section{Introduction}
A {\it manifold partial differential equation} is a partial differential equation (PDE) whose solution is restricted to lie on a manifold $\S$. Such manifold PDEs arise naturally in many fields, including applied mathematics, mathematical physics, image processing, computer vision, fluid dynamics, and computer graphics. We focus on geometry processing, where a numerical solution is typically sought by approximating the manifold as a mesh and discretizing the PDE using finite element or discrete exterior calculus techniques. 
However, the introduction of a mesh entails some drawbacks. One must perform mesh generation if the input manifold is not given as a mesh. The mesh quality also strongly influences the resulting solution and therefore remeshing is required if the input mesh is of low quality or inappropriate resolution. Both mesh generation and remeshing are nontrivial tasks. Finally, depending on the chosen numerical method, the discretization of a particular manifold PDE can differ significantly from the corresponding discretized PDE on Cartesian domains; further analysis can be needed to derive an appropriate convergent scheme for the manifold case.

\begin{figure}
     \begin{subfigure}[b]{0.47\textwidth}
         \begin{subfigure}[b]{0.32\textwidth}
             \centering
            \includegraphics[width=0.95\textwidth]{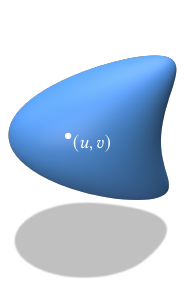}
            \vspace{-4pt}
         \end{subfigure}
         \hfill
         \begin{subfigure}[b]{0.32\textwidth}
             \centering
             \includegraphics[width=0.9\textwidth]{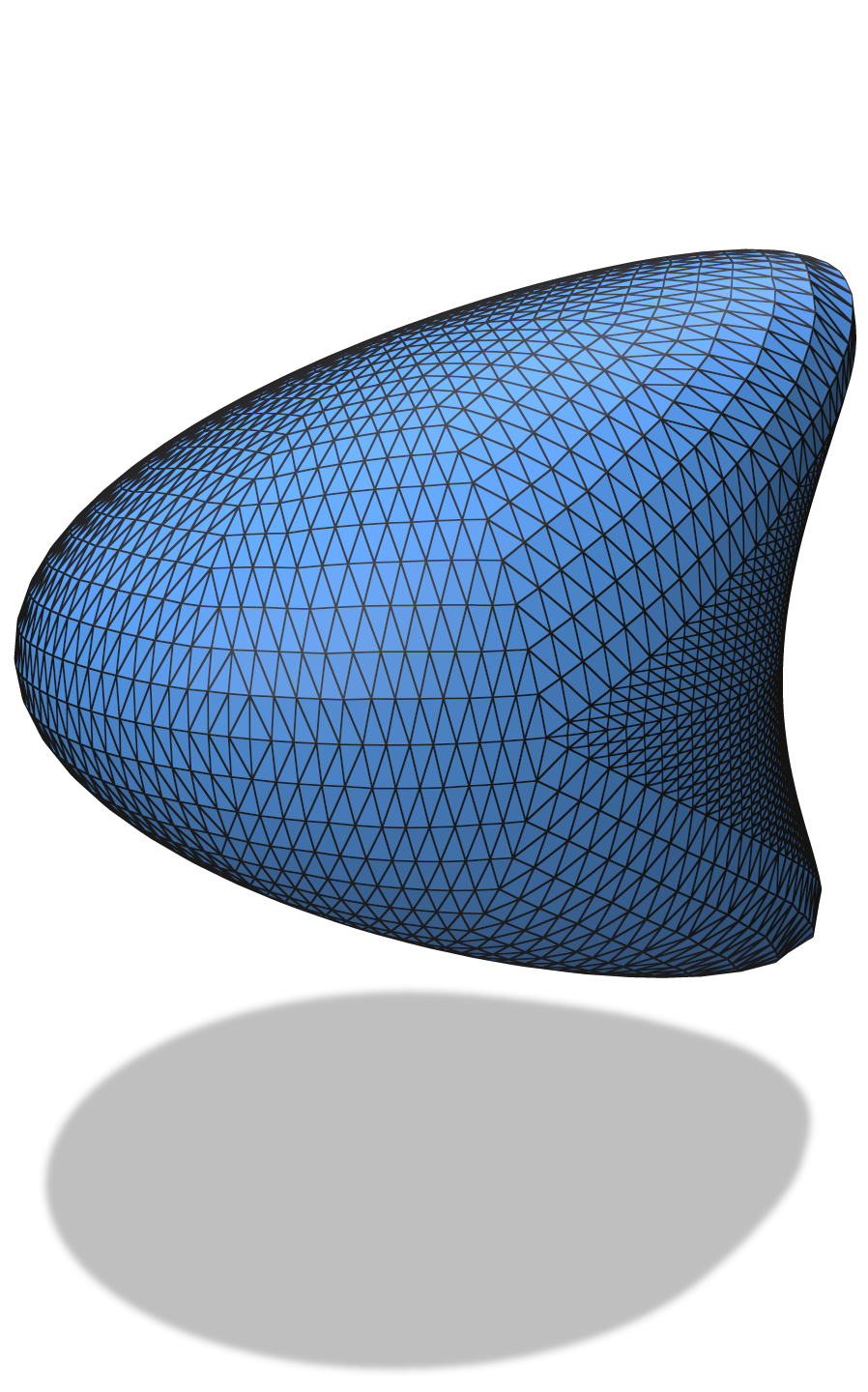}
         \end{subfigure}   
        \hfill
         \begin{subfigure}[b]{0.32\textwidth}
             \centering
             \includegraphics[width=0.9\textwidth]{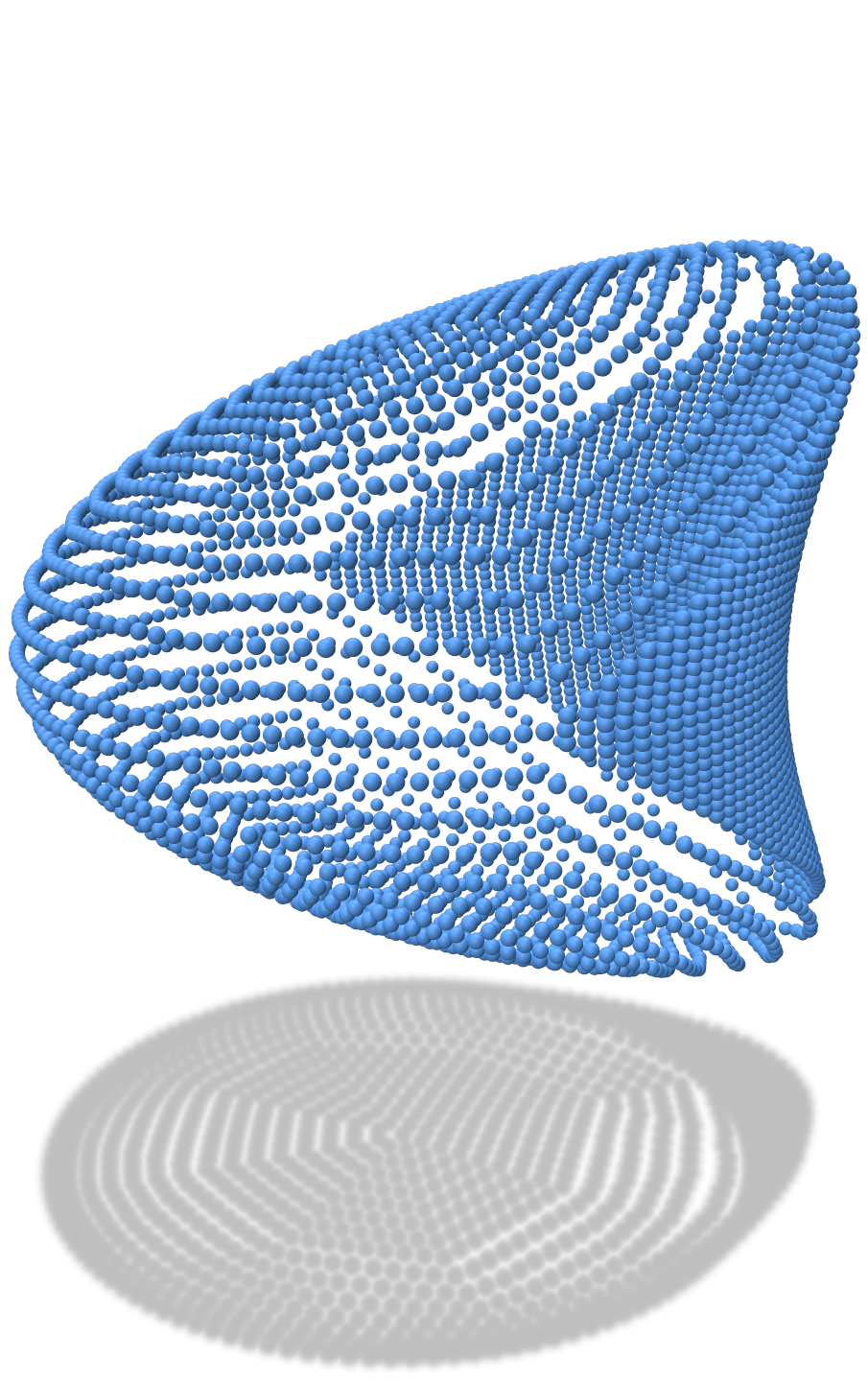}
         \end{subfigure}
        \end{subfigure}

        \begin{subfigure}[b]{0.47\textwidth}
         \begin{subfigure}[b]{0.49\textwidth}
             \centering
             \includegraphics[width=0.85\textwidth]{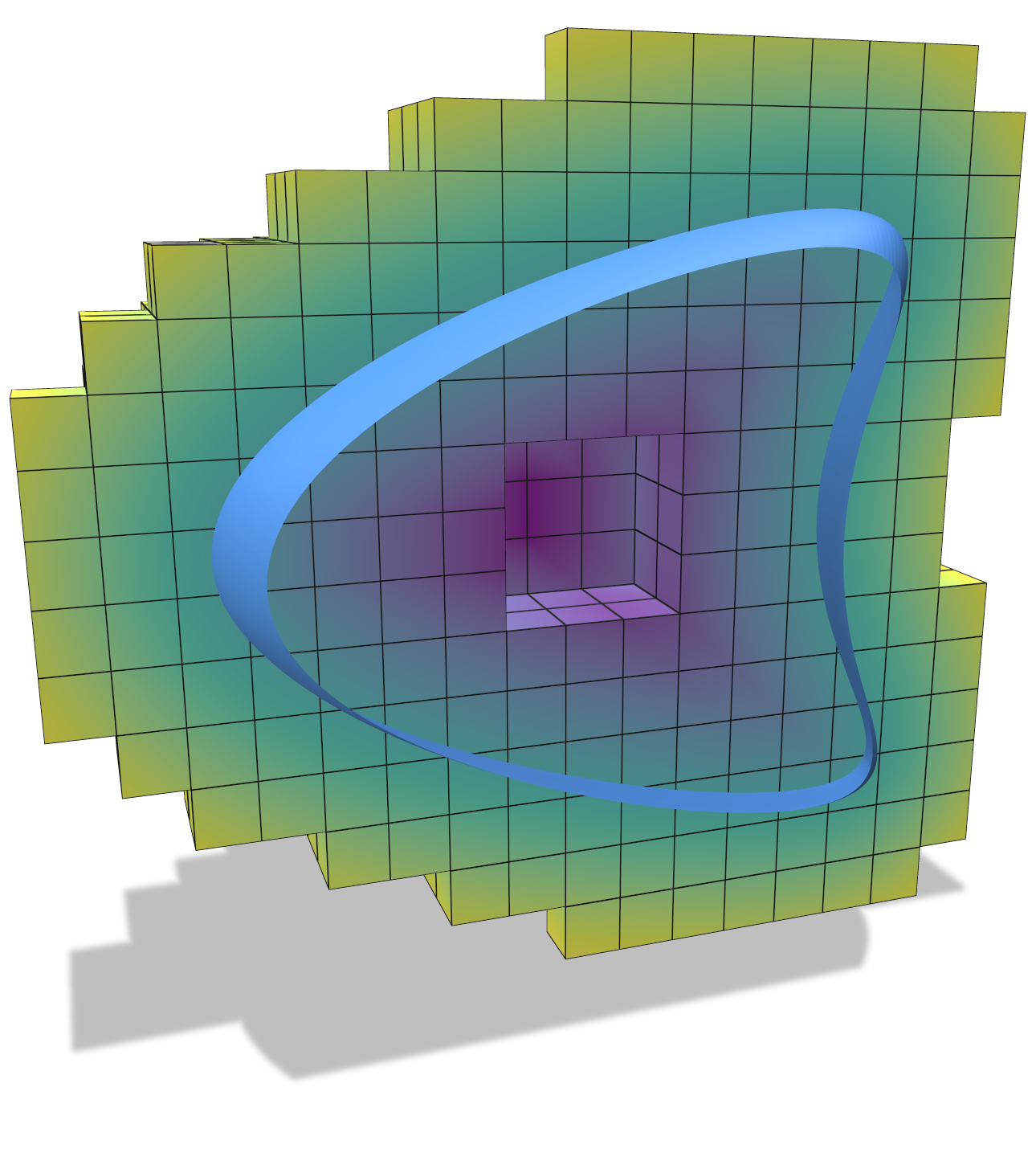}
         \end{subfigure}
     \hfill
         \begin{subfigure}[b]{0.49\textwidth}
             \centering
             \includegraphics[width=0.85\textwidth]{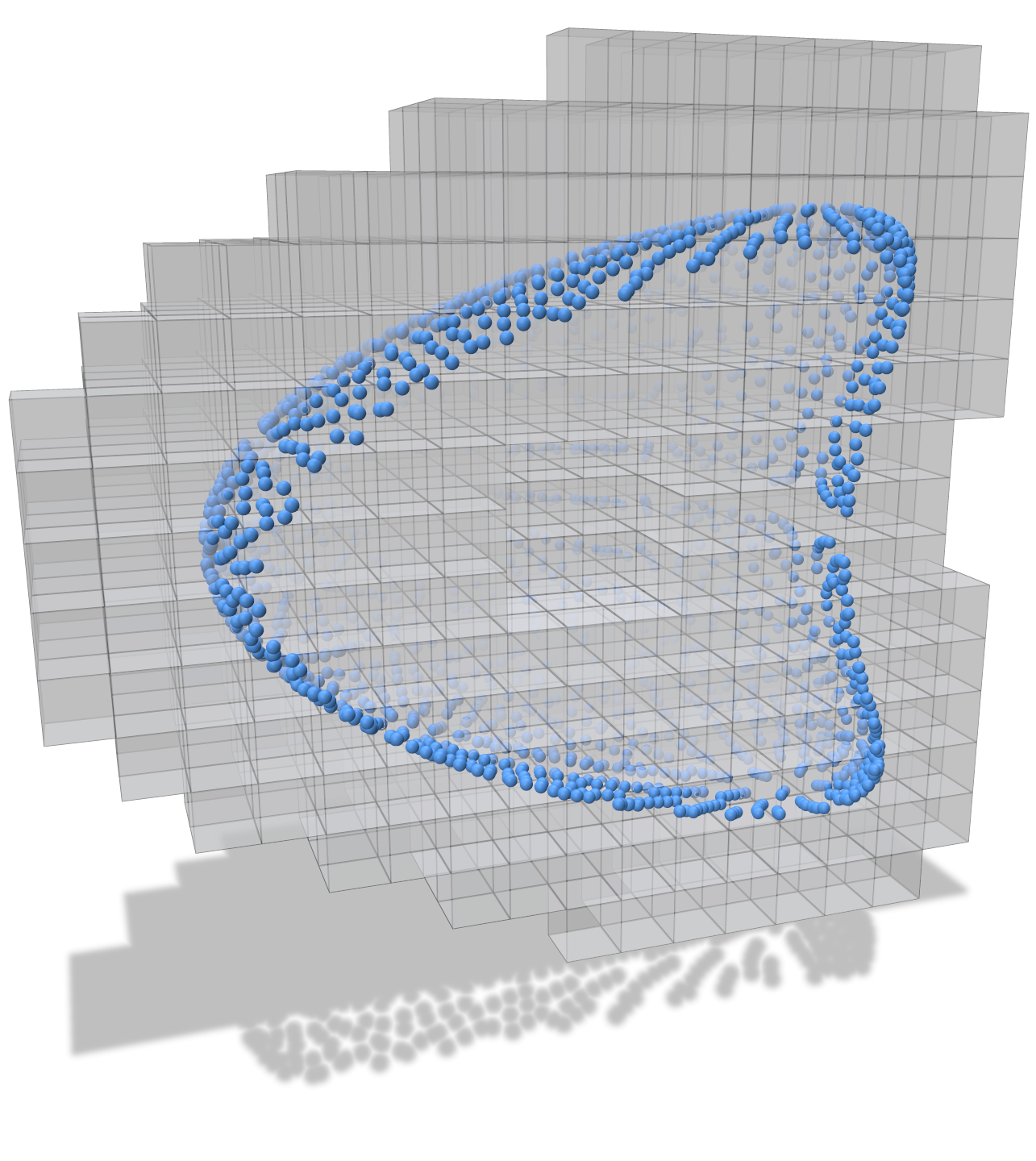}
         \end{subfigure}
    \end{subfigure}
        \caption{CPM can be applied to any manifold representation that supports closest point queries, including parametrizations, meshes, and point clouds, as well as discrete or continuous level sets and closest point functions.}
        \label{fig:manifold-representations}
\end{figure}

A powerful alternative is the use of embedding techniques, which solve the manifold problem by embedding it into a surrounding higher-dimensional Cartesian space. The \emph{closest point method} (CPM) \cite{Ruuth2008} is an especially attractive instance of this strategy, as it offers a remarkable combination of simplicity and generality. Its simplicity lies in its ability to leverage standard Cartesian numerical methods in the embedding space to solve the desired manifold problem, given only a closest point function for the manifold. Its generality lies in its support for diverse manifold characteristics, manifold representations, and manifold PDEs. 

Requiring only a closest point function allows input manifolds to be open or closed, orientable or not, and of any codimension or even mixed codimension. Closest point queries are available for many common manifold representations (as highlighted by \citet{Sawhney2020}), and therefore CPM can be applied to meshes, level sets, point clouds, parametric manifolds, constructive solid geometry, neural implicit surfaces, etc. (see Figure~\ref{fig:manifold-representations}). Such generality is appealing given the increasing demand for algorithms that can ingest general ``in-the-wild'' and high-order geometries (\cite{Hu2018,Barill2018,Sawhney2020,marschner2021sum}). Furthermore, the \emph{embedding PDE} solved on the Cartesian domain is often simply the Cartesian analog of the desired manifold PDE. Thus, CPM has been applied to the heat equation, Poisson and screened-Poisson equations, Laplace-Beltrami eigenproblem, biharmonic equation, advection-diffusion and reaction-diffusion equations, Hamilton-Jacobi equation, Navier-Stokes equation, Cahn-Hilliard equation, computation of ($p$-)harmonic maps, and more.

Yet, despite the desirable properties of CPM and its adoption in applied mathematics, CPM has only infrequently been employed by computer graphics researchers, and almost exclusively for fluid animation \cite{Hong2010,Auer2012, auer2013semi,Kim2013,Morgenroth2020}. 
In the present work, we demonstrate CPM's wider potential value for computer graphics problems by extending CPM to handle several applications in geometry processing: diffusion curves on surfaces, geodesic distance, tangent vector field design, harmonic maps with feature (landmark) points and curves, and reaction-diffusion textures.

However, a crucial limitation of the existing CPM stands in the way of the objective above. CPM supports standard boundary conditions on the geometric (exterior) boundary of an open manifold, $\partial \S$, but it does not yet support accurate \emph{interior boundary conditions} (IBCs), i.e., boundary conditions at manifold points or curves away from $\partial \S$. CPM's use of the embedding space makes enforcing IBCs nontrivial, but they are vital for the applications above. For example, the curves in diffusion curves or the source points for geodesic distance computation generally lie on the interior of $\S$. Therefore, we propose a novel mechanism that enables accurate IBC enforcement for CPM in $\mathbb{R}^2$ and $\mathbb{R}^3$, while retaining its simplicity and generality.

To scale up to surfaces with finer details, we further develop a tailored numerical framework and solver. The computational domain is only required near $\S$, so we use a sparse grid structure to improve memory efficiency. We then develop a custom preconditioned BiCGSTAB solver for solving the linear system that also better utilizes memory. The combination of the sparse grid structure near $\S$ and the custom solver allows us to efficiently scale to tens of millions of degrees of freedom. To foster wider adoption of CPM, our code has been released publicly at \url{https://github.com/nathandking/cpm-ibc}. 

In summary, the key contributions of our work are to: 
\begin{itemize}
\item introduce a novel treatment of interior boundary conditions for CPM with up to second-order accuracy; 
\item employ a sparse grid structure and develop a custom solver for memory efficiency, which enables scaling to tens of millions of degrees of freedom; and
\item demonstrate the effectiveness of our new CPM scheme for several geometry processing tasks.

\end{itemize}

\section{Related Work}
\label{sec:related-work}

\subsection{CPM in Applied Mathematics}
CPM was introduced by \citet{Ruuth2008}, who applied it to diffusion, advection, advection-diffusion, mean curvature flow of curves on surfaces, and reaction-diffusion. They drew inspiration from earlier embedding methods based on level sets \cite{Bertalmio2001,Greer2006}, while eliminating the restriction to closed manifolds, supporting more general PDEs, and allowing for narrow-banding without loss of convergence order. Subsequently, CPM has been shown to be effective for a wide range of additional PDEs including the screened-Poisson (a.k.a. positive-Helmholtz) equation \cite{Chen2015,May2020}, Hamilton-Jacobi equations/level-set equations \cite{Macdonald2008}, biharmonic equations \cite{Macdonald2010}, Cahn-Hilliard equation \cite{Gera2017}, Navier-Stokes equation \cite{Auer2012, Yang2020}, construction of ($p$-)harmonic maps \cite{King2017}, and more. Despite being initially designed for manifold PDEs, CPM can additionally be applied to volumetric (codimension-0) problems and surface-to-bulk coupling scenarios \cite{Macdonald2013}. Related closest point mapping approaches have also been used to handle integral equations \cite{Kublik2013,Kublik2016,Chen2017,Chu2018}. 

Some prior work on CPM has focused on problems of relevance to geometry processing. For example, \citet{Macdonald2011} computed eigenvalues and eigenfunctions of the Laplace-Beltrami operator via CPM, and the resulting eigenvalues of surfaces were used by \citet{Arteaga2015} to compute the `Shape-DNA' \cite{Reuter2006} for clustering similar surfaces into groups. Segmentation of data on surfaces was demonstrated by \citet{Tian2009} who adapted the Chan-Vese algorithm common in image processing. Different approaches to compute normals and curvatures were discussed in the appendix of the original CPM paper \cite{Ruuth2008}.

CPM has mostly been used on static manifolds with a uniform grid in the embedding space as the computational domain. However, \citet{Petras2016} combined CPM with a grid-based particle method to solve PDEs on moving surfaces. A mesh-free CPM approach was investigated in \cite{piret2012orthogonal, cheung2015localized, Petras2018, Petras2019, Petras2022} using radial-basis functions. 

The CutFEM family of methods \cite{burman2015cutfem} represent another embedding approach. They use finite elements (rather than finite differences) on a non-conforming simplicial embedding mesh. They have been used to solve various manifold PDEs (e.g., Laplace-Beltrami \cite{burman2015stabilized}, convection \cite{burman2019stabilized}).

\subsection{CPM in Computer Graphics}
Embedding methods similar to CPM have also been proposed and used in the computer graphics community. Perhaps most closely related is the work of \citet{chuang2009} who solved Poisson problems using the finite element method over a function space consisting of 3D grid-based B-spline basis functions restricted to the shape's surface. They demonstrated geometry processing applications such as texture back-projection and curvature estimation. They also showed that the observed eigenspectra are much less dependent on the surface triangulation than with standard mesh-based methods. While their approach has some conceptual connections to CPM, it does not possess the same degree of simplicity or generality as CPM, nor does it support IBCs.
The thesis by~\citet{chuang2013grid} further demonstrates an extension of this approach to use locally non-manifold grids to address narrow bottlenecks, where two pieces of a surface are close in Euclidean distance but far apart in geodesic distance. Our work also introduces a non-manifold grid structure, but with the distinct aim of handling IBCs.
 
CPM itself has been applied in computer graphics, primarily for fluid animation. \citet{Hong2010} used a modified CPM to evolve and control the motion of flame fronts restricted to surfaces.
The work of \citet{Kim2013} increased the apparent spatial resolution of an existing volumetric liquid simulation by solving a wave simulation on the liquid surface. The surface wave equation and Navier-Stokes equations were solved by \citet{Auer2012} with a real-time implementation on the GPU. 
\citet{auer2013semi} subsequently extended this work to support deforming surfaces given by a sequence of time-varying triangle meshes (predating the moving surface work of \citet{Petras2016} in computational physics).  \citet{Morgenroth2020} employed CPM for one-way coupling between a volumetric fluid simulation and a surface fluid simulation for applications such as oil films spreading on liquid surfaces. 

\citet{wang2020codimensional} coupled moving-least-squares approximations on codimension-1 and 2 objects with grid-based approximations for codimension-0 operators in surface-tension driven Navier-Stokes systems. The ability of CPM to handle mixed-codimension objects makes it an ideal candidate for a unified solver. 

\subsection{Interior Boundary Conditions on Manifolds}
\label{sec:IBC_related_work}
Existing numerical methods for manifold PDEs support IBCs in various ways depending on the chosen manifold representation and method of discretization. In the Dirichlet case, the nearest degrees of freedom (DOFs) to the interior boundary can often simply be assigned the desired Dirichlet value. For example, on a point cloud representation, the nearest interior points in the cloud could be set to the Dirichlet value, similar to how \emph{exterior} Dirichlet BCs have been handled in point clouds \cite{Liang2013}. With triangle mesh-based discretizations (finite element, discrete exterior calculus, etc.) one can similarly enforce the Dirichlet condition at the nearest surface vertices to the interior boundaries. However, enforcing the IBC at the nearest DOF is inaccurate if the DOF does not lie exactly on the interior boundary $\C$ (i.e., the mesh does not precisely conform to $\C$). Specifically, an error of $O(\|\h\|)$ is introduced where $\|\h\|$ is the distance between the nearest DOF and $\C$.  Moreover, only Dirichlet conditions can be treated in this manner; depending on the chosen manifold representation and/or discretization, it can be nontrivial to enforce Neumann boundary conditions.

For Dirichlet IBCs in CPM, Auer et al.~\shortcite{Auer2012, auer2013semi} fixed all the nearest DOFs in the embedding space within a ball centred around $\C$ (considering only the case when $\C$ is a point). This again is only first-order accurate, incurring an $O(\Delta x)$ error, where $\Delta x$ is the grid spacing in the embedding space. Enforcing the IBC over a ball effectively inflates the boundary region to a wider area of the surface. That is, a circular region of the surface around the point $\C$ will be fixed with the prescribed condition. We show in Section~\ref{sec:conv-studies} that this approach can also be applied to boundary curves, but the observed error is much larger compared to our proposed method. Moreover, it cannot be applied when Dirichlet values differ on each side of $\C$.

With a surface triangulation, a more accurate approach is to remesh the surface with constrained Delaunay refinement (possibly with an intrinsic triangulation) so that vertices or edges of the mesh conform to $\C$, as discussed for example by \citet{Sharp2020Flip}. However, this necessarily introduces remeshing as an extra preprocess. Another mesh-based approach, which avoids remeshing, is the extended finite element method  \cite{Moes1999XFEM,Kaufmann2009}, which uses modified basis functions to enforce non-conforming boundaries or discontinuities.

Most similar to our approach is the method of \citet{Shi2007} who enforced Dirichlet IBCs for a manifold PDE method based on level sets. As with CPM, solving surface PDEs with level sets \cite{Bertalmio2001} involves extending the problem to the surrounding embedding space. For such embedding methods, it is crucial not only to account for the interior boundary itself but also its influence into the associated embedding space. To do so, the approach of \citet{Shi2007} explicitly constructs a triangulation to represent a normal manifold $\S_{\perp}$ (see~\eqref{eqn:Sperp}) extending outwards from the interior boundary curve $\C$ (notably contrasting with the implicit nature of level-sets). They then perform geometric tests to determine if stencils intersect $\S_{\perp}$ and modify the discretization locally. We instead introduce a simple triangulation-free approach to determine if stencils cross $\S_{\perp}$ that only involves closest points, bypassing explicit construction of $\S_{\perp}$. Moreover, such level-set approaches necessarily require a well-defined inside and outside, which makes handling open manifolds, nonorientable manifolds, and manifolds of codimension-two or higher impossible with a single level set. 

Our proposed CPM extension overcomes several limitations of the existing CPM (Dirichlet-only) IBC treatment of Auer et al.~\citeyearpar{Auer2012}. We demonstrate that our method can easily be extended to second-order, for both Dirichlet and zero-Neumann cases. It can also handle jump discontinuities in Dirichlet values across interior boundary curves. Furthermore, our approach supports what we call \emph{mixed} boundary conditions, e.g., Dirichlet on one side and Neumann on the other. Both jump discontinuities and mixed IBCs are useful for various applications, such as diffusion curves \cite{Orzan2008}.

The key attribute of our IBC approach that allows the above flexibility for BC types is the introduction of new DOFs near $\C$. This idea shares conceptual similarities with virtual node algorithms \cite{molino2004virtual}, which have been used for codimension-zero problems \cite{Bedrossian2010,Hellrung2012,azevedo2016preserving}. It is also similar to the CPM work of ~\citet{cheung2015localized}, who used new DOFs near sharp features of $\S$ (albeit with the radial-basis function discretization of CPM).

\subsection{Efficiency of CPM}
\label{sec:cpm-related-works-efficiency}
CPM involves constructing a computational domain $\Omega(\S)$ in the embedding space $\mathbb{R}^d$ surrounding $\S$. Linear systems resulting from the PDE discretization on $\Omega(\S)$ must then be solved. For large systems (usually resulting from problems with $d\geq 3$) memory consumption is dominated by the storage of $\Omega(\S)$. However, computation time is dominated by the linear system solve. 

CPM naturally allows $\Omega(\S)$ to occupy only a narrow tubular region of the embedding space near $\S$, analogous to narrow banding for level-set techniques \cite{Adalsteinsson1995}. Therefore, the number of unknowns scales with ${\rm dim}(\S)$ rather than $d$. Note that ${\rm dim}(\S) \leq d$ for manifold PDEs. The linear system solve will be faster with fewer unknowns, so it is important that the construction of the computational domain be carried out local to $\S$ only. \citet{Ruuth2008} used a simple procedure to construct $\Omega(\S)$ that involved storing a uniform grid in a bounding box of $\S$ and computing the closest point for every grid point in the bounding box. Finally, an indexing array was used to label which grid points are within a distance $r_{\Omega(\S)}$ of $\S,$ where $r_{\Omega(\S)}$ is the computational tube-radius (see~\eqref{eqn:bandwidth}). 

The procedure of \citet{Ruuth2008} gives linear systems that scale with ${\rm dim}(\S)$, but memory usage and closest point computation still scale with $d$. \citet{Macdonald2010} used a breadth-first-search (BFS), starting at a grid point near $\S$, that allows the number of closest points computed to scale with ${\rm dim}(\S)$. We use a similar BFS when constructing $\Omega(\S)$; see Section~\ref{sec:imp} for details.  However, \citet{Macdonald2010} still required storing the grid in the bounding box of $\S$, while we adopt sparse grid structures which achieve efficient memory use by allocating only grid points of interest instead of the full grid.

\citet{May2020} overcame memory restrictions arising from storing the full bounding-box grid by using domain decomposition to solve the PDE with distributed memory parallelism. The code detailed by \citet{may2022closest} is publicly available but requires specialized hardware to exploit distributed memory parallelism.

\citet{Auer2012} also used specialized hardware, i.e., their CPM-based fluid simulator was implemented on a GPU. However, they employed a two-level sparse block structure for memory-efficient construction of $\Omega(\S)$ that is also suitable for the CPU. A coarse-level grid in the bounding box of $\S$ is used to find blocks of the fine-level grid (used to solve the PDE) that intersect $\S$. Thus, the memory usage to construct the fine-level grid $\Omega(\S)$ scales with ${\rm dim}(\S)$, as desired. The coarse-level grid still scales with $d,$ but does not cause memory issues because its resolution is much lower than the fine-level one. We adopt a similar approach for constructing $\Omega(\S)$, although our implementation is purely CPU-based.

There has also been work on efficient linear system solvers for CPM. \citet{Chen2015} developed a geometric multigrid solver for the manifold screened-Poisson equation. \citet{May2020, may2022closest} proposed Schwarz-based domain decomposition solvers and preconditioners for elliptic and parabolic manifold PDEs. We implement a custom BiCGSTAB solver (with OpenMP parallelism), as detailed in Section~\ref{sec:partially-mat-free-solver}, that avoids explicit construction of the full linear system. Our solver is more efficient, with respect to memory and computation time (see Section~\ref{sec:iterative-solver-comparison}), compared to Eigen's SparseLU and BiCGSTAB implementations \cite{eigenweb}. Moreover, it circumvents the intricacies associated with implementing multigrid or domain decomposition techniques.

\section{Closest Point Method and Exterior Boundary Conditions}
\label{sec:cpm}

\subsection{Continuous Setting}
\label{sec:cpm-continuous}
Consider a manifold $\S$ embedded in $\mathbb{R}^d,$ where $d \geq {\rm dim}(\S)$. The closest point method uses a closest point (CP) representation of $\S$, which is a mapping from points $\x\in \mathbb{R}^d$ to points $\cp_{\S}(\x)\in \S$. The point $\cp_{\S}(\x)$ is  defined as the closest point on $\S$ to $\x$ in Euclidean distance, i.e.,  
\begin{equation*}
\cp_{\S}(\x) = \argmin_{\mathbf{y}\in\S} \|\x-\y\|.
\end{equation*}
A CP representation can be viewed as providing both implicit and explicit representations. The mapping $\cp_{\S}: \mathbb{R}^d \rightarrow \S$ represents $\S$ implicitly: a traditional scalar (though unsigned) implicit manifold can be recovered by computing the distance ${\|\x - \cp_{\S}(\x)\|}$. Meanwhile, the closest points themselves give an explicit representation of $\S$, albeit without connectivity (i.e., a point cloud).

CPM embeds the manifold problem into the space surrounding $\S$. Consider a tubular neighbourhood defined as 
\begin{equation*}
\N(\S) = \left\{\x\in\mathbb{R}^d \;\Big|\; \|\x - \cp_{\S}(\x)\| \leq r_{\N(\S)}\right\},
\end{equation*}
\begin{wrapfigure}{r}{0.12\textwidth} 
 \vspace{-8pt}
    \centering
    \hspace*{-1\columnsep}
        \begin{minipage}[b]{0.15\textwidth}
            \includegraphics{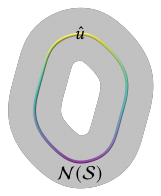}
         \end{minipage}
         \centering
         \hspace*{-1\columnsep}
         \begin{minipage}[b]{0.15\textwidth}
            \includegraphics{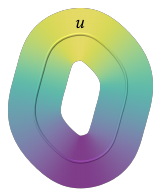}
         \end{minipage}
\end{wrapfigure}
where $r_{\N(\S)}$ is called the \emph{tube radius}. The inset (top) shows an example of a tube $\N(\S)$ (gray) around a 1D curve $\S$ (coloured) embedded in $\mathbb{R}^2$. To solve manifold PDEs with CPM an {\it embedding PDE} is constructed on $\N(\S)$, whose solution agrees with the solution of the manifold PDE at points $\y\in\S$. Let $\hat{u}(\y),$ for $\y\in\S,$ and $u(\x)$, for $\x\in \N(\S),$ denote the solutions to the manifold PDE and embedding PDE, respectively. Fundamentally, CPM is based on extending manifold data $\hat{u}$ from $\S$ onto $\N(\S)$ such that the data is constant in the normal direction of $\S$. This task is accomplished using the {\it closest point extension}, which is the composition of $\hat{u}$ with $\cp_{\S}$, i.e., we take $u(\x) = \hat{u}(\cp_{\S}(\x))$ for all $\x\in \N(\S)$. The inset (bottom) visualizes $u\in \N(\S)$ resulting from the CP extension of $\hat{u}\in\S$ (inset, top). 

Crucially, \citet{Ruuth2008} observed that this extension allows manifold differential operators $\L_{\S}$ on $\S$ to be replaced with Cartesian differential operators $\L$ on $\N(\S)$. Since the function $u$ on $\N(\S)$ is constant in the normal direction, $u$ only changes in the tangential direction of $\S$. Hence, Cartesian gradients on $\N(\S)$ are equivalent to manifold gradients for points on the manifold. By a similar argument, manifold divergence operators can be replaced by Cartesian divergence operators on $\N(\S)$. Higher order derivatives are handled by combining these gradient and divergence principles with CP extensions onto $\N(\S)$.

In this section, we illustrate CPM for solving the manifold Poisson equation $\Delta_{\S} \hat{u} = \hat{f}$, with the embedding PDE $\Delta \hat{u}(\cp_{\S}(\x)) = \hat{f}(\cp_{\S}(\x))$ or equivalently $\Delta u(\x) = f(\x).$ (Technically, this embedding PDE is ill-posed because $f(\x)$ is constant in the normal direction of $\S$, but $\Delta u(\x)$ is not. It is used here for ease of exposition. \citet[Section 2.3]{Chen2015} and \citet{Macdonald2011} discuss the well-posed version which modifies $\Delta u(\x)$. The well-posed version is used in our numerical examples, see Section~\ref{sec:op-dis}.)

\subsection{Discrete Setting}
\label{sec:discrete_setting}
In the discrete setting, the computational domain is a collection of Cartesian grid points $\Omega(\S) \subseteq \N(\S)$ with uniform spacing $\Delta x$. The closest point $\cp_{\S}(\x_i)$ to each grid point $\x_i \in \Omega(\S)$ is computed and stored. Discrete approximations of the CP extension and differential operators are needed to solve the embedding PDE. For our example Poisson equation, $\Delta u(\x) = f(\x),$ we need to approximate the CP extensions $u(\x) = \hat{u}(\cp_{\S}(\x))$ and $f(\x) = \hat{f}(\cp_{\S}(\x))$, as well as the Laplacian $\Delta$. Interpolation is used to approximate the CP extension and finite-differences (FDs) are used for differential operators.

The CP extension requires interpolation since $\cp_{\S}(\x_i)$ is generally not a grid point in $\Omega(\S)$. Thus, the manifold value $\hat{u}(\cp_{\S}(\x_i))$ is approximated by interpolating from discrete values $u_i \approx u(\x_i)$ stored at grid points $\x_i \in \Omega(\S)$ surrounding $\cp_{\S}(\x_i)$. The interpolation degree should be chosen such that interpolation error does not dominate the solution. Throughout we use barycentric-Lagrange interpolation with polynomial degree $p$ \cite{Berrut2004}. This is an efficient form of Lagrange interpolation for CPM~\cite[Section 2.5]{Ruuth2008}. (Manifold data given in the manifold PDE problem, e.g., the function $\hat{f}$ or an initial condition for time-dependent problems, is extended onto $\Omega(\S)$ in a different way that depends on the data representation. See Section~\ref{sec:extending-given-data} for details.)

For a given grid point $\x_k \in \Omega(\S),$ we have the following approximation of the closest point extension:
\begin{equation}
    \hat{u}(\cp_{\S}(\x_k)) = u(\x_k) \approx \sum_{j \in \I_k} w_j^k u_j,
    \label{eqn:interp-stencil}
\end{equation}
where $\I_k$ denotes the set of indices corresponding to grid points in the interpolation stencil for the query point $\cp_{\S}(\x_k)$ and $w_j^k$ are the barycentric-Lagrange interpolation weights corresponding to each grid point in $\I_k$.

FD discretizations on $\Omega(\S)$ are used to approximate a Cartesian differential operator $\L$ as 
\begin{equation}
    \L u(\x_i) \approx \sum_{k \in \D_i} l_k^i u_k, 
    \label{eqn:FD-stencil}
\end{equation}
where $\D_i$ denotes the set of indices corresponding to grid points in the FD stencil centred at the grid point $\x_i$. The FD weights are denoted $l_k^i$ for each $\x_k$ with $k\in\D_i$. For example, the common second-order centred-difference for the discrete Laplacian has weights $1/(\Delta x)^2$ if $k\neq i$ and $-2d/(\Delta x)^2$ if $k = i$.

With these CP extension and differential operator approximations, the Laplace-Beltrami operator $\Delta_{\S} \hat{u}$ is approximated on $\Omega(\S)$ as
\begin{equation}
    \Delta_{\S} \hat{u}(\cp_{\S}(\x_i)) \approx \sum_{k \in \D_i} l_k^i \left( \sum_{j \in \I_k} w_j^k u_j \right). 
    \label{eqn:discrete-Laplace-Beltrami}
\end{equation}
Hence, to solve the discrete embedding PDE (for $\Delta_{\S}\hat{u} = \hat{f}$) we form a linear system using the equation 
\begin{equation*}
    \sum_{k \in \D_i} l_k^i \left( \sum_{j \in \I_k} w_j^k u_j \right) = f_i, 
    \label{eqn:discrete-Poisson}
\end{equation*}
to solve for unknowns $u_i$ at grid points $\x_i\in\Omega(\S)$. Finally, the solution to the original manifold PDE can be recovered at any $\y\in\S$ by interpolation as needed. The reader may refer to prior CPM work \cite{Ruuth2008,Macdonald2010,Macdonald2011} for further background.

\subsubsection*{Tube Radius of the Computational Domain}
\label{sec:banding-S}
One could use a grid $\Omega(\S)$ that completely fills $\mathbb{R}^d$, but this choice is inefficient since only a subset of those points (i.e., those near $\S$) affect the numerical solution on the manifold. It is only required that all grid points within the interpolation stencil of any point on the manifold have accurate approximations of the differential operators. Barycentric-Lagrange interpolation uses a hypercube stencil of ${p+1}$ grid points in each dimension. Consider a hyper-cross FD stencil that uses $q$ grid points from the centre of the stencil in each dimension. An upper bound estimate of the computational tube-radius, $r_{\Omega(\S)}$, for the computational domain $\Omega(\S)$ is \cite{Ruuth2008} 
\begin{equation}
r_{\Omega(\S)} = \Delta x \sqrt{(d-1) \left(\frac{p+1}{2} \right) ^2 + \left(q+\frac{p+1}{2}\right)^2}.
\label{eqn:bandwidth}
\end{equation}
Therefore, our computational domain $\Omega(\S)$ consists of all grid points $\x_i$ satisfying 
$\|\x_i - \cp_{\S}(\x_i)\| \leq r_{\Omega(\S)}.$
Explicit construction of $\Omega(\S)$ is discussed in Section~\ref{sec:comp-dom-setup}.

\subsection{Exterior Boundary Conditions for Open Manifolds}
\label{sec:BCs}
When the manifold $\S$ is open (i.e., its geometric boundary $\partial \S \neq \emptyset$) some choice of boundary condition (BC) must usually be imposed on $\partial \S$ (e.g., Dirichlet, Neumann, etc.). We will refer to these as {\it exterior} boundary conditions. In many applications, however, similar types of boundary conditions may be needed at locations on the \emph{interior} of $\S$, irrespective of $\S$ being open or closed. In this case, {\it interior boundary conditions} (IBCs) should be enforced on a subset $\C \subset \S$, which typically consists of points $\C$ on a 1D curve $\S$, or points and/or curves $\C$ on a 2D surface $\S$. Our proposed approach for IBCs in Section~\ref{sec:interior-constraints} builds on existing CPM techniques for applying exterior BCs at open manifold boundaries, which we review below.

\begin{figure}[b]
\includegraphics{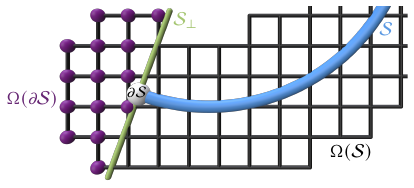}
\caption{The boundary subset $\Omega(\partial \S)$ (purple points) for a curve $\S$ (blue) comprises those grid points in $\Omega(\S)$ (black grid) whose closest point is on the boundary $\partial \S$ (white point). The points $\x_i\in\Omega(\partial \S)$ are those past the normal manifold $\S_{\perp}$ based at $\partial \S$ (green).}
\label{fig:Boundary}
\end{figure}

A subset $\Omega(\partial \S) \subset \Omega(\S)$ of grid points called the boundary subset is used to enforce exterior BCs. It consists of all $\x_i$ satisfying $\cp_{\S}(\x_i) \in \partial \S$, i.e., grid points whose closest manifold point is on the boundary of $\S$. Equivalently, 
\begin{equation}
\Omega(\partial \S) = \big\{\x_i \in \Omega(\S) \;\big|\; \cp_{\S}(\x_i) = \cp_{\partial \S}(\x_i) \big\},
\label{eqn:boundary_set}
\end{equation}
where $\cp_{\partial \S}$ is the closest point function to $\partial \S$.
Geometrically, $\Omega(\partial \S)$ is a half-tubular region of grid points past $\partial \S$, halved by the manifold orthogonal to $\S$ at $\partial \S$ defined by
\begin{equation}
 \S_{\perp} = \{\x\in \N(\S) \;|\; \x = \y + t\, \n_{\S}(\y), \;\y\in\partial\S, \; |t| \leq r_{\Omega(\S)} \},
 \label{eqn:Sperp}
\end{equation} 
when $\S$ is codimension one. The manifold normal at $\y\in\partial \S$ is defined as the limiting normal $\n_{\S}(\y) = \lim_{\mathbf{z}\rightarrow\y}\n_{\S}(\mathbf{z})$, where $\mathbf{z}\in\S$ and $\n_{\S}(\z)$ is the unit normal of $\S$ at $\z$. Figure~\ref{fig:Boundary} illustrates this for a 1D curve embedded in $\mathbb{R}^2$.

CPM naturally applies first-order homogeneous Neumann BCs, $\nabla_{\S} \hat{u} \cdot \n_{\partial \S} = 0$, where $\n_{\partial \S}$ is the unit conormal of $\partial \S$. The conormal is a vector normal to $\partial \S$, tangential to $\S$, and oriented outward \cite{dziuk2007surface}. Therefore, $\n_{\partial \S}(\y) \neq \n_{\S}(\y)$ for $\y\in\partial\S,$ and $\n_{\partial \S}(\y)$ is orthogonal to $\n_{\S}(\y)$ since $\n_{\partial \S}(\y)$ is in the tangent space of $\S$. The CP extension propagates manifold data constant in both $\n_{\S}$ and $\n_{\partial \S}$ at $\partial \S$. Hence, finite differencing across the boundary subset $\Omega(\partial \S)$ will measure zero conormal derivatives \cite{Ruuth2008} and the discretization of the manifold differential operator can be used without any changes at $\x_i \in \Omega(\partial \S)$.   

However, to enforce first-order Dirichlet BCs on $\partial \S$, the CP extension step must be changed. The prescribed Dirichlet value at the closest point of $\x_i \in \Omega(\partial \S)$ is extended to $\x_i$ (instead of the interpolated value in~\eqref{eqn:interp-stencil}). That is, the CP extension assigns $u_i = \hat{u}(\cp_{\S}(\x_i))$ for all $\x_i \in \Omega(\partial \S)$, where $\hat{u}(\cp_{\S}(\x_i))$ is the Dirichlet value at $\cp_{\S}(\x_i)\in \partial \S$. Only this extension procedure changes; the FD discretization is unchanged for all exterior BC types and orders. 

For improved accuracy, second-order Dirichlet and zero-Neumann exterior BCs were introduced by \citet{Macdonald2011} using a simple modification to the closest point function.
The closest point function is replaced with
\begin{equation}
    \overline{\cp}_{\S}(\x) = \cp_{\S}(2\cp_{\S}(\x) - \x).
    \label{eqn:cp-bar}
\end{equation}
Effectively, rather than finding the closest point, this expression determines a ``reflected'' point, and returns \emph{its} closest point instead.

Observe that $\overline{\cp}_{\S}$ satisfies $\overline{\cp}_{\S}(\x_j) = \cp_{\S}(\x_j)$ if $\x_j\not\in\Omega(\partial\S)$
\begin{wrapfigure}{r}{0.155\textwidth} 
\vspace{-12pt}
    \centering
    \hspace*{-1\columnsep}
    \includegraphics{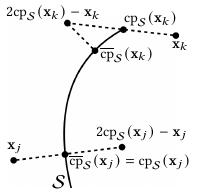}
\vspace{-16pt}
\end{wrapfigure}
(and $\cp_{\S}(\x)$ is unique). Therefore, no change occurs to CPM on the interior of $\S$ (see inset, bottom), so we continue to use $\cp_{\S}(\x)$ for $\x\in\Omega(\S)\setminus\Omega(\partial \S)$. However, for boundary points $\x_k \in \Omega(\partial\S)$, we have $\overline{\cp}_{\S}(\x_k) \neq \cp_{\S}(\x_k),$ since $\overline{\cp}_{\S}(\x_k)$ is a point on the interior of $\S$ while $\cp_{\S}(\x_k)$ is a point on $\partial \S$ (see inset, top). Hence, for a flat manifold, $\hat{u}(\overline{\cp}_{\S}(\x_k))$ gives the interior mirror value for $\x_k$. For a general, curved manifold $\hat{u}(\overline{\cp}_{\S}(\x_k))$ gives an approximate mirror value.

Thus, replacing $\cp_{\S}$ with $\overline{\cp}_{\S}$ will naturally apply second-order homogeneous Neumann exterior BCs: approximate mirror values are extended to $\x_k \in  \Omega(\partial\S)$, so the effective conormal derivative becomes zero at $\partial \S$. This approach generalizes  popular methods for codimension-zero problems with embedded boundaries, where mirror values are also assigned to ghost points  (see e.g., Section 2.12 of \cite{Leveque2007}). In practice, the only change required is to replace $\I_k$ and corresponding weights in~\eqref{eqn:interp-stencil} with those for $\overline{\cp}_{\S}(\x_k)$.

Second-order Dirichlet exterior BCs similarly generalize their codimension-zero counterparts, e.g., the ghost fluid method \cite{Gibou2002} that fills ghost point values by linear extrapolation. The CP extension at $\x_k \in \Omega(\partial\S)$ becomes $u(\x_k) = 2 \hat{u}(\cp_{\S}(\x_k)) - u(\overline{\cp}_{\S}(\x)),$ where $\hat{u}(\cp_{\S}(\x_k))$ is the prescribed Dirichlet value on $\partial \S$. Hence, for $\x_k \in \Omega(\partial\S)$ we change~\eqref{eqn:interp-stencil} to 
\begin{equation}
u_k = 2\hat{u}(\cp_{\S}(\x_k)) - \sum_{j \in \overline{\I}_k} \overline{w}_j^k u_j,
\label{eqn:exterior-2nd-order-Dirichlet}
\end{equation}
where $\overline{\I}_k$ and $\overline{w}_j^k$ are the interpolation stencil indices and weights for $\overline{\cp}_{\S}(\x_k)$, respectively.

Remark that $\S$ can have multiple boundaries, so there may be multiple $\Omega(\partial \S)$ regions where this BC treatment must be applied.

\section{Interior Boundary Conditions}
\label{sec:interior-constraints}
As discussed in Section~\ref{sec:cpm}, the discrete setting of CPM involves two main operations: interpolation for CP extensions and finite differences (FDs) for differential operators. Exterior BCs are handled by modifying the CP extension interpolation while keeping the finite differencing the same (Section~\ref{sec:BCs}). Below we describe our proposed technique to extend CPM with support for  interior BCs, which consists of two key changes: adding new degrees of freedom (DOFs) and carefully altering both the interpolation and FD stencils.

\begin{figure}
    \centering
     \begin{subfigure}[b]{0.48\textwidth}
         \begin{subfigure}[b]{0.48\textwidth}
             \centering
            \includegraphics[scale=0.98]{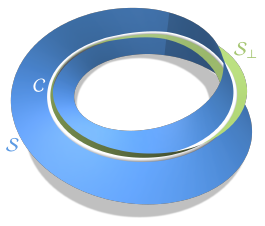}
         \end{subfigure}
         \hspace{1pt}
         \begin{subfigure}[b]{0.48\textwidth}
             \centering
             \includegraphics[width=0.98\textwidth]{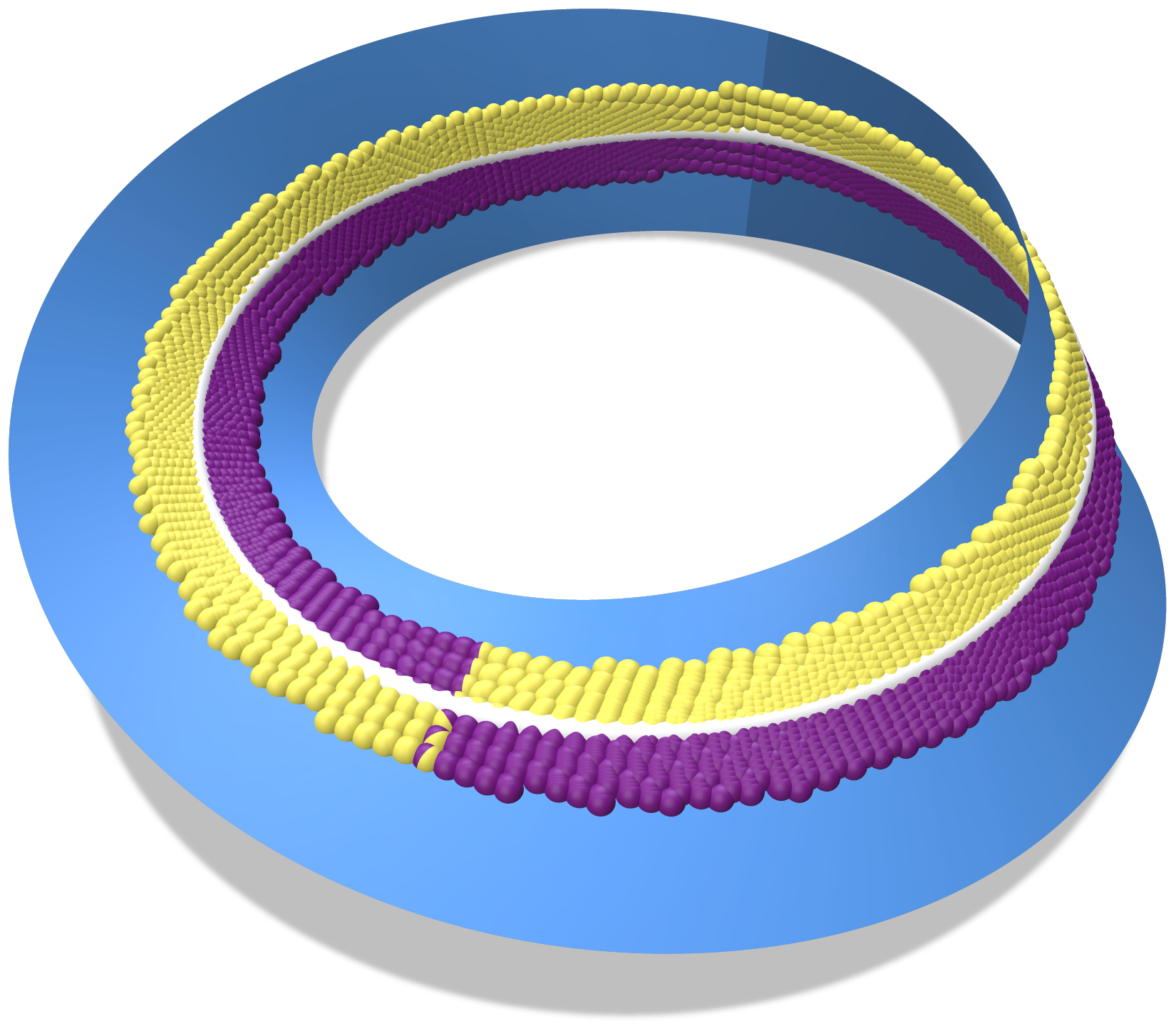}
        \end{subfigure}
    \end{subfigure}
\caption{On the left, a normal manifold $\S_\perp$ (green) extends perpendicularly outwards from a curve $\C$ (white) where an IBC is to be applied. On the right, closest points $\cp_{\S}(\x_i)$ for $\x_i \in \Omega(\C)$ (yellow and purple) cannot be globally partitioned into two disjoint sets by $\C$ on a nonorientable $\S$ (blue).}
\label{fig:nonorientable-partition}
\end{figure}

Table~\ref{tab:symbols} summarizes important notation.  For the rest of this paper we focus on the cases where the manifold $\S$ is a curve embedded in $\mathbb{R}^2$ or a surface embedded in $\mathbb{R}^3.$ Let $\C\subset \S$ denote the interior region where the BC is to be applied, which can be a point (in 2D or 3D) or an open or closed curve (in 3D). Since CPM is an embedding method we must consider the influence of $\C$ on the embedding space $\N(\S)$. Let $\S_{\perp}$ denote a (conceptual) manifold orthogonal to $\S$ along $\C$, i.e., analogous to $\S_{\perp}$ defined in~\eqref{eqn:Sperp} for the exterior boundary case, but with $\partial \S$ replaced by $\C$. See Figure~\ref{fig:nonorientable-partition} (left) for an example curve $\C$ on a surface $\S$ and its normal manifold $\S_{\perp}$ at $\C$.

\begin{table}
    \caption{A summary of symbols used in this paper.}
    \centering
    \small
        \begin{tabular}{ll}
            \hline
            {\bf Symbol} & {\bf Description} \\\hline
            \myrc $\S$ & Manifold \\
            $\C$ & Subset of $\S$ where IBC is enforced \\
            \myrc ${\rm dim(\S)}$ & Dimension of manifold $\S$\\
            $d$ & Dimension of embedding space surrounding $\S$\\
            \myrc $\hat{u}$ & Manifold intrinsic function \\
            $u$ & Function in embedding space $\mathbb{R}^d$\\
            \myrc $\N(\S)$ & Tubular neighbourhood surrounding $\S$ \\
            $\n_{\S}$ & Unit manifold normal vector\\
            \myrc $\n_{\partial \S}$ & Unit conormal vector along $\partial \S$\\
             $\S_{\perp}$ & Manifold orthogonal to $\S$ along $\C$\\
             \myrc $\cp_{\S}(\x)$ & Closest point in $\S$ to $\x \in \mathbb{R}^d$\\
             $\cp_{\C}(\x)$ & Closest point in $\C$ to $\x \in \mathbb{R}^d$\\ 
            \myrc  $\cp_{\S-\C}(\x)$ & Difference between closest point to $\S$ and $\C$\\
             $\Omega(\S)$ & Grid surrounding  $\S$ (subset of $\N(\S)$)\\ 
             \myrc $\Omega(\C)$ & Interior boundary subset of $\Omega(\S)$\\         
            $\Omega(\partial \S)$ & (Exterior) boundary subset of $\Omega(\S)$ \\
            \myrc $\Omega(\partial \C)$ & Boundary subset of interior boundary subset $\Omega(\C)$\\
            $r_{\N(\S)}$ & Tube radius of $\N(\S)$\\
            \myrc $r_{\Omega(\S)}$ & Computational tube-radius\\
            $N_{\S}$ & Number of grid points in $\Omega(\S)$\\ 
            \myrc $N_{\C}$ & Number of grid points in $\Omega(\C)$\\
            $J_{\S}$ & Set of indices for $\x_i\in\Omega(\S)$\\ 
            \myrc $J_{\C}$ & Set of indices for $\x_{\alpha}\in\Omega(\C)$\\
            $i$ & Index in $J_{\S}$\\ 
            \myrc $\alpha$ & Index in $J_{\C}$\\            
            $\x_i$ & Grid point in $\Omega(\S)$ \\
            \myrc $\x_{\alpha}$ & Grid point in $\Omega(\C)$\\
            $\D_i$ & Indices of grid points in finite-difference stencil of $\x_i$ \\
            \myrc $\I_i$ & Indices of grid points in interpolation stencil of $\cp_{\S}(\x_i)$\\
            \hline
        \end{tabular}
    \label{tab:symbols}
\end{table}

\subsection{Adding Interior Boundary DOFs}

Exterior BCs incorporate the BC using grid points $\x_i \in \Omega(\partial \S)$ as defined in~\eqref{eqn:boundary_set}. These grid points $\x_i \in \Omega(\partial \S)$ are only needed to enforce the exterior BC since they lie on the opposite side of $\S_{\perp}$ from $\S$. Therefore, CP extension stencils for $\x_i \in \Omega(\partial \S)$ can be safely modified to enforce exterior BCs. 

For interior BCs, the situation is more challenging. Similar to $\Omega(\partial \S)$, a new \emph{interior} boundary subset $\Omega(\C) \subset \Omega(\S)$ is defined as
\begin{equation}
\Omega(\C) = \{\x_i \in \Omega(\S) \;|\; \|\x_i - \cp_{\C}(\x_i) \| \leq r_{\Omega(\S)} \},
\end{equation}
where $\cp_{\C}$ is the closest point function of $\C$. Comparing with \eqref{eqn:boundary_set}, the subsets $\Omega(\partial \S)$ and $\Omega(\C)$ are defined in the same way, except $\Omega(\partial \S)$ has the extra property $\cp_{\S}(\x_i) = \cp_{\partial \S}(\x_i)$ for all $\x_i \in \Omega(\partial \S)$; i.e., points in the exterior boundary subset have a closest manifold point that is \emph{also} their closest boundary point. Grid points in the interior boundary subset do not: $\x_i \in \Omega(\C)$ will in general have $\cp_{\S}(\x_i) \neq \cp_{\C}(\x_i)$ unless the point $\x_i \in \S_{\perp}$. 

Ideally, we would use the grid points $\x_i \in \Omega(\C)$ to enforce the IBC, analogous to the exterior case. 
However, the tubular volume surrounding $\C$, $\{\x\in\N(\S) \;|\; \|\x-\cp_{\C}(\x)\| \leq r_{\Omega(\S)}\},$
which contains $\Omega(\C)$, also intersects with $\S$. 
Therefore, we cannot simply repurpose and modify CP extension stencils for $\x_i \in \Omega(\C)$, since they are needed to solve the manifold PDE on $\S \setminus \C$.

\begin{figure*}
\centering
     \begin{subfigure}[b]{\textwidth}
         \begin{subfigure}[b]{0.32\textwidth}
            \includegraphics{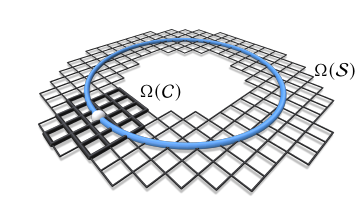}
         \end{subfigure}
         \hspace{-5pt}
         \begin{subfigure}[b]{0.32\textwidth}
            \includegraphics{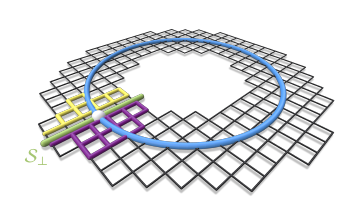}
         \end{subfigure}
          \hspace{-6pt}
         \begin{subfigure}[b]{0.32\textwidth}
            \includegraphics{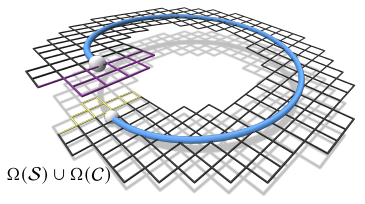}
         \end{subfigure}
     \end{subfigure}
     \hfill
          \begin{subfigure}[b]{\textwidth}
         \begin{subfigure}[b]{0.32\textwidth}
            \includegraphics{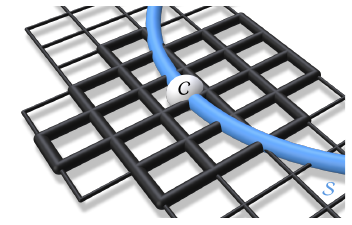}
         \end{subfigure}
         \begin{subfigure}[b]{0.32\textwidth}
            \includegraphics{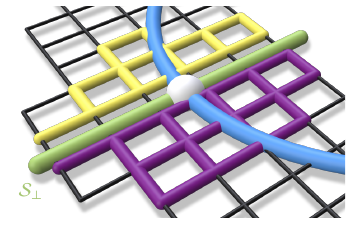}
         \end{subfigure}
         \begin{subfigure}[b]{0.32\textwidth}
            \includegraphics{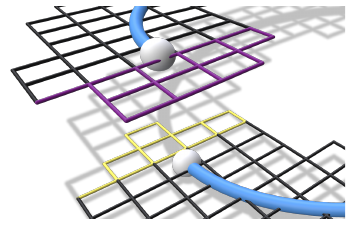}
         \end{subfigure}
     \end{subfigure}
\caption{A conceptual illustration of our approach to interior boundaries for a point $\C$ (white) on a curve $\S$ (blue) in $\mathbb{R}^2$. Left column: Duplicated BC DOFs are generated in the boundary subset $\Omega(\C)$ around $\C$ (thick black grid). Middle column: The normal manifold $\S_\perp$ (green) locally partitions the grid into two sides (yellow, purple). Right column: The modified grid connectivity is illustrated by warping it into $\mathbb{R}^3$.}
\label{fig:nonmanifold-grid}
\end{figure*}

We propose to add a second set of spatially colocated DOFs, called the {\it BC DOFs}, at all $\x_i \in \Omega(\C)$. The BC DOFs allow us to apply similar techniques for interior BCs as was done for exterior BCs. 
Specifically, given a computational domain $\Omega(\S)$ of $N_{\S}$ grid points and the subset $\Omega(\C)$ of $N_{\C}$ grid points, the discrete linear system to be solved will now involve $N_{\S} + N_{\C}$ DOFs. We order the BC DOFs after the original {\it PDE DOFs}. That is, indices in the set $J_{\S} = \{j \in \mathbb{N}\;|\; 0 \leq j < N_{\S} \}$ give $\x_j\in\Omega(\S)$ while indices in the set $J_{\C} = \{\alpha \in \mathbb{N}\;|\; N_{\S} \leq \alpha < N_{\S} + N_{\C} \}$ give $\x_{\alpha} \in \Omega(\C).$ Throughout we use Greek letters to denote indices in $J_{\C}$ to clearly distinguish from indices in $J_{\S}.$ Note that for every BC DOF $\alpha \in J_{\C}$ there is a corresponding PDE DOF $j\in J_{\S}$ such that $\x_{\alpha} = \x_j.$ The key question then becomes: when do we use PDE DOFs versus BC DOFS? 

Intuitively, the answer is simple: interpolation and FD stencils ($\I_{i}$ and $\D_{i}$ from~\eqref{eqn:interp-stencil} and~\eqref{eqn:FD-stencil}) must only use manifold data $\hat{u}$ from the same side of $\S_{\perp}$ that the stencil belongs to. Therefore, if a stencil involves manifold data on the opposite side of $\S_{\perp}$, the IBC must be applied using the BC DOFs. 

Figure~\ref{fig:nonmanifold-grid} gives a conceptual illustration of the process for a point $\C$ on a circle $\S$ embedded in $\mathbb{R}^2$. Both BC DOFs and PDE DOFs are present in the region of $\Omega(C)$. The BC DOFs are partitioned into one of two sets depending on which side of $\S_{\perp}$ the closest point $\cp_{\S}(\x_i)$ is on. The original grid $\Omega(\S)$ and duplicated portion $\Omega(\C)$ are cut, and each half of $\Omega(\C)$ is joined to the opposing side of $\Omega(\S)$.

The same treatment of BCs as in the exterior case is then applied on this nonmanifold grid $\Omega(\S) \cup \Omega(\C)$. That is, the required modifications to the CP extension interpolation stencils in Section~\ref{sec:BCs} are applied. Unlike the exterior BC case, however, changes to FD stencils do occur for IBCs since $\Omega(\C)$ and $\Omega(\S)$ are cut and joined to opposite sides of each other. 

If $\S$ is {\it orientable} then this intuitive picture in Figure~\ref{fig:nonmanifold-grid} is an accurate depiction of the necessary grid connectivity. That is, near $\C$ we must duplicate DOFs and cut and join opposite pieces of $\Omega(\S)$ and $\Omega(\C)$ to produce regions (similar to $\Omega(\partial \S)$) where BCs can be imposed. However, if $\S$ is nonorientable the closest points $\cp_{\S}(\x_i)$ for $\x_i\in\Omega(\C)$ cannot be globally partitioned into two sides. For example, on the M\"obius strip in Figure~\ref{fig:nonorientable-partition} (right), an apparent flip in the partitioning of $\cp_{\S}(\x_i)$ is unavoidable as one moves along a curve $\C$ that loops around the whole strip.

Fortunately, IBCs can still be enforced on nonorientable manifolds because the manifold can be oriented \emph{locally}. The interpolation and FD stencils only perform operations in a small local region of $\Omega(\S)$, so locally orienting the manifold is sufficient to enforce IBCs.

\subsection{$\S_{\perp}$ Crossing Test}
\label{sec:sperp-crossing-test}
We must keep computation local to each stencil to handle nonorientable manifolds. Therefore, first consider testing if any two closest points of $\x_1, \x_2 \in \N(\S)$ are on opposite sides of $\S_{\perp}$. A naive approach would be to construct $\S_{\perp}$ explicitly, e.g., with a surface triangulation (as was done by \citet{Shi2007}), and then test if the line segment between $\cp_{\S}(\x_1)$ and $\cp_{\S}(\x_2)$ intersects the triangulation. However, building an explicit surface is counter to the implicit spirit of CPM. 

Determining if $\cp_{\S}(\x_1)$ and $\cp_{\S}(\x_2)$ are on opposite sides of $\S_{\perp}$ can instead be accomplished based on closest points on $\C$. Let $\cp_{\C}(\x_1)$ and $\cp_{\C}(\x_2)$ be the closest points to $\x_1$ and $\x_2$ on $\C$, respectively.  
Define the vector $\cp_{\S-\C}(\x)$ as 
\begin{equation}
    \cp_{\S-\C}(\x) \equiv \cp_{\S}(\x) - \cp_{\C}(\x).
    \label{eqn:cpdiff}
\end{equation}
Denote the locally-oriented unit normal to $\S_{\perp}$ at $\y\in\C$ as $\n_{\S_{\perp}}(\y)$. The function
\begin{equation}
    F(\x) \equiv \cp_{\S-\C}(\x) \cdot \n_{\S_{\perp}} (\cp_{\C}(\x)) 
    \label{eqn:direct_test}
\end{equation}
 will have different signs for $F(\x_1)$ and $F(\x_2)$ if $\cp_{\S}(\x_1)$ and $\cp_{\S}(\x_2)$ are on different sides of $\S_{\perp}$, or equivalently $F(\x_1) F(\x_2) < 0$. However, this direct test would require computing $\n_{\S_{\perp}}$ along $\C$ and locally orienting that normal vector. 

\begin{figure}
\centering
\includegraphics{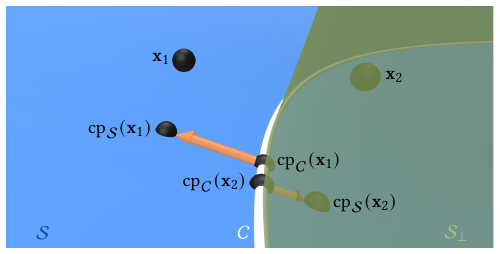}
\caption{For two points $\x_1, \x_2 \in \N(\S)$, we can determine if the closest points, $\cp_{\S}(\x_1)$, $\cp_{\S}(\x_2)$, lie on opposite sides of $\C$ based on their orientations relative to the corresponding closest points on $\C$, $\cp_{\C}(\x_1)$, $\cp_{\C}(\x_2)$.}
\label{fig:CP-diff}
\end{figure}

Instead of checking the directions $\cp_{\S-\C}$ relative to the locally oriented normals $\n_{\S_{\perp}}$, we can check the directions of $\cp_{\S-\C}(\x_1)$ and $\cp_{\S-\C}(\x_2)$ relative to each other. As illustrated in Figure~\ref{fig:CP-diff}, if $\cp_{\S}(\x_1)$ and $\cp_{\S}(\x_2)$ are on opposite sides of $\S_{\perp}$ the associated $\cp_{\S-\C}(\x)$ vectors will point in opposing directions; thus, we can simply check if their dot product is negative:
\begin{equation}
\cp_{\S-\C}(\x_1) \cdot \cp_{\S-\C}(\x_2) < 0.
\label{eqn:two-point-crossing-test}
\end{equation}
In practice, we find~\eqref{eqn:two-point-crossing-test} sufficient to obtain second-order accuracy in the convergence studies of Section~\ref{sec:conv-studies} on smooth $\S$ and $\C$.

When $\x$ is close to $\S_{\perp}$ the vector $\cp_{\S-\C}(\x) \approx \mathbf{0},$ which can result in an inaccurate classification of which side $\cp_{\S}(\x)$ is on. Therefore, if $\|\cp_{\S - \C}(\x)\| = \mathcal{O}(\Delta x^2)$ the point $\cp_{\S}(\x)$ is considered to lie on $\C$ and can be safely assigned to either side, while maintaining second-order accuracy. In practice, we consider $\cp_{\S}(\x)$ to lie on $\C$ if $\|\cp_{\S - \C}(\x)\| < 0.1\Delta x^2$.

As we have noted, the locality of this $\S_{\perp}$ crossing test allows it to handle nonorientable manifolds with CPM and IBCs. 
However, on orientable manifolds one can still globally orient stencils in $\Omega(\C)$ to impose different values or types of IBCs on either side of $\C$. For example, different prescribed Dirichlet values on each side of $\C$ are useful for vector field design. Mixing Dirichlet and Neumann  IBCs on $\C$ in this way can also be useful for diffusion curves.

\subsection{Stencil Modifications}
\label{sec:IBCs}
In this section, we describe how to use the $\S_{\perp}$ crossing test to impose IBCs by altering interpolation and FD stencils. The $\S_{\perp}$ crossing test~\eqref{eqn:two-point-crossing-test} allows us to determine if any two points $\x_1,\x_2 \in \N(\S)$ have closest points $\cp_{\S}(\x_1), \; \cp_{\S}(\x_2)$ on opposite sides of $\S_{\perp}$. Ultimately, we employ this test to determine if the closest points $\cp_{\S}(\x_j)$ for $j \in \I_i \; {\rm or} \; \D_i$ are on the opposite side of $\S_{\perp}$ relative to a stencil for $\x_i$, so the stencil can use the correct PDE vs.\ BC data.

A stencil is itself assigned to a particular side of $\S_{\perp}$  based on the location of an associated point on $\S$ that we call the {\it stencil director}, denoted $\y^{\star}$. For the FD stencil of $\x_i$ the stencil director is $\y^{\star}_i = \cp_{\S}(\x_i)$, since grid data at $\x_i$ corresponds to manifold data at $\cp_{\S}(\x_i)$. For the interpolation stencil of $\x_i$ used for the CP extension, the stencil director is the interpolation query point $\cp_{\S}(\x_i)$, i.e., $\y^{\star}_i = \cp_{\S}(\x_i)$. Each stencil director also has a corresponding {\it stencil direction} denoted $\mathbf{d}^{\star}.$ For FD and CP extension interpolation stencils $\mathbf{d}^{\star}_i = \cp_{\S - \C}(\x_i) = \y^{\star}_i - \cp_{\C}(\x_i).$

It is, however, not always the case that $\y^{\star}_i = \cp_{\S}(\x_i).$ Interpolation of the solution on the grid $\Omega(\S) \cup \Omega(\C)$ can also be used to obtain the final solution at \emph{any} set of manifold points. For example, if one desires to transfer the solution to a mesh or a point cloud (e.g., for display or downstream processing), interpolation can be used to obtain the solution on vertices of the mesh or points in the cloud (see Section~\ref{sec:visualization}). In this case, the stencil director is just the interpolation query point $\y^{\star} = \y_q \in \S$ and the stencil direction is $\mathbf{d}^{\star} = \y^{\star} - \cp_{\C}(\y^{\star})$.

\subsubsection*{PDE DOF Modifications}
The first step to incorporate  IBCs is to alter the stencils for the PDE DOFs in $J_{\S}$. The computation in both~\eqref{eqn:interp-stencil} and~\eqref{eqn:FD-stencil} for $i \in J_{\S}$ has the form
$$u_i = \sum_{j\in\G_i} c^i_j u_j,$$
where $\G_i\subset J_{\S}$ are indices corresponding to grid points in the stencil for $i$ (i.e., $\G_i = \I_i$ or $\G_i = \D_i$) and $c^i_j$ are corresponding weights.

To incorporate IBCs, the index $j\in\G_i$ is replaced with its corresponding BC DOF index $\alpha \in J_{\C}$ if data at $\x_j$ comes from the opposite side of $\S_{\perp}$. The corresponding stencil weight $c^i_j$ remains unchanged. Using the $\S_{\perp}$ crossing test~\eqref{eqn:two-point-crossing-test}, for all $j\in\G_i$, we replace $j\in J_{\S}$ with its corresponding $\alpha\in J_{\C}$ if
\begin{equation}
\mathbf{d}^{\star}_i \cdot \cp_{\S-\C}(\x_j) < 0.
\label{eqn:stencil-crossing-test}
\end{equation}

If our equations are written in matrix form, these modifications to the PDE DOFs above would change $N_{\S} \times N_{\S}$ matrices to be size $N_{\S} \times (N_{\S} + N_{\C})$. The next step is to add the BC equations for the BC DOFs in $J_{\C}$, resulting in square matrices again of size $(N_{\S} + N_{\C}) \times (N_{\S} + N_{\C})$. 

\subsubsection*{BC DOF Modifications}
Finite-difference stencils are added for the BC DOFs with $\alpha\in J_{\C}$ and modified in a similar way to the PDE DOFs above. The same grid connectivity is present in $\Omega(\C)$ as the corresponding portion of $\Omega(\S)$ (except at the boundary of $\Omega(\C)$). Therefore, the same FD stencils on $\Omega(\S)$ are used on $\Omega(\C)$ except with indices $\beta \in J_{\C}$ (and indices not present in $\Omega(\C)$, i.e., grid points in $\Omega(\S)$ around the edge of $\Omega(\C)$, are removed). Hence, using the $\S_{\perp}$ crossing test~\eqref{eqn:two-point-crossing-test} for all $\beta\in\D_{\alpha}$, the index $\beta\in J_{\C}$ is replaced with its corresponding $j\in J_{\S}$ if
\begin{equation}
\mathbf{d}^{\star}_{\alpha} \cdot \cp_{\S-\C}(\x_{\beta}) < 0.
\label{eqn:FD-crossing-test}
\end{equation}

The CP extension BC equations discussed in Section~\ref{sec:BCs} for exterior BCs are used on the BC DOFs with $\alpha \in J_{\C}$. However, first-order zero-Neumann IBCs are no longer automatically imposed as in Section~\ref{sec:BCs}. Instead, for first-order zero-Neumann IBCs, the CP extension extends manifold data $\hat{u}$ at $\cp_{\C}(\x_{\alpha})$ for $\x_{\alpha} \in \Omega(\C)$, i.e., 
\begin{equation*}
    \hat{u}(\cp_{\C}(\x_{\alpha})) = u(\x_{\alpha}) \approx  \sum_{\beta \in \I_{\alpha}} w_{\beta}^{\alpha} u_{\beta}.
    \label{eqn:IBC-1st-order-Neumann}
\end{equation*}
Once again the $\S_{\perp}$ crossing test~\eqref{eqn:two-point-crossing-test} is used to ensure DOFs are used from the correct sides of $\S_{\perp}$. In this case, the stencil director (interpolation query point) is $\y^{\star}_{\alpha} = \cp_{\C}(\x_{\alpha})$, which gives $\mathbf{d}^{\star}_{\alpha} = \mathbf{0}$ since $\y^{\star}_{\alpha}$ is on both $\C$ and $\S$. However, the vector $\mathbf{d}^{\star}_{\alpha} \equiv \cp_{\S}(\x_{\alpha}) - \cp_{\C}(\x_{\alpha})$ gives the correct direction to define which side of $\S_{\perp}$ the interpolation stencil belongs to. Then, for all $\beta\in\I_{\alpha}$, we replace $\beta\in J_{\C}$ with its corresponding $j\in J_{\S}$ if~\eqref{eqn:FD-crossing-test} holds. 

For second-order zero-Neumann IBCs, the only modification required is to replace $\cp_{\C}(\x)$ with
\begin{equation}
    \overline{\cp}_{\C}(\x) = \cp_{\S}(2\cp_{\C}(\x) - \x).
    \label{eqn:cpc-bar}
\end{equation}
Note that~\eqref{eqn:cpc-bar} is different from the form used for exterior BCs in~\eqref{eqn:cp-bar}, as it involves both $\cp_{\S}$ and $\cp_{\C}$. However, the purpose of this modified closest point function~\eqref{eqn:cpc-bar} remains the same, i.e., the point $\overline{\cp}_{\C}(\x)$ is an approximate mirror location.

The CP extension equations for BC DOFs, with $\alpha\in J_{\C},$ to enforce Dirichlet IBCs are analogous to Section~\ref{sec:BCs}. The prescribed Dirichlet value, $\hat{u}$ on $\C$, is extended for first-order Dirichlet IBCs, i.e., $u(\x) = \hat{u}(\cp_{\C}(\x))$ or in the discrete setting $u_{\alpha} = \hat{u}(\cp_{\C}(\x_{\alpha})).$ For second-order Dirichlet IBCs, the extension is $u(\x) = 2 \hat{u}(\cp_{\C}(\x)) - u(\overline{\cp}_{\C}(\x)),$ which becomes analogous to~\eqref{eqn:exterior-2nd-order-Dirichlet} in the discrete setting.  

\subsection{Open Curves $\C$ in $\mathbb{R}^3$}
\label{sec:open-curves}
Past the endpoints of an open curve $\C$ the PDE should be solved without the IBC being enforced. However, the set $\Omega(\C)$ includes half-spherical regions of grid points past the boundary point $\partial \C$. These half-spherical regions are analogous to the exterior boundary subsets $\Omega(\partial \S)$ in Section~\ref{sec:BCs} and are defined as
\begin{equation}
\Omega(\partial \C) = \{\x_{\alpha} \in \Omega(\C) \;|\;  \cp_{\C}(\x_{\alpha}) = \cp_{\partial \C}(\x_{\alpha}) \}.
\label{eqn:IBC_boundary_set}
\end{equation}
We do not perform the modifications of Section~\ref{sec:IBCs} for points $\x_{\alpha}\in\Omega(\partial \C)$ since this would enforce the IBC where only the PDE should be solved. In other words, the BC DOFs in $\Omega({\partial \C})$ are not added to the linear system. 



\subsection{Points $\C$ in $\mathbb{R}^3$}

Remarkably, and unlike for open curves, when $\C$ is a point on $\S$ embedded in $\mathbb{R}^3$ no change to the stencil modification procedure in Section~\ref{sec:IBCs} is needed. To understand why, consider two simpler options. First, without \emph{any} boundary treatment whatsoever near $\C$ the PDE is solved but the IBC is ignored. Second, a naive first-order treatment simply sets either the nearest grid point or a ball of grid points around $\C$ to the Dirichlet value; however, at those grid points the PDE is now ignored. Instead, the grid points near $\C$ should be influenced by the IBC at $\C$, while also satisfying the PDE.

Under the procedure of Section~\ref{sec:IBCs}, the $\cp_{\S-\C}(\x_j)$ and $\mathbf{d}^{\star}_i$ vectors will point radially outward from the point $\C$ (approximately in the tangent space of $\S$ at $\C$). The $\S_{\perp}$ crossing test~\eqref{eqn:two-point-crossing-test} becomes a half-space test, where the plane $P$ partitioning the space goes through $\C$ with its normal given by the stencil direction vector, $\mathbf{d}^{\star}_i.$ In the stencil for $\y^{\star}_i$, points on the same side of $P$ as $\y^{\star}_i$ are treated as PDE DOFs, while points on the opposite side receive the IBC treatment (either first or second-order as desired). However, the direction of $\mathbf{d}^{\star}_i$, and hence the half-space, changes for each grid point's stencil (radially around $\C$). The $\mathbf{d}^{\star}_i$ changes because the location of $\y^{\star}_i$ changes for each $i$ with $\cp_{\C}(\x_i)$ fixed at $\C$. This spinning of $P$ radially around $\C$ allows the PDE and the IBC to be enforced simultaneously since both PDE and IBC equations are added to the linear system for all points $\x_i\in\Omega(\C)$.

Therefore, for a point $\C\in\S\subset \mathbb{R}^3$, our first-order Dirichlet IBC method acts as an improvement of the approach of \citet{Auer2012}, where only points $\x_j\in\Omega(\C)$ on one side of $P$ (which revolves around $\C$) are fixed with the prescribed Dirichlet value. We observe that this reduces the error constant compared to \citet{Auer2012} in convergence studies in Section~\ref{sec:conv-studies}. Furthermore, our approach in Section~\ref{sec:IBCs} allows us to achieve second-order accuracy, whereas the method of \citet{Auer2012} is restricted to first-order accuracy. Neumann IBCs at a point $\C$ are not well-defined since there is no preferred direction conormal to $\C$.

\subsection{Localizing Computation Near $\C$}
Computation to enforce IBCs should only be performed locally around $\C$ for efficiency. The new BC DOFs satisfy this requirement since they are only added at grid points $\x_i$ within a distance $r_{\Omega(\S)}$ of $\C$. This banding of $\Omega(\C)$ is possible for the same reason it is possible to band $\Omega(\S)$ (see Section~\ref{sec:banding-S}): grid points are only needed near $\S$ and $\C$ because accurate approximations of differential operators are only needed at grid points within interpolation stencils.

The use of the $\S_{\perp}$ crossing test~\eqref{eqn:two-point-crossing-test} has been discussed in terms of checking all interpolation and FD stencils in $\Omega(\S)$ and $\Omega(\C)$ above. For efficiency, we would rather only check if $\cp_{\S}(\x_1)$ and $\cp_{\S}(\x_2)$ are on different sides of $\S_{\perp}$ if $\x_1$ and $\x_2$ are near $\C$. However, depending on the geometry of $\S$ and $\C,$ points $\x_i\notin{\Omega(\C)}$ can have stencils for interpolating at $\cp_{\S}(\x_i)$ that cross $\S_{\perp}$, so testing only points  $\x_i\in{\Omega(\C)}$ does not suffice.

We therefore check stencils that include grid points $\x_i\in \Omega(\S)$ with $\|\x_i - \cp_{\C}(\x_i)\| < 2 r_{\Omega(\S)}$ for all the examples in this paper. The closest points $\cp_{\C}(\x_i)$ are needed to compute $\|\x_i - \cp_{\C}(\x_i)\|$. Computation of $\cp_{\C}$ for all $\x_i \in \Omega(\S)$ is avoided using a similar breadth-first search to the one used in the construction of $\Omega(\S)$ (see Algorithm~\ref{alg:BFS} discussed in Section~\ref{sec:imp}). 

\subsection{Improving Robustness of $\S_{\perp}$ Crossing Test}
\label{sec:robust_cp_diff}
In practice, manifolds with small bumps of high curvature relative to the grid resolution can cause the $\S_{\perp}$ crossing test~\eqref{eqn:two-point-crossing-test} to be inaccurate. For example, the headdress of the Nefertiti mesh in Figure~\ref{fig:teaser}(a) has many small bumps, which causes the $\cp_{\S-\C}$ and $\mathbf{d}^{\star}$ vectors to be far from orthogonal to $\S_{\perp}$ and $\C$. The closest points near $\C$ are then misclassified as being on the wrong side of $\S_{\perp}$.

To make~\eqref{eqn:two-point-crossing-test} more robust, we modify the $\cp_{\S-\C}$ and $\mathbf{d}^{\star}$ vectors to be orthogonal to $\S_{\perp}$ and $\C$ before computing the dot product. We illustrate this for a surface (2D manifold) embedded in $\mathbb{R}^3$ throughout this section. For this case, \eqref{eqn:two-point-crossing-test} is used with $\cp_{\S-\C}(\x)$ replaced by
\begin{equation}
    \cp^{\perp}_{\S-\C}(\x) = \left(\mathbf{I} - \n_{\S} \n_{\S}^T - \t_{\C} \t_{\C}^T\right)\cp_{\S-\C}(\x), 
    \label{eqn:projected_cp_diff}
\end{equation}
(and similarly for $\mathbf{d}^{\star}$) where $\mathbf{I}$ is the identity matrix and $\t_{\C}$ is the unit tangent vector along $\C$. The manifold normal $\n_{\S}$ and tangent $\t_{\C}$ are evaluated at $\cp_{\C}(\x)$. Projecting out the $\n_{\S}$ and $\t_{\C}$ components is equivalent to projecting $\cp_{\S - \C}(\x)$ onto $\n_{\S_{\perp}}(\cp_{\C}(\x)).$ Therefore, the $\S_{\perp}$ crossing test~\eqref{eqn:two-point-crossing-test} becomes equivalent to the direct test that checks if $F(\x_1) F(\x_2) < 0$ (see Section~\ref{sec:sperp-crossing-test}), but without needing to orient $\n_{\S_{\perp}}$.
The vectors $\n_{\S}$ and $\t_{\C}$ must be evaluated at $\cp_{\C}(\x)$ since the vector $\cp_{\S-\C}(\x)$ starts at $\cp_{\C}(\x)$ (and goes to $\cp_{\S}(\x)$). When $\C$ is a single point the tangent direction is undefined, so only the $\n_{\S}$ component is projected out in this case. Let us now consider how to compute $\n_{\S}$ and $\t_{\C}$.

For a codimension-one manifold $\S$ the Jacobian of the closest point function, $\J_{\cp_{\S}},$ is the projection operator onto the tangent space of $\S$ for points on the manifold \cite{Marz2012, King2017}. Therefore, for a surface in $\mathbb{R}^3$, the eigenvectors of $\J_{\cp_{\S}}$ are the manifold normal $\n_{\S}$ and two tangent vectors. However, two arbitrary tangent vectors of $\S$ will not suffice; we need the tangent $\t_{\C}$ along $\C$. The curve $\C\in\mathbb{R}^3$ has codimension two. The corresponding Jacobian for $\C$, $\J_{\cp_{\C}}$, is likewise equivalent to a projection operator onto the tangent space of $\C$~\cite{Kublik2016}. However, the eigenvectors of $\J_{\cp_{\C}}$ only provide a unique tangent vector $\t_{\C}$, since the normal and binormal to $\C$ can freely rotate around $\t_{\C}$. Hence, we compute the manifold normal $\n_{\S}$  from the eigendecomposition of $\J_{\cp_{\S}},$ while $\t_{\C}$ is computed from the eigendecomposition of $\J_{\cp_{\C}}.$

Second-order centred FDs in $\Omega(\S)$ are used to compute  $\J_{\cp_{\S}}$. The Jacobian $\J_{\cp_{\S}}$ is only equivalent to the tangent space projection operator at points on $\S$. Therefore, a CP extension must be performed to obtain the projection operator at all points $\x_i \in \Omega(\S)$, i.e., $\J_{\cp_{\S}}(\x_i) = \J_{\cp_{\S}}(\cp_{\S}(\x_i))$. In the discrete setting, the CP extension is computed with the same interpolation discussed in Section~\ref{sec:discrete_setting}. The Jacobian of $\cp_{\C}$ is computed similarly over $\Omega(\C).$

From the above computation of $\J_{\cp_{\S}}$ and $\J_{\cp_{\C}}$, the projection operators are known at points $\cp_{\S}(\x_i)$ and $\cp_{\C}(\mathbf{x}_i)$, respectively. However, since the $\n_{\S}$ vectors are computed from $\J_{\cp_{\S}}$, they are not yet available at $\cp_{\C}(\x_i)$ where we need them. The $\n_{\S}$ vectors are therefore computed at $\cp_{\C}(\x_i)$ via barycentric-Lagrange interpolation (with the same degree $p$ polynomials as the CP extension). Interpolating $\n_{\S}$ vectors requires some care since they are {\it unoriented} manifold normals. We adapt a technique proposed by \citet{Auer2012}: when interpolating $\n_{\S},$ given at points $\x_i\in\Omega(\S),$ we locally orient the vectors within each interpolation stencil by negating vectors satisfying $$\n_{\S}(\x_i) \cdot \n_{\S}(\tilde{\x}) < 0,$$ where $\tilde{\x}$ is a single, fixed grid point in the interpolation stencil.


\subsection{A Nearest Point Approach for Dirichlet IBCs}
\label{sec:nearest-point}
It is also interesting to consider a {\it nearest point} approach for handling Dirichlet IBCs at $\C$, similar to techniques discussed in Section~\ref{sec:IBC_related_work} for other manifold representations. That is, simply fix the grid points $\x_i\in\Omega(\S)$ nearest to $\C$ with the prescribed Dirichlet value, and remove them as DOFs. If $\C$ is a point, a single grid point is assigned the Dirichlet value and removed as a DOF. If $\C$ is a curve, a set of nearest grid points is obtained (i.e., a raster representation of $\C$) and removed as DOFs by assigning Dirichlet values. To our knowledge, this approach has not been used with CPM in any previous work.

This nearest point approach is attractive since new BC DOFs are unnecessary, i.e., $\Omega(\C)$ is not needed. However, it can only be used for Dirichlet IBCs with the same value on both sides of $\C$. That is, two-sided Dirichlet IBCs cannot be imposed with the nearest point approach, nor can Neumann IBCs. The nearest point approach is also only first-order accurate since the nearest point can be $\Delta x \sqrt{d} / 2$ away from $\C$. In Section~\ref{sec:conv-studies}, we observe that the nearest point approach has a better error constant than the method of \citet{Auer2012}, but a similar or worse error constant than our first-order IBC approach above (see Figure~\ref{fig:comparison}(d)). 

\section{Implementation Aspects}
\label{sec:imp}

\subsection{Closest Points and Computational Domain Setup}
\label{sec:comp-dom-setup}
The method of computing closest points, and its cost, will depend on the underlying manifold representation. In Appendix~\ref{sec:cp_computation}, we discuss the computation of closest points for some popular representations, including parameterized manifolds, triangulated surfaces, point clouds, signed-distance functions, and more general level-set functions (i.e., implicit manifolds).

To solve PDEs with CPM, the first step is to construct the computational domain $\Omega(\S)$ around $\S$. 
We use a breadth-first search (BFS) procedure to only compute $\cp_{\S}$ near $\Omega(\S)$. We adopt a sparse-grid data structure and allocate memory for it only as needed during the BFS. The BFS can be started from any grid point $\x_0$ within $r_{\Omega(\S)}$ distance to the manifold. The BFS for $\Omega(\S)$ construction is detailed in Algorithm~\ref{alg:BFS}. A similar BFS to Algorithm~\ref{alg:BFS} is used to construct $\Omega(\C)$ around $\C$. The use of a BFS could fail if $\S$ is composed of disjoint pieces. However, PDEs are only solved on a single, connected manifold throughout this paper. Since IBCs can consist of multiple $\C$, we perform a BFS for each $\C$ independently.

\begin{algorithm}
\caption{BFS to construct $\Omega(\S)$}\label{alg:BFS}
Given $\x_0$ near $\S$, i.e., with $\|\x_0 - \cp_{\S}(\x_0)\|\leq r_{\Omega(\S)}$\\
Add $\x_0$ to $\Omega(\S)$ and store $\cp_{\S}(\x_0)$\\
Add $\x_0$ to the queue $Q$\\
\While{$Q\neq \emptyset$}{
    Set $\x_{{\rm current}} \leftarrow Q$.front()\\
    \For{each neighbour $\x_{{\rm nbr}}$ of $\,\x_{{\rm current}}$}{
        \If{$\x_{{\rm nbr}}$ has not been visited}{
            Compute $\cp_{\S}(\x_{{\rm nbr}})$\\
            \If{ $\|\x_{{\rm nbr}} - {\rm cp}_{\S}(\x_{{\rm nbr}})\| \leq r_{\Omega(\S)}$ }{
                Add $\x_{{\rm nbr}}$ to $\Omega(\S)$ and store $\cp_{\S}(\x_{{\rm nbr}})$\\
                Add $\x_{{\rm nbr}}$ to $Q$\\
            }
        }
    }
    Pop front of $Q$\\
}
\end{algorithm}

The computational tube-radius $r_{\Omega(\S)}$ given by~\eqref{eqn:bandwidth} is an upper bound on the grid points needed in $\Omega(\S)$. The {\it stencil set} approach to construct $\Omega(\S)$ given by \citet{Macdonald2008, Macdonald2010} can reduce the number of DOFs by including only the strictly necessary grid points for interpolation and FD stencils. It was shown by \citet{Macdonald2008} that the reduction in the number of DOFs is between 6-15\% for $\S$ as the unit sphere.  We opted for implementation simplicity over using the stencil set approach due to this low reduction in the number of DOFs.

\subsection{Specifying Initial and Boundary Data}
\label{sec:extending-given-data}
Manifold PDEs generally involve some given data on the manifold, for initial or boundary conditions, that must first be extended onto $\Omega(\S)$ or $\Omega(\C)$. Examples include $\hat{f}$ in Poisson problems $\Delta_{\S} \hat{u} = \hat{f}$, initial conditions $\hat{u}(t = 0)$ for time-dependent problems, or Dirichlet IBC values on $\C$. The necessary extension procedure depends on the specific representation of the manifold and the data, e.g., an analytical function on a parameterization or discrete data on mesh vertices. However, the extension must still be a CP extension: data at $\cp_{\S}(\x_i)$ (or $\cp_{\C}(\x_i)$) is assigned to $\x_i\in\Omega(\S)$ (or $\in\Omega(\C)$).

\subsection{Operator Discretization}
\label{sec:op-dis}
With the initial data on $\Omega(\S)$ and $\Omega(\C)$, the PDE is then discretized using the equations given in Sections~\ref{sec:cpm} and~\ref{sec:interior-constraints}. Matrices  $\mathbf{E}$ and $\mathbf{L}$ are constructed for the CP extension and discrete Laplacian, respectively. The standard 7-point discrete Laplacian in $\mathbb{R}^3$ (5-point in $\mathbb{R}^2$) is used.
In our implementation $\mathbf{E}$ and $\mathbf{L}$ are constructed as discussed by \citet{Macdonald2010}. Constructing the (sparse) matrices amounts to storing stencil weights for DOF $i$ in the columns of row $i$.  Instead of $\tilde{\mathbf{M}} = \mathbf{L} \mathbf{E}$, we use the more numerically stable CPM approximation of the Laplace-Beltrami operator \cite{Macdonald2010, Macdonald2011} 
$$\mathbf{M} = {\rm diag}(\mathbf{L}) + (\mathbf{L} - {\rm diag}(\mathbf{L})) \mathbf{E}.$$

\subsection{Linear Solver}
\label{sec:partially-mat-free-solver}
The linear system resulting from CPM could be solved with direct solvers, e.g., Eigen's SparseLU was used in Section~\ref{sec:conv-discont}, but they are only appropriate for smaller linear systems (usually obtained from 1D curves embedded in $\mathbb{R}^2$). Iterative solvers are preferred for larger linear systems (as noted in \cite{Macdonald2010, Chen2015}), particularly from problems involving 2D surfaces embedded in $\mathbb{R}^3$ or higher. The linear system is non-symmetric due to the closest point extension, therefore Eigen's BiCGSTAB is an option for larger systems. However, we show in Section~\ref{sec:iterative-solver-comparison} that using Eigen's BiCGSTAB with the construction of the full matrix system can be too memory intensive. 

To efficiently accommodate large-scale problems, we have designed a custom BiCGSTAB solver tailored to CPM. Our implementation closely follows Eigen's BiCGSTAB solver\footnote{\href{https://eigen.tuxfamily.org/dox/BiCGSTAB\_8h\_source.html}{https://eigen.tuxfamily.org/dox/BiCGSTAB\_8h\_source.html}}, with key differences for memory-efficiency and parallelization. This is achieved by exploiting a key property of iterative Krylov solvers: explicit construction of the system matrix is not required (in contrast to direct solvers). For iterative Krylov solvers, only the \emph{action} of the matrix on a given input vector is required (i.e., the matrix-vector product).

Specifically, we implemented our solver with the goal of solving linear systems $\A\u=\mathbf{f}$ with
\begin{equation*}
    \A = m\II + n\left[{\rm diag}(\LL) + \left(\LL-{\rm diag}(\LL)\right)\EE\right],
\end{equation*}
where $m\in\{0,1\}$ and $n\in\{1,-\Delta t,-\Delta t/2\}$. This generalized form for $\A$ supports the applications described in Sections \ref{sec:conv-studies} and \ref{sec:applications}. For example, setting $m=n=1$ results in the linear system for the screened-Poisson problem described in Section~\ref{sec:conv-shifted-poisson}. The matrices $\EE$ and $\LL$ are stored explicitly, as discussed in Section~\ref{sec:op-dis}, and the matrix-vector product $\A\u$ is computed as follows:
\begin{enumerate}
    \item Compute $\mathbf{a} = \EE\u$.
    \item Compute $\mathbf{b} = (\LL-{\rm diag}(\LL))\mathbf{a}$.
    \item Compute $\mathbf{a} = {\rm diag}(\LL)\u$.
    \item Return $\mathbf{v} = m\u + n\mathbf{a} + n\mathbf{b}$.
\end{enumerate}
OpenMP is used for parallelizing each of the steps over the DOFs. 

In addition, iterative Krylov solvers allow for a \emph{preconditioner} (i.e., approximate inverse operator) for improving convergence of the linear solver. The preconditioner step requires solving the equation $\mathbf{M}\z=\rr$, where $\mathbf{M}$ is an approximation to $\A$ and $\rr$ is the residual vector. Depending on the particular problem, we either use a diagonal preconditioner or a damped-Jacobi preconditioner. Computing the diagonal entries of $\A$ would require extra computations since the full matrix is not constructed. In practice, however, we found that the diagonal values of $m\II + n{\rm diag}(\LL)$ are a good enough approximation. (In our experiments, we have verified that the infinity norm of the error matches the result produced by Eigen's solver.) For damped-Jacobi preconditioning, the iteration $\u\leftarrow\u + \omega{\rm diag}(\LL)^{-1}\rr$ is applied for a fixed number of iterations with $\omega=2/3$.

\subsection{Visualization}
\label{sec:visualization}
The solution can be visualized in multiple ways. \citet{Demir2015} proposed a direct raycasting approach based on the closest points $\cp_{\S}(\x_i)$ for $\x_i\in\Omega(\S)$. The set of $\cp_{\S}(\x_i)$ can also be considered a point cloud and visualized as such. Lastly, interpolation allows the solution to be transferred to any explicit representation, e.g., triangle mesh, point cloud, etc.

For convenience, we visualize the surface solution at points $\cp_{\S}(\x_i)$ (e.g., Figure~\ref{fig:harmonic-map}) or interpolate onto a triangulation. If the given surface $\S$ is provided as a triangulation we use it; if a surface can be described by a parameterization, we connect evenly spaced points in the parameter space to create a triangulation. Both point clouds and triangulations are visualized using \texttt{polyscope} \cite{polyscope}.

\section{Convergence Studies}
\label{sec:conv-studies}
We begin our evaluation by verifying that our proposed IBC schemes achieve the expected convergence orders on various analytical problems. We also compare our approach with the existing CPM approach of \citet{Auer2012}, the nearest point approach, as well as a standard mesh-based method for reference. Lastly, we compare our partially matrix-free solver against Eigen's SparseLU and BiCGSTAB implementations \cite{eigenweb} as well as Intel MKL PARDISO. All error values are computed using the max-norm. Throughout the rest of the paper, the hat symbol has been dropped from manifold functions (e.g., $\hat{u}$), since it is apparent from the context.

\subsection{Poisson Equation with Discontinuous Solution}
\label{sec:conv-discont}

\begin{figure*}
\centering
    \begin{subfigure}[b]{0.98\textwidth}
            \centering

            \includegraphics{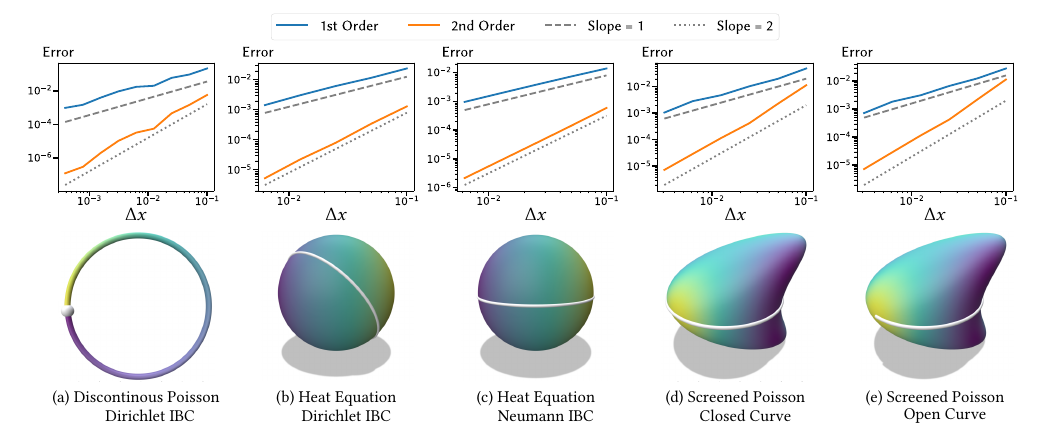}
    \end{subfigure}
\caption{Convergence studies and associated geometries for the model problems in Sections ~\ref{sec:conv-discont}-\ref{sec:conv-shifted-poisson}. The plots show results for our CPM approach using first (blue) and second (orange) order IBCs, along with lines of slopes 1 (gray, dashed) and 2 (gray, dotted). In (a)-(c) analytical $\cp_{\S}$ are used, while (d) and (e) compute $\cp_{\S}$ from the level-set representation of $\S$. All examples use analytical $\cp_{\C}$.}
\label{fig:conv-studies}
\end{figure*}

Consider the Poisson equation 
\begin{equation*}
\begin{aligned}
    -\frac{\partial^2 u}{\partial \theta^2} &= 2\cos(\theta - \theta_{\C}),\\
    u(\theta_{\C}^-) &= 2,\\
    u(\theta_{\C}^+) &= 22,
\end{aligned}
\end{equation*}
on the unit circle parameterized by $\theta$. The right-hand-side expression is found by differentiating the exact solution
\begin{equation*}
u(\theta) = 2\cos(\theta - \theta_{\C}) + \frac{10}{\pi} (\theta - \theta_{\C}),
\end{equation*}
where $\theta_{\C}$ is the location of the Dirichlet IBC. The Dirichlet IBC is two-sided and thus discontinuous at the point $\theta_{\C}$, with $u = 2$ as $\theta\rightarrow\theta_{\C}^-$ and $u = 22$ as $\theta\rightarrow\theta_{\C}^+$. We use $\theta_{\C} = 1.022\pi$; no grid points coincide with the IBC location. 

Eigen's SparseLU is used to solve the linear system for this problem on the circle embedded in $\mathbb{R}^2$. Figure~\ref{fig:conv-studies}(a) shows that the first and second-order IBCs discussed in Section~\ref{sec:interior-constraints} achieve the expected convergence rates. Neither the nearest point approach (Section~\ref{sec:nearest-point}) nor the method of Auer et al.~\shortcite{Auer2012} can handle discontinuous IBCs.

\subsection{Heat Equation}
\label{sec:heat-conv}
CPM can also be applied to time-dependent problems. Consider the heat equation
\begin{equation}
    \frac{\partial u}{\partial t} = \Delta_{\S} u,
    \quad {\rm with} \quad
\begin{cases}
    u = g, \;{\rm or}\\
    \nabla_{\S} u \cdot \b_{\C} = 0,
\end{cases}
{\rm on}\; \C,
\label{eqn:heat-conv}
\end{equation}
where $\b_{\C}$ is the binormal direction to $\C$ that is also in the tangent plane of $\S,$ i.e., $\b_{\C} = \n_{\S} \times \t_{\C}$ (see Section~\ref{sec:robust_cp_diff}). If imposing the Dirichlet IBC, the exact solution, $g,$ is used as the prescribed function on $\C$. Here we solve the heat equation on the unit sphere with the exact solution 
\begin{equation*}
    g(\theta, \phi, t) = e^{-2t}\cos(\phi),
\end{equation*}
where $\theta$ is the azimuthal angle and $\phi$ is the polar angle. The IBC is imposed with $\C$ as a circle defined by the intersection of a plane with $\S$. The initial condition is taken as $g(\theta, \phi, 0) = \cos(\phi).$

Crank-Nicolson time-stepping \cite{Leveque2007} (i.e., trapezoidal rule) is used with $\Delta t = 0.1 \Delta x$ until time $t = 0.1.$ Figure~\ref{fig:conv-studies} (b) and (c) show convergence studies for \eqref{eqn:heat-conv} with Dirichlet and zero-Neumann IBCs imposed, respectively. The expected order of accuracy for first and second-order IBCs is achieved for both the Dirichlet and zero-Neumann cases. Recall that the nearest point approach and the method of \citet{Auer2012} cannot handle Neumann IBCs. 

\subsection{Screened-Poisson Equation}
\label{sec:conv-shifted-poisson}
Exact solutions for manifold PDEs can also be derived on more complex manifolds defined as level sets. Consider the screened-Poisson problem in Section 4.6.5 of \cite{Chen2015}, which was inspired by an example by \citet{dziuk1988finite}. The surface is defined as $\S = \{\x \in \mathbb{R}^3 \;|\; (x_1 - x_3^2)^2 + x_2^2 + x_3^2 = 1 \},$ which we refer to as the Dziuk surface. 

The screened-Poisson equation we solve is
\begin{equation}
    \begin{aligned}
        -\Delta_{\S} u  + u &= f,\\
        \nabla_{\S} u \cdot \b_{\C} &= 0,
    \end{aligned}
    \label{eqn:Dziuk-Neumann}
\end{equation}
with exact solution $u(\x) = x_1 x_2.$ Although the solution is simple, the function $f$ is complicated; we  derived it by symbolic differentiation using the formulas in \cite{Chen2015, dziuk1988finite}. 

The zero-Neumann IBC of~\eqref{eqn:Dziuk-Neumann} is satisfied on the intersection of $\S$ with the $x_1 x_2$-plane. From the definition of $\S$, this intersection is the unit circle in the $x_1 x_2$-plane. Figure~\ref{fig:conv-studies} (d) and (e) show convergence studies imposing the zero-Neumann IBC on the full circle (closed curve) and the arc  with $\theta \in [-\frac{3\pi}{4}, \frac{\pi}{4}]$ (open curve), respectively. The expected order of accuracy is observed for the implementations of first and second-order IBCs. 

\subsection{Different CPM approaches vs.\ a Mesh-Based Method}
CPM is principally designed to solve problems on \emph{general} manifolds, given by their closest point functions. The closest point function can be thought of as a black box allowing many manifold representations to be handled in a unified framework. Hence, we emphasize that one should not expect CPM to universally surpass specially tailored, well-studied approaches for particular manifold representations, such as finite elements on (quality) triangle meshes. Nevertheless, mesh-based schemes provide a useful point of reference for our evaluation. CPM also retains some advantages even for triangle meshes, such as mesh-independent behaviour.

With the above caveat in mind, we compare the various CPM approaches to the standard cotangent Laplacian \cite{pinkall1993computing, dziuk1988finite} that approximates the Laplace-Beltrami operator on a triangulation of the surface. We use the implementation from \texttt{geometry-central} \cite{geometrycentral}, adapted slightly to include IBCs. The Poisson equation $-\Delta_{\S} u = f$ is solved on the Dziuk surface defined in Section~\ref{sec:conv-shifted-poisson}. The same exact solution $u(\x) = x_1 x_2$ is used, but Dirichlet IBCs are imposed using this exact solution. 

\begin{figure}
     \begin{subfigure}[b]{0.45\textwidth}
     \centering
         \begin{subfigure}[b]{0.4\textwidth}
             \centering
             \caption*{$\mathcal{T}_g$}
             \includegraphics[width=\textwidth]{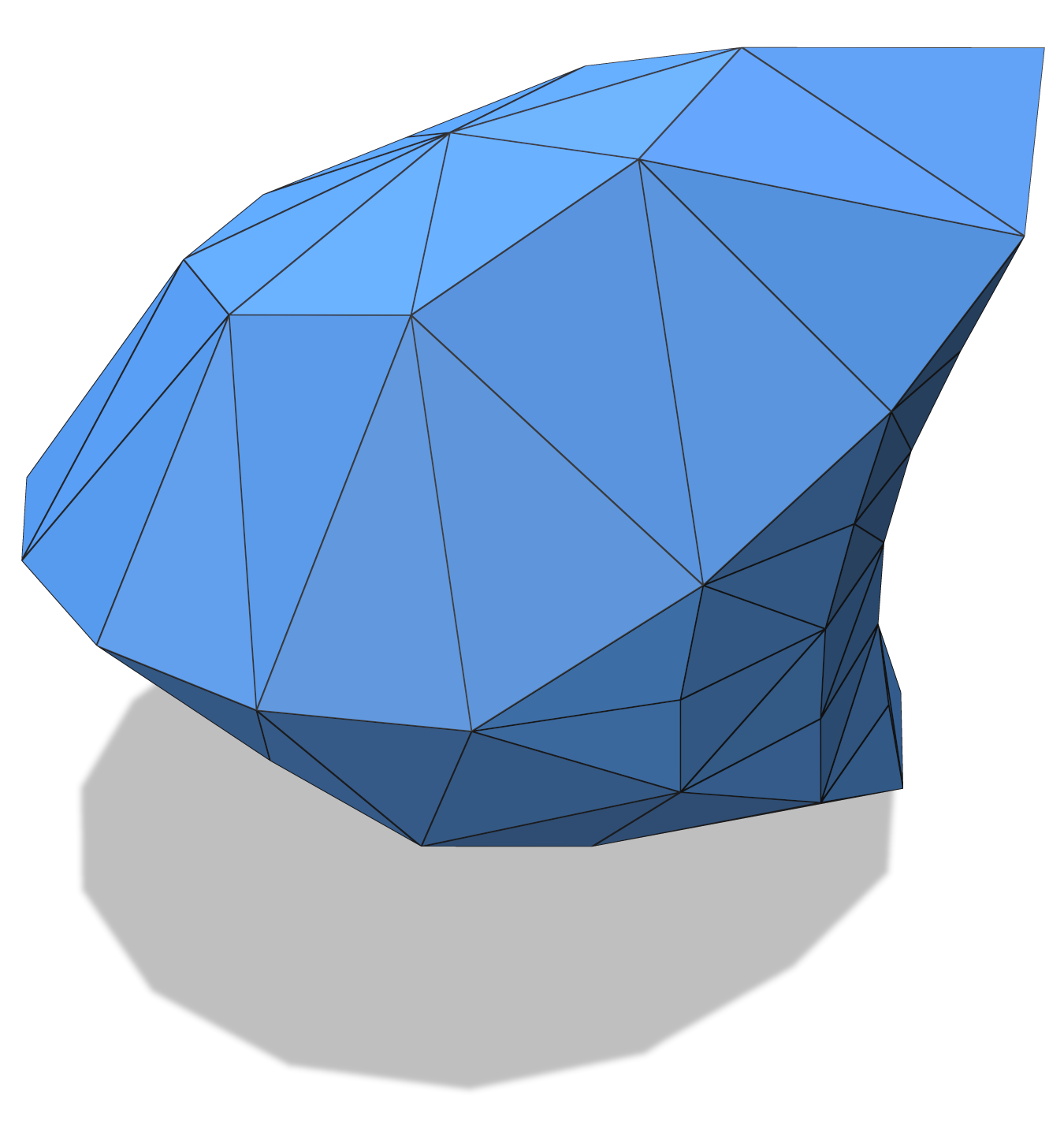}
         \end{subfigure}
         \vspace{-10pt}
         \begin{subfigure}[b]{0.4\textwidth}
             \centering
             \caption*{$\mathcal{T}_b$}
             \includegraphics[width=\textwidth]{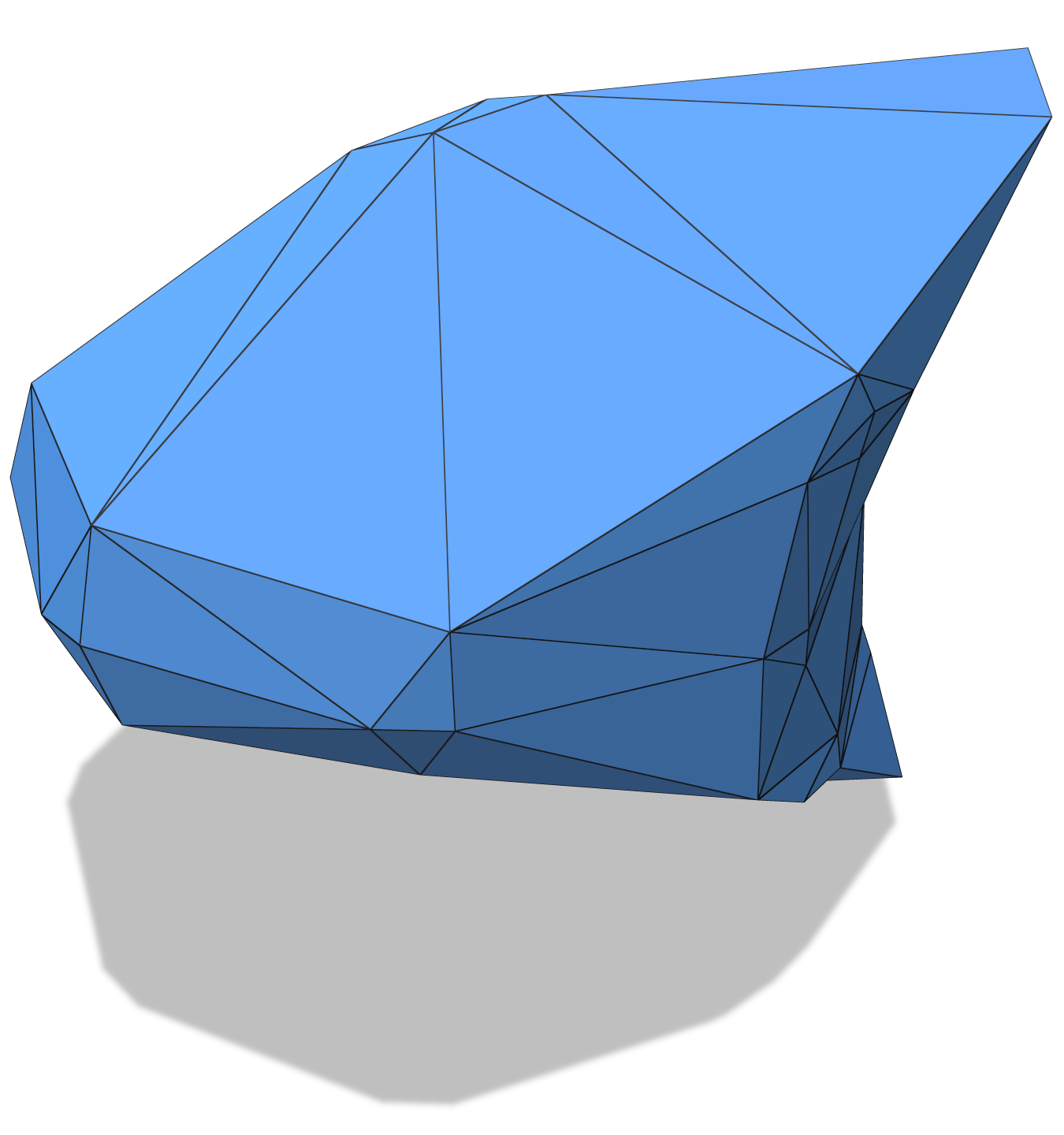}
        \end{subfigure}
    \end{subfigure}
    \begin{subfigure}[b]{0.45\textwidth}
    \centering
        \begin{subfigure}[b]{0.4\textwidth}
             \centering
             \includegraphics[width=\textwidth]{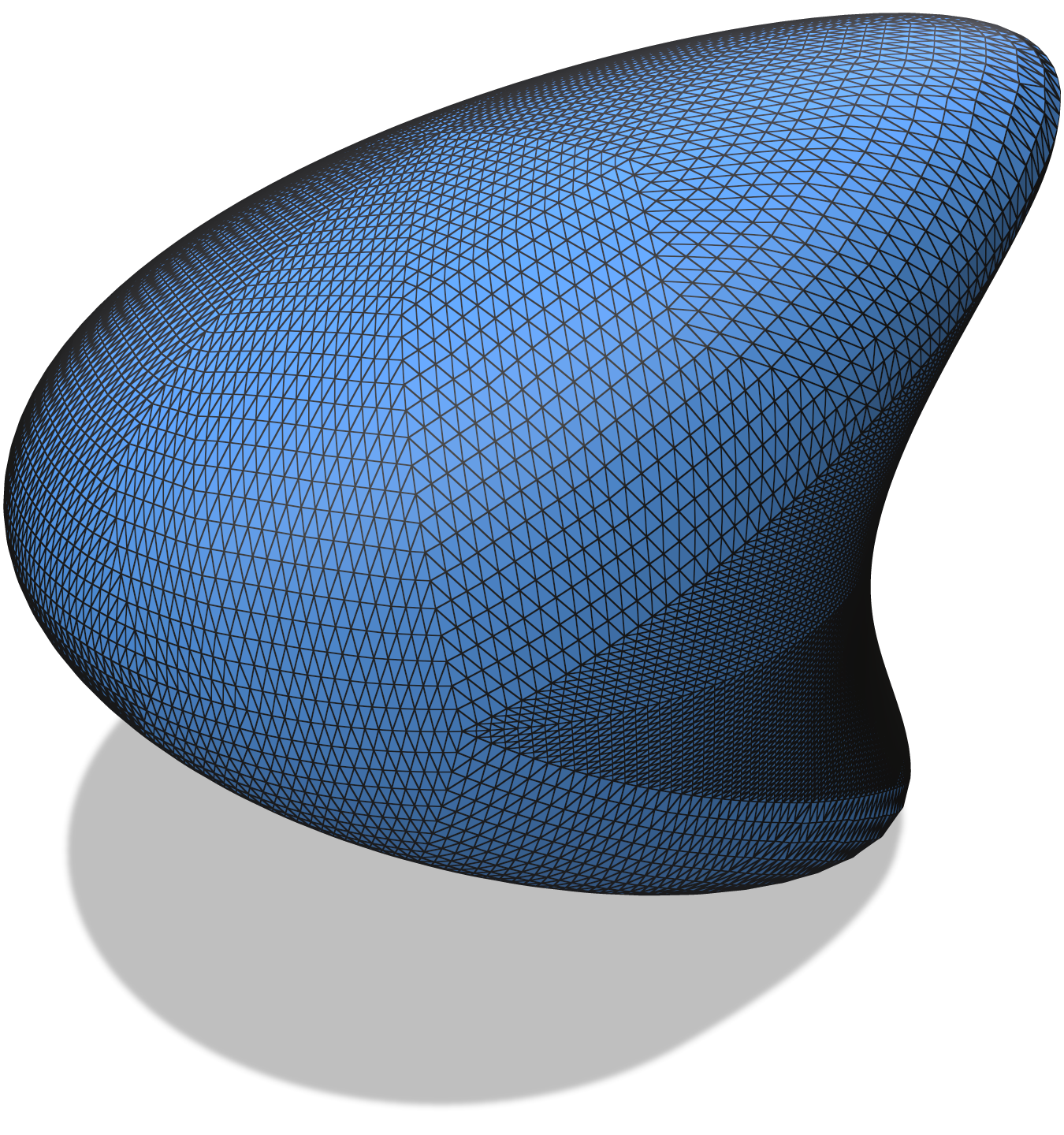}
         \end{subfigure}
        \begin{subfigure}[b]{0.4\textwidth}
             \centering
             \includegraphics[width=\textwidth]{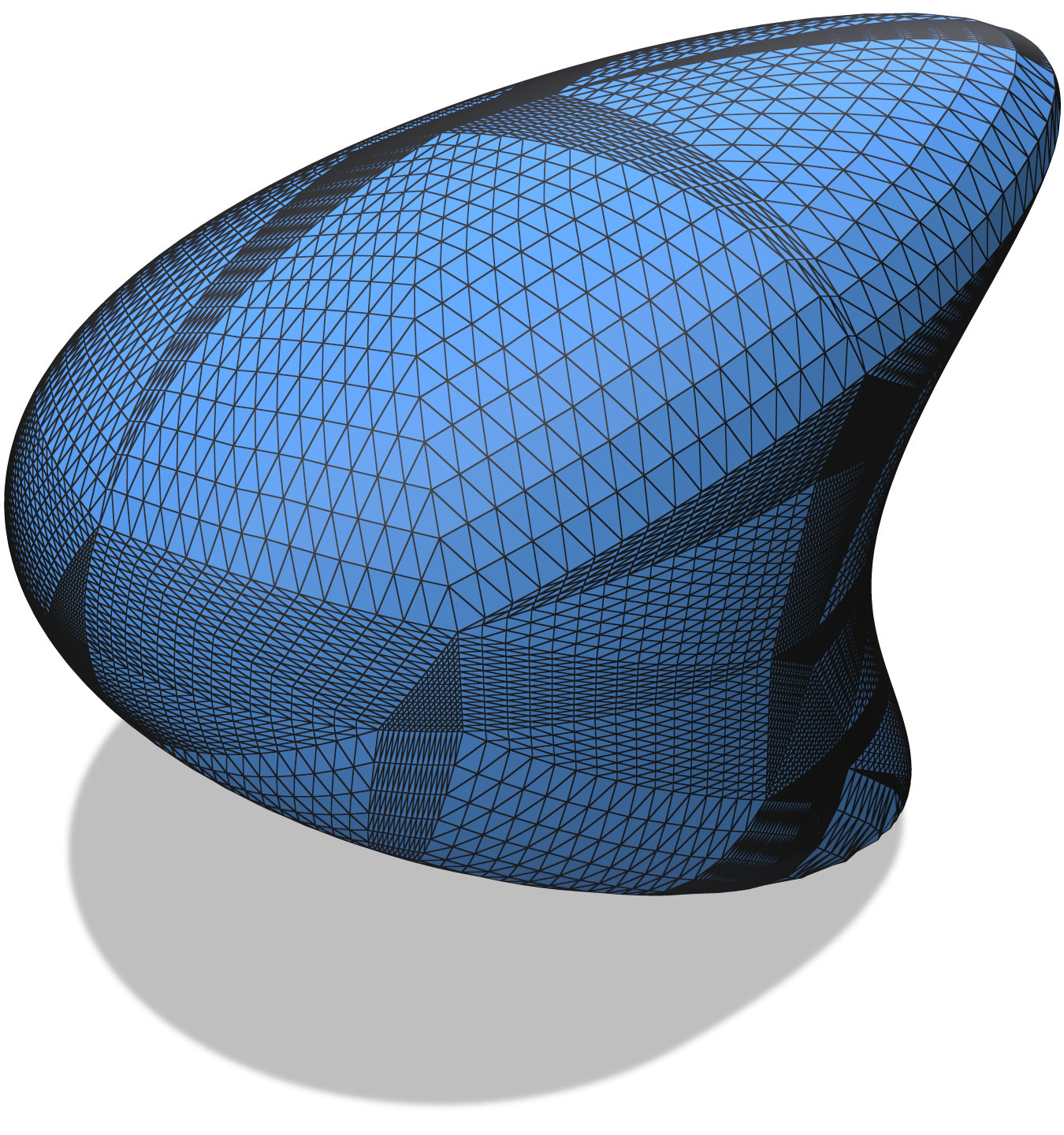}
         \end{subfigure}
     \end{subfigure}
\caption{Triangulations of the Dziuk surface used for testing. Top-left: Good-quality base triangulation, $\mathcal{T}_g$. Top-right: Low-quality base triangulation, $\mathcal{T}_b$. Bottom row: The same triangulations after four rounds of refinement.}
\label{fig:Dziuk_triangulations}
\end{figure}

``Good'' and ``bad'' triangulations of the Dziuk surface, denoted $\mathcal{T}_{g}$ and $\mathcal{T}_{b}$ respectively,  are used to illustrate the dependence of the mesh-based method on triangulation quality (Figure \ref{fig:Dziuk_triangulations}). Both triangulations are constructed starting from six vertices on $\S$ as in \cite{dziuk1988finite}. An initial round of 1:4 subdivision is performed by adding new vertices along each edge, at the midpoint for $\mathcal{T}_g$ and at the 20\% position for $\mathcal{T}_b$, to induce skinnier triangles in the latter. The new vertices are projected to their closest points on $\S$. 

Evaluations under refinement for the mesh-based method are performed starting with the above first-level $\mathcal{T}_g$ and $\mathcal{T}_b$. We refine with uniform 1:4 subdivision, for both $\mathcal{T}_g$ and $\mathcal{T}_b,$ by adding new vertices at midpoints of edges and then projecting them onto $\S$ (see Figure~\ref{fig:Dziuk_triangulations}). Delaunay edge flips are also performed to improve the quality of $\mathcal{T}_g$ at each refinement level. 

Triangle mesh resolution is measured as the mean edge-length in $\mathcal{T}_g$ or $\mathcal{T}_b$, whereas for CPM resolution is measured as the uniform $\Delta x$ used in the computational-tube $\Omega(\S)$. This core incompatibility makes it inappropriate to use resolution as the independent variable for comparative evaluations of error, computation time, or memory usage.
A more equitable comparison is to investigate computation time versus error and memory versus error. Computation times for CPM include the construction of $\Omega(\S)$ and $\Omega(\C)$ (which involves computing $\cp_{\S}$ and $\cp_{\C}$) and the time for constructing and solving the linear system. Computation times for the mesh-based method include the triangulation refinement and the construction and solution of the linear system. Separate evaluations are performed with $\C$ as a closed curve, an open curve, and a point, since CPM IBC enforcement is slightly different for each type of $\C$. 

\subsubsection*{Closed Curve IBC}
The boundary curve $\C$ is constructed using the flip geodesics algorithm in \texttt{geometry-central} \cite{geometrycentral}. The resulting $\C$ is represented as a polyline $\mathcal{P}$, which in general does \emph{not} conform to edges or vertices of $\mathcal{T}$. For IBC enforcement, the nearest vertex in the triangulation $\mathcal{T}$ to each vertex in $\mathcal{P}$ is assigned the prescribed Dirichlet value.

This treatment of Dirichlet IBCs for the mesh-based method is first-order accurate in general. More accurate (and involved)  Dirichlet IBC approaches could be used as discussed in Section~\ref{sec:IBC_related_work}. However, we set these options aside, as the goal of this comparison is simply to show that CPM with our first and second-order IBC approaches gives comparable results to basic mesh-based methods, that is, mesh-based methods where the representations of $\S$ and $\C$ are held fixed, e.g., no (extrinsic or intrinsic) remeshing is performed.

\begin{figure*}
\centering

\includegraphics{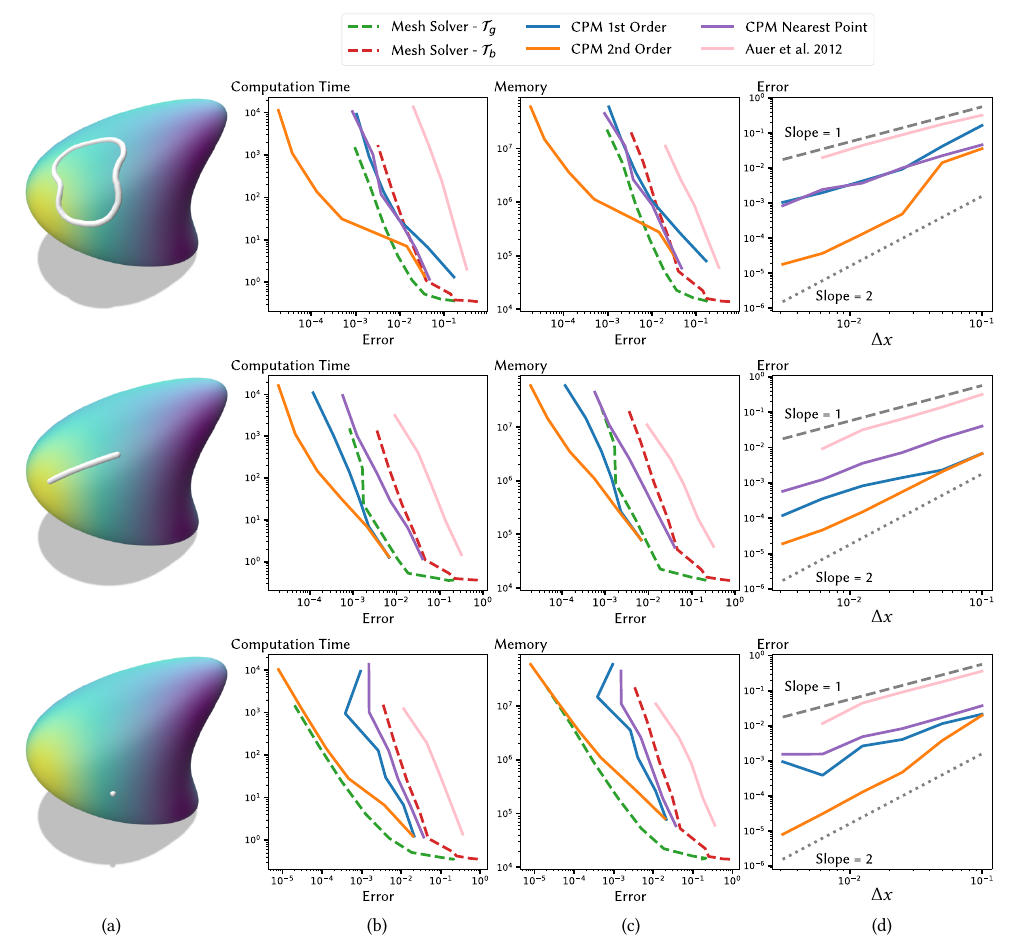}
\caption{A comparison of CPM vs.\ the mesh-based cotangent Laplacian for the Poisson equation with Dirichlet IBC. Top row: Closed curve $\C$. Middle row: Open curve $\C$. Bottom row: Point $\C$. Columns (b) and (c) show computation time vs.\ error and memory vs.\ error, respectively. Mesh results are shown separately for the $\mathcal{T}_g$ and $\mathcal{T}_b$ triangulations. Column (d) illustrates the convergence behaviour of error vs.\ $\Delta x$ for only the CPM schemes. The $\cp_{\S}$ are computed from a level-set representation, while $\cp_{\C}$ are computed from polyline representations for curves $\C$ and exactly for the point $\C$.}
\label{fig:comparison}
\end{figure*}

Figure~\ref{fig:comparison} (top row) compares all types of CPM IBC approaches against the mesh-based method on $\mathcal{T}_g$ and $\mathcal{T}_b$ in columns (b) and (c). CPM with second-order IBCs achieves the lowest error for the same computation time and memory usage as other approaches. The mesh-based method with $\mathcal{T}_g$ outperforms the use of $\mathcal{T}_b$, as expected. CPM with first-order IBCs and nearest point approaches are similar and lie between the mesh-based method with $\mathcal{T}_g$ and $\mathcal{T}_b$. The method of \citet{Auer2012} has the largest error compared to all others. The expected order of convergence is seen for all CPM IBC approaches in the error versus $\Delta x$ plot of Figure~\ref{fig:comparison} (top row, (d)).

\subsubsection*{Open Curve IBC}
The open curve $\C$ is also constructed using the flip geodesics algorithm in \texttt{geometry-central} \cite{geometrycentral}. The Dirichlet IBC is enforced in the mesh-based solver in the same way as the closed curve above. Figure~\ref{fig:comparison} (middle row) shows the same ranking of the methods as in the closed curve case, except CPM with first-order IBCs now outperforms both triangulations and the nearest point CPM approach. The expected order of convergence is seen for all CPM IBC approaches in Figure~\ref{fig:comparison} (middle row, (d)).

\subsubsection*{Point IBC}
The point $\C$ is intentionally chosen as one of the vertices in the base triangulation so that it is present in all refinements of $\mathcal{T}_g$ and $\mathcal{T}_b$. The Dirichlet IBC at $\C$ is imposed by replacing the vertex DOF in $\mathcal{T}$ with the prescribed Dirichlet value. Figure~\ref{fig:comparison} (bottom row) shows the results for a point $\C$. 

The mesh-based solver on $\mathcal{T}_g$ converges with second-order accuracy (since the IBC is a vertex), but only first-order accuracy on $\mathcal{T}_b$. Therefore, the mesh-based method with $\mathcal{T}_g$ outperforms CPM with second-order IBCs in the larger error regime. In the lower error regime, the latter methods are similar. All other methods show the same ranking as the open curve case. 

The expected order of convergence is seen for all CPM IBC approaches in Figure~\ref{fig:comparison} (bottom row, (d)). Notably, the second-order IBC version of CPM exhibits slightly higher than expected errors at the finest grid resolution for the closed and open curve IBCs (see Figure~\ref{fig:comparison}, top and middle rows, (d)). This is caused by the resolution of the polyline representation of $\C$: at fine grid resolutions, the inherent sharp features of the coarse polyline $\C$ begin to be resolved more fully by the discrete CP function. Accordingly, no such reduction in convergence order is seen for the point IBC.

\subsection{Linear System Solvers}
\label{sec:iterative-solver-comparison}

\begin{figure*}
\centering
\includegraphics{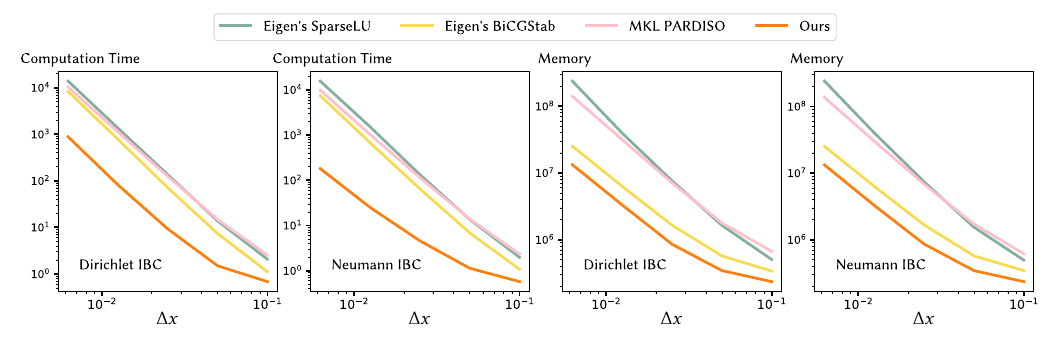}
\caption{Left pair: Computation time vs.\ $\Delta x$ plots for the heat equation~\eqref{eqn:heat-conv} with Dirichlet and zero-Neumann IBCs with four solver options. Right pair: Memory vs.\ $\Delta x$ plots for the same problems and solvers. Our solver (orange) achieves the lowest computation time and memory costs.}
\label{fig:mfmb_comp}
\end{figure*}

Our partially matrix-free BiCGSTAB solver (see Section~\ref{sec:partially-mat-free-solver}) is faster and more memory efficient than Eigen's SparseLU and BiCGSTAB implementations \cite{eigenweb} as well as the Intel MKL PARDISO. An example of the improved efficiency is shown in Figure~\ref{fig:mfmb_comp} for the heat problem in Section~\ref{sec:heat-conv} with Dirichlet and zero-Neumann IBCs. Solving the heat equation involves multiple linear system solves (i.e., one for each time step). SparseLU requires the most computation time, even though it prefactors the matrix once and just performs forward/backward solves for each time step. SparseLU also uses the most memory, as expected. PARDISO facilitates parallelism during factorization, enhancing the speed of the initialization process compared to Eigen's SparseLU. However, the forward/backward solves are still conducted sequentially, limiting the magnitude of the performance improvement.

Table~\ref{tab:solver-comparison} gives the max and average computation time speedup, $T_{{\rm spdup}}$, and memory reduction, $M_{{\rm red}}$, for the results in Figure~\ref{fig:mfmb_comp}. The computation time speedup compared to Eigen's SparseLU (similarly for BiCGSTAB and PARDISO) is computed as $$T_{{\rm spdup}} = T({\rm SparseLU})/T({\rm Ours}),$$ where $T({\rm SparseLU})$ and $T({\rm Ours})$ are the computation times of SparseLU and our solver, respectively. The memory reduction factor is calculated in an analogous manner with computation times replaced by memory consumption. The max and average $T_{{\rm spdup}}$ and $M_{{\rm red}}$ are computed over all $\Delta x$. 
\begin{table}
    \caption{Ratios of computation time $T_{{\rm spdup}}$ and memory usage $M_{{\rm red}}$ for Eigen's SparseLU and BiCGSTAB as well as PARDISO as compared to our tailored BiCGSTAB solver, for the experiments of Figure~\ref{fig:mfmb_comp}.}
    \centering
        \begin{tabular}{ccrrrr}
            \hline
            \multirow{2}{*}{Solver} & \multirow{2}{*}{IBC} & \multicolumn{2}{c}{$T_{{\rm spdup}}$} & \multicolumn{2}{c}{$M_{{\rm red}}$}\\
             &  & Max & Avg. & Max & Avg.\\
             \multirow{2}{*}{Eigen's SparseLU} & \mycc Dirichlet & \mycc 16.6 &\mycc 11.8 &\mycc 17.9 &\mycc 9.1\\
                                       & Neumann  & 86.2 & 38.3 & 18.1 & 9.1\\       
             \multirow{2}{*}{Eigen's BiCGSTAB} &\mycc Dirichlet &\mycc 9.5  &\mycc 6.6 &\mycc 1.9  &\mycc 1.8\\
                                       & Neumann  & 40.9 & 18.0 & 1.9  & 1.8\\
            \multirow{2}{*}{MKL PARDISO} & \mycc Dirichlet & \mycc 13.7 &\mycc 10.5 &\mycc 10.6 &\mycc 7.3\\
                                       & Neumann  & 54.2 & 27.3 & 10.3 & 7.0\\       
             
            \hline
        \end{tabular}
    \label{tab:solver-comparison}
\end{table}

The speedup of our solver is significant compared to Eigen's SparseLU and BiCGSTAB as well as PARDISO. The memory reduction of our method is significant compared to Eigen's SparseLU and PARDISO, but less significant compared to Eigen's BiCGSTAB. The speedup exhibits problem-dependence since $T_{{\rm spdup}}$ factors in Table~\ref{tab:solver-comparison} are larger for the zero-Neumann IBC compared to the Dirichlet IBC. However, as expected, $M_{{\rm red}}$ is not problem-dependent.

\section{Applications}
\label{sec:applications}

We now show the ability of our CPM approach to solve PDEs with IBCs that are common in applications from geometry processing: diffusion curves, geodesic distance, vector field design, harmonic maps, and reaction-diffusion textures.

\begin{figure}
\centering
\includegraphics[scale=0.97]{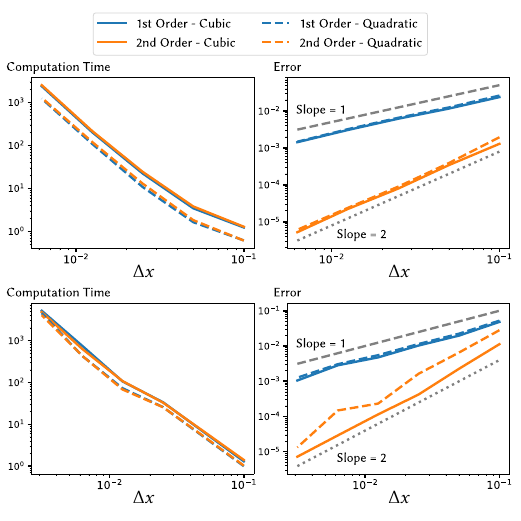}
\caption{A comparison of CPM with quadratic vs.\ cubic interpolation stencils for the heat (top row) and Poisson (bottom row) problems of Figure~\ref{fig:conv-studies} (b) and (d). Comparable results are achieved, but quadratic is often faster while cubic typically exhibits more regular convergence.}
\label{fig:interp-comparison}
\end{figure}

Quadratic polynomial interpolation, i.e., $p = 2$, is used for all the examples in this section. Current CPM theory suggests that only first-order accuracy can be expected with quadratic polynomial interpolation, but CPM has been observed to give second-order convergence numerically (see \cite{Macdonald2010}, Section 4.1.1). This behaviour is confirmed with IBCs in Figure~\ref{fig:interp-comparison}. 

The main motivation for choosing quadratic interpolation is to obtain smaller computational tube-radii, $r_{\Omega(\S)}$, which allows higher curvature $\S$ and $\C$ to be handled with larger $\Delta x$. The resulting $\Omega(\S)$ and $\Omega(\C)$ contain fewer DOFs and therefore the computation is more efficient. Furthermore, Figure~\ref{fig:interp-comparison} shows that, for the same $\Delta x$, quadratic interpolation has lower computation times. Quadratic interpolation is 1.1-2.1 times faster than cubic interpolation in Figure~\ref{fig:interp-comparison}. We used $p = 3$ in the convergence studies of Section~\ref{sec:conv-studies} because the error for second-order BCs with $p = 2$ can sometimes be less regular (i.e., decreasing unevenly or non-monotonically) than with $p = 3$ (Figure~\ref{fig:interp-comparison}, bottom right).

CPM with first-order IBCs is used in all the examples in this section. The geodesic distance, vector field design, and harmonic map algorithms used here are themselves all inherently first-order accurate; hence using second-order IBCs would only improve accuracy near $\C$. Second-order IBCs could have been used for diffusion curves and reaction-diffusion textures, but the first-order method was used for consistency. Note also that any surface represented as a mesh is scaled (with fixed aspect ratio) to fit in $[-1,1]^3$.

\subsection{Diffusion Curves}
Diffusion curves offer a sparse representation of smoothly varying colours for an image \cite{Orzan2008} or surface texture \cite{jeschke2009rendering}. Obtaining colours over all of $\S$ requires solving the Laplace-Beltrami equation with IBCs:    
\begin{equation}
 \Delta_{\S} u^i = 0, \quad {\rm with} \quad
\begin{cases}
    u^i = g^i, \;{\rm or}\\
    \nabla_{\S} u^i \cdot \b_{\C} = 0.
\end{cases}
{\rm on} \; \C.
    \label{eqn:diff-curves}
\end{equation}

The Laplace-Beltrami equation~\eqref{eqn:diff-curves} is solved for each colour channel $u^i$ independently with CPM. The colour vector is composed of all the colour channels, e.g., for RGB colours $\u = [u^1, u^2, u^3]^T$. Dirichlet IBCs, $u^i = g^i$ on $\C$, are used to specify the colour values at sparse locations on $\S$. These colours spread over all of $\S$ when the Laplace-Beltrami equation is solved. Zero-Neumann IBCs can be used to treat $\C$ as a passive barrier that colours cannot cross. Two-sided IBCs along $\C$ are also easily handled, and can even be of mixed Dirichlet-Neumann type (not to be confused with Robin BCs).

The surface of the Nefertiti bust~\cite{Nefertiti-mesh} is coloured by solving the Laplace-Beltrami equation with CPM with $\Delta x = 0.00315$ and IBCs specified by diffusion curves in Figure~\ref{fig:teaser} (a). IBC curves are polylines created using the flip geodesics algorithm in \texttt{geometry-central} \cite{geometrycentral}. Most curves are two-sided Dirichlet IBCs (white curves, Figure~\ref{fig:teaser} (a) left). However, the red and green band on the headdress is created using two-sided red-green Dirichlet IBCs vertically and two-sided Neumann-Dirichlet IBCs horizontally (black curves, Figure~\ref{fig:teaser} (a) left).

\subsubsection*{Mixed-Codimensional Objects} 
The generality of CPM allows PDEs on mixed-codimensional objects to be solved. The theoretical assumption that $\cp_{\S}$ is unique is violated in this case (near pieces of differing codimension). However, CPM gives the expected result in practice on mixed-codimensional objects (e.g., Figure 4.4 of~\cite{Macdonald2010}). 

\begin{figure}
     \centering
    \includegraphics{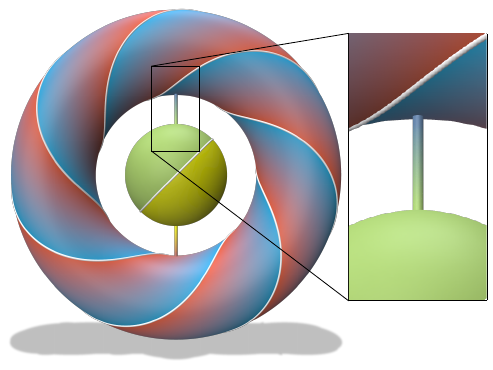}
\caption{Diffusion curves on a nonmanifold object of mixed codimension. Line segments connect the torus to the sphere, which are all represented with analytical $\cp_{\S}$. The $\cp_{\C}$ for the circle on the sphere is computed analytically, while $\cp_{\C}$ for the torus knot is computed from a parametrization.}
\label{fig:mixed-codim-diff-curves}
\end{figure}

Figure~\ref{fig:mixed-codim-diff-curves} shows a diffusion curves example (with $\Delta x = 0.05$) featuring a mixed 1D and 2D object embedded in $\mathbb{R}^3$. This mixed-codimensional $\S$ is created using analytical closest point functions for the torus, sphere, and line segment. The torus has minor radius $r=1$ and major radius $R=3$, while the sphere is of radius 1.25. The closest point to $\S$ is determined by computing the closest point to each of the torus, sphere, and line segments, then taking the closest of all four. The two curves $\C$ in this example are two-sided Dirichlet IBCs. $\C$ on the torus is a torus knot specified by the parametric equation
\begin{equation}
x(s) = v(s) \cos(a s),\quad y(s) = v(s) \sin(a s),\quad z(s) = \sin(b s),
\label{eqn:torus-knot}
\end{equation}
with $v(s) = R + \cos(b s),$ $a = 3$, $b = 7,$ and $s\in[0, 2\pi]$. Closest points for the torus knot are computed using the optimization problem discussed in Appendix~\ref{sec:cp_computation}. $\C$ on the sphere is an analytical closest point function for a circle defined as the intersection of the sphere and a plane. Notice the colour from the torus to the sphere blends across the line segments as expected (see Figure~\ref{fig:mixed-codim-diff-curves} zoom).

\subsubsection*{Codimension-Zero Manifolds}
Interestingly, CPM can also be applied with codimension-zero manifolds (see Section 6.2.4 of~\cite{Macdonald2013}). A codimensional-zero manifold is a solid object that is a subset of $\mathbb{R}^{{\rm dim}(\S)}$. 
Consider a codimension-zero $\S$, with a boundary $\partial \S$. The computational domain $\Omega(\S)$ consists of all grid points $\x_i\in\S$ (having $\cp_{\S}(\x_i) = \x_i$) plus a layer of grid points outside $\S$ where $\cp_{\S}(\x_i) \in \partial \S$ and $\|\x_i - \cp_{\S}(\x_i)\| \leq r_{\Omega(\S)}.$ 

\begin{figure}
     \begin{subfigure}[b]{0.47\textwidth}
     \hspace{2pt}
         \begin{subfigure}[b]{0.48\textwidth}
             \centering
             \includegraphics[width=\textwidth]{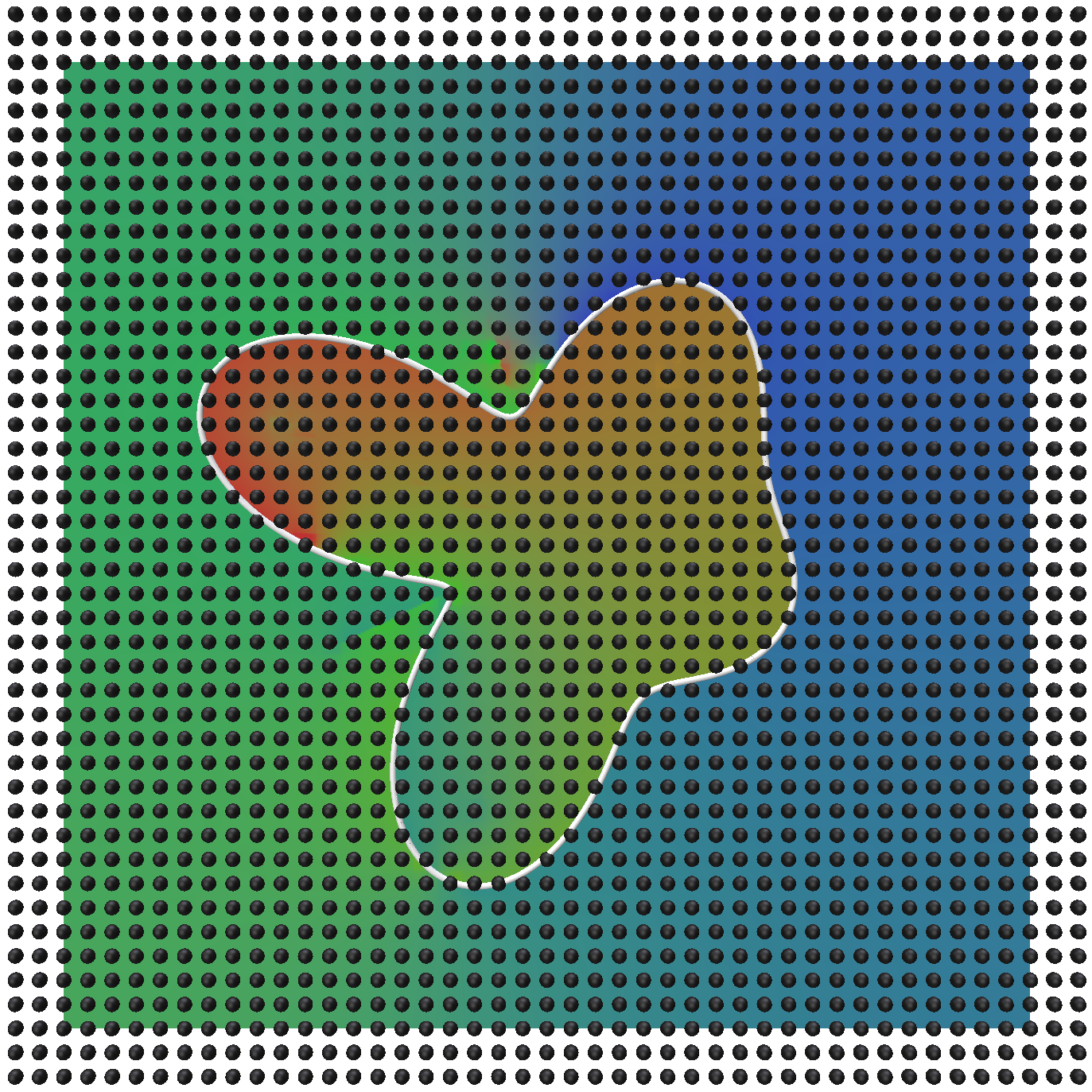}
         \end{subfigure}
         \hspace{-4.5pt}
         \begin{subfigure}[b]{0.48\textwidth}
             \centering
             \raisebox{-0.5pt}{\includegraphics[width=\textwidth]{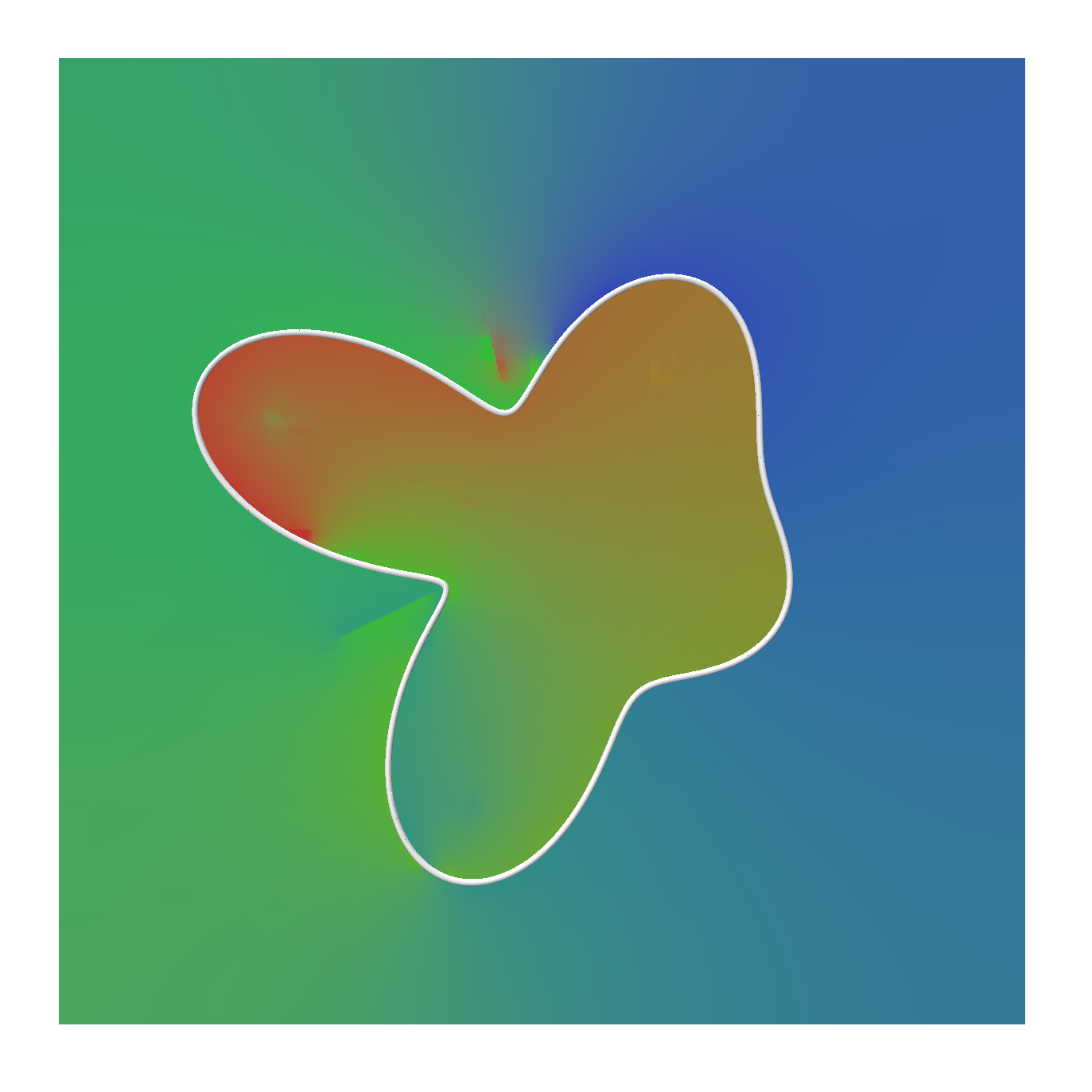}}
        \end{subfigure}
    \end{subfigure}
    \begin{subfigure}[b]{0.47\textwidth}
    \hspace{6.8pt}
        \begin{subfigure}[b]{0.44\textwidth}
             \centering
             \raisebox{3.4pt}{\includegraphics[width=\textwidth]{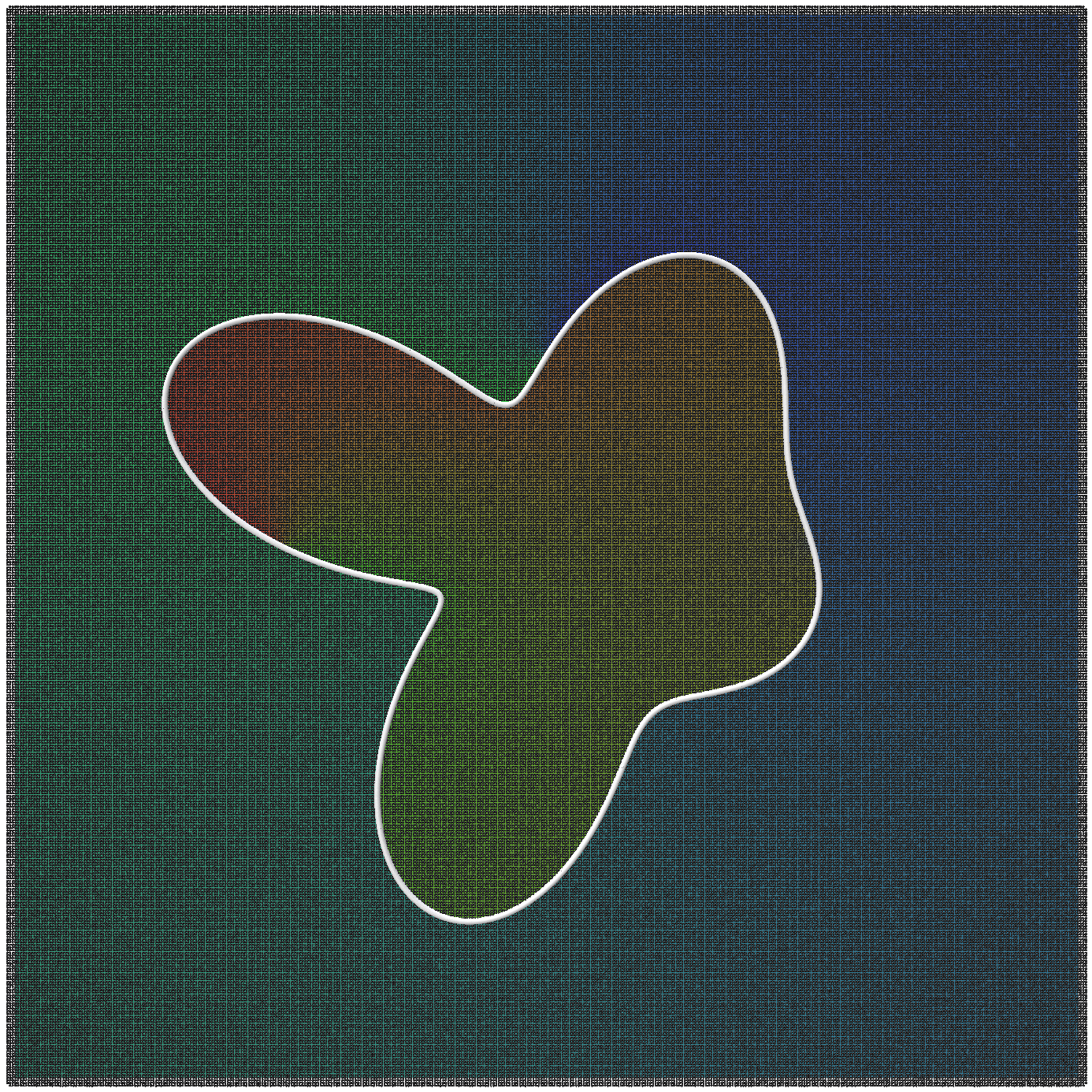}}
         \end{subfigure}
             \hspace{1.15pt}
        \begin{subfigure}[b]{0.4645\textwidth}
             \centering
             \includegraphics[width=\textwidth]{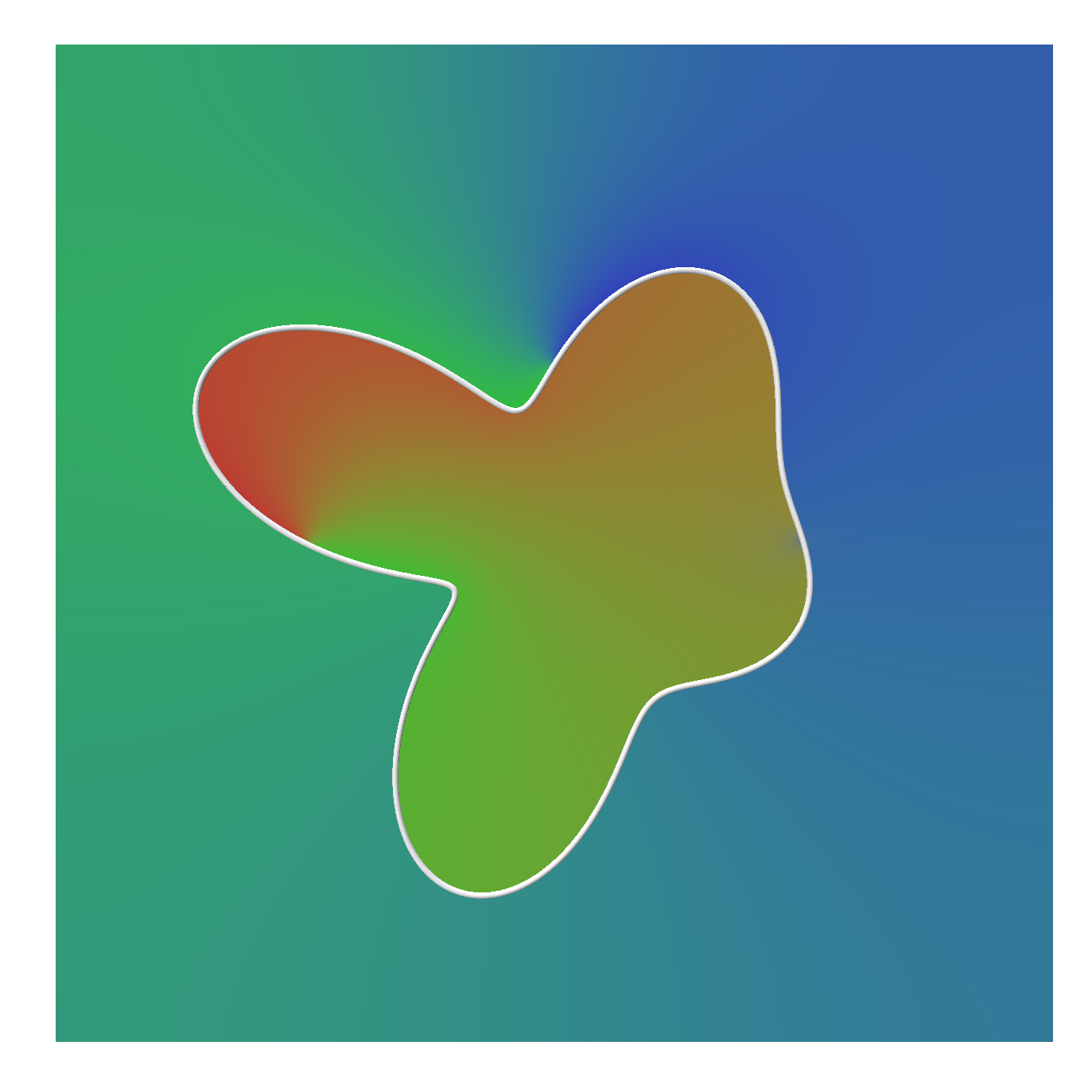}
         \end{subfigure}
     \end{subfigure}
\caption{CPM applied to a codimension-zero diffusion curve problem, with the Dirichlet colour value varying along the white IBC curve. Top row: At an insufficient grid resolution of $\Delta x = 0.05$ (left), high curvature regions exhibit errors near the curve's medial axis (right). Bottom row: A high-resolution grid with $\Delta x = 0.005$ (left) resolves the artifacts (right). The $\cp_{\S}$ are computed analytically and $\cp_{\C}$ are computed from a parametric representation.}
\label{fig:volumetric-diff-curves}
\end{figure}

Figure~\ref{fig:volumetric-diff-curves} shows an example of applying CPM to the diffusion curves problem with $\S$ as the square $ [-1,1]^2$ and $\Omega(\S) \subset \mathbb{R}^2.$  

A parametric curve on the interior of $\S$ defines a diffusion curve $\C$ as a two-sided Dirichlet IBC, given by
\begin{equation}
x(s) = v(s)\cos(s) + c, \quad y(s) = v(s)\sin(s) + c,
\label{eqn:planar-param-curve}
\end{equation}
where 
\begin{equation*}
        v(s) = \frac{\cos(s) \left(\frac{1}{2}(a + b) + \sin(a s) + \sin(b s)\right) + \frac{1}{2}(a+b)}{a + b},
\end{equation*}
 with $a = 3$, $b = 4$, $c = -\frac{1}{2},$ and $s\in[0,2\pi]$. Note that the colour varies along $\C$ from red to green inside $\C$ and blue to green outside $\C$. (Such colour variations along boundaries $\C$ can also easily be applied to problems where the embedding domain has higher dimension than $\S$.) First-order zero-Neumann exterior BCs are applied on $\partial \S$ naturally by CPM, which enforces no (conormal, i.e., normal to $\partial S$ and in the tangent space of $\S$) colour gradient at $\partial \S$.


The grid spacing $\Delta x$ needs to be fine enough near $\C$ to give an accurate solution.
Artifacts can occur if stencils undesirably cross the medial axis of $\C$ when $\Delta x$ is too large (cf. Figure~\ref{fig:volumetric-diff-curves} top and bottom rows). A promising direction of future work is therefore to explore the use of adaptive grids based on the geometry of $\C$. Adaptivity would reduce the total number of DOFs in the linear system and thus improve efficiency. Adaptive grids based on the geometry of $\S$ would also improve efficiency when ${\rm codim}(\S) > 0$. 

Applying CPM with ${\rm codim}(\S) = 0$ represents an alternative to (or generalization of) various existing embedded boundary methods for irregular domains, e.g., \cite{Gibou2002,ng2009efficient,schwartz2006cartesian}. Advantages and disadvantages of this approach should be explored further in future work. One advantage shown by \citet{Macdonald2013} is the ability to couple volumetric and surface PDEs in a unified framework. 


\subsection{Geodesic Distance}
The heat method for geodesic distance computation \cite{Crane2013} has been implemented on many surface representations, including polygonal surfaces, subdivision surfaces \cite{de2016subdivision}, spline surfaces \cite{nguyen2016c1}, tetrahedral meshes \cite{Belyaev2015}, and point clouds \cite{Crane2013}, with each requiring nonnegligible tailoring and implementation effort. By introducing our Dirichlet IBC treatment for CPM, we enable a single implementation covering all these cases, since closest points can be computed to these and many other manifold representations.

The heat method approximates the geodesic distance $\phi$ using the following three steps:

\begin{enumerate}
\item Solve $\frac{\partial u}{\partial t} = \Delta_{\S} u$ to give $u_t$ at time $t,$
\item Evaluate the vector field $\mathbf{X} = -\nabla_{\S} u_t / \| \nabla_{\S} u_t \|,$
\item Solve $\Delta_{\S} \phi = \nabla_{\S} \cdot \mathbf{X}$ for $\phi$.
\end{enumerate}
Step (1) uses a Dirac-delta heat source for a point $\C$ or a generalized Dirac distribution over a curve $\C$ as the initial condition. The time discretization of step (1) employs implicit Euler, for one time-step, which is equivalent (up to a multiplicative constant) to solving
\begin{equation}
\begin{aligned}
(\mathbf{I} - t \Delta_{\S}) v_t &= 0 \; \text{ on } \; \S\backslash \C,\\
v_t &= 1 \; \text{ on } \; \C.
\end{aligned}
\label{eqn:bvp}
\end{equation}
The discrete system for~\eqref{eqn:bvp} can be written as $\mathbf{A} \mathbf{v} = \mathbf{f}$, where $\mathbf{A}\in \mathbb{R}^{(N_{\S} + N_{\C}) \times (N_{\S} + N_{\C})}$ and $\mathbf{v},\mathbf{f}\in \mathbb{R}^{N_{\S} + N_{\C}}$. 

Imposing first-order IBCs involves the Heaviside step function for $\mathbf{f}$. That is, $\mathbf{f}_i = 0$ if $i$ is in the PDE DOF set ($i\in J_{\S}$) and $\mathbf{f}_i = 1$ if $i$ is in the BC DOF set ($i\in J_{\C}$). When imposing this IBC in~\eqref{eqn:bvp}, CPM can experience Runge's phenomenon due to the polynomial interpolation used for the CP extension. Therefore, we approximate the Heaviside step function with a smooth approximation as
$$ \mathbf{f}_i = \frac{1}{2}{\rm tanh}\left(-k \|\cp_{\mathcal{S}-\mathcal{C}}(\mathbf{x}_i)\|\right) + \frac{1}{2}, \quad {\rm with } \quad k = \frac{{\rm atanh}(1-\epsilon)}{e}.$$
The parameters $e$ and $\epsilon$ correspond to the “extent” $[-e, e]$ and the maximum error of the approximation outside of the extent, respectively. That is, when $\|\cp_{\mathcal{S}-\mathcal{C}}(\mathbf{x}_i)\| = e$, the error in approximating the Heaviside function is $\epsilon$ and the error becomes smaller further outside of $[-e, e]$. We choose $e = r_{\Omega(\S)}$ and $\epsilon = \Delta x$ for our results. 

Step (3) of the heat method also involves a Dirichlet IBC, $\phi = 0$ on $\C$, since the geodesic distance is zero for points on $\C$. No special treatment is required for this IBC. To improve accuracy, steps (2) and (3) are applied iteratively as discussed by \citet{Belyaev2015}. Two extra iterations of steps (2) and (3) are applied in all our examples of the CPM-based heat method. 

\begin{figure*}
     \centering
     \begin{subfigure}[b]{0.33\textwidth}
     \caption*{Exact Polyhedral}
         \begin{subfigure}[b]{0.49\textwidth}
             \centering
             \includegraphics[width=0.9\textwidth]{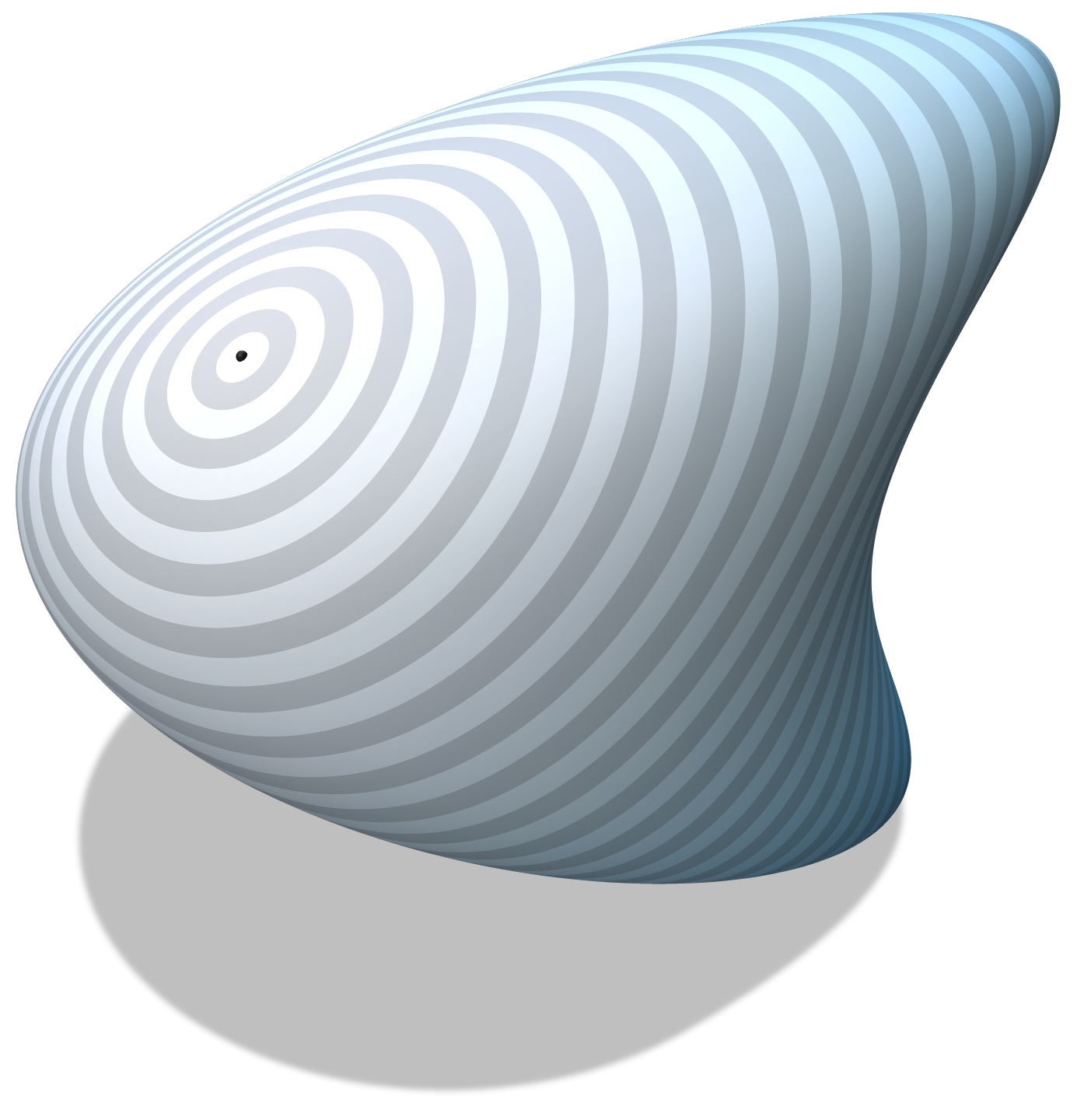}
         \end{subfigure}
         \hfill
         \begin{subfigure}[b]{0.49\textwidth}
             \centering
             \includegraphics[width=0.8\textwidth]{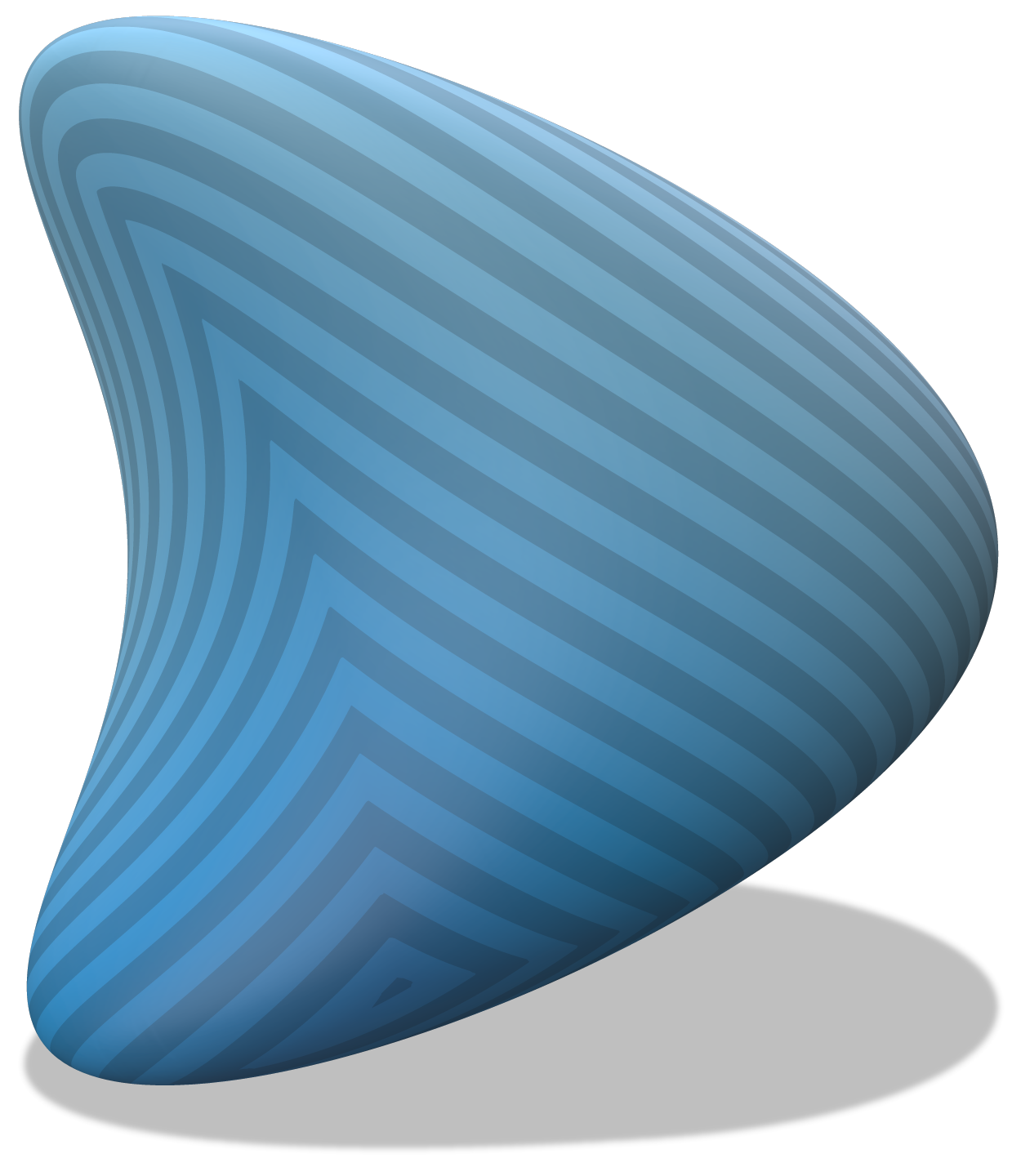}
         \end{subfigure}
     \end{subfigure}     
     \hfill
     \begin{subfigure}[b]{0.33\textwidth}
     \caption*{CPM Heat Method (Ours)}
         \begin{subfigure}[b]{0.49\textwidth}
             \centering
             \includegraphics[width=0.9\textwidth]{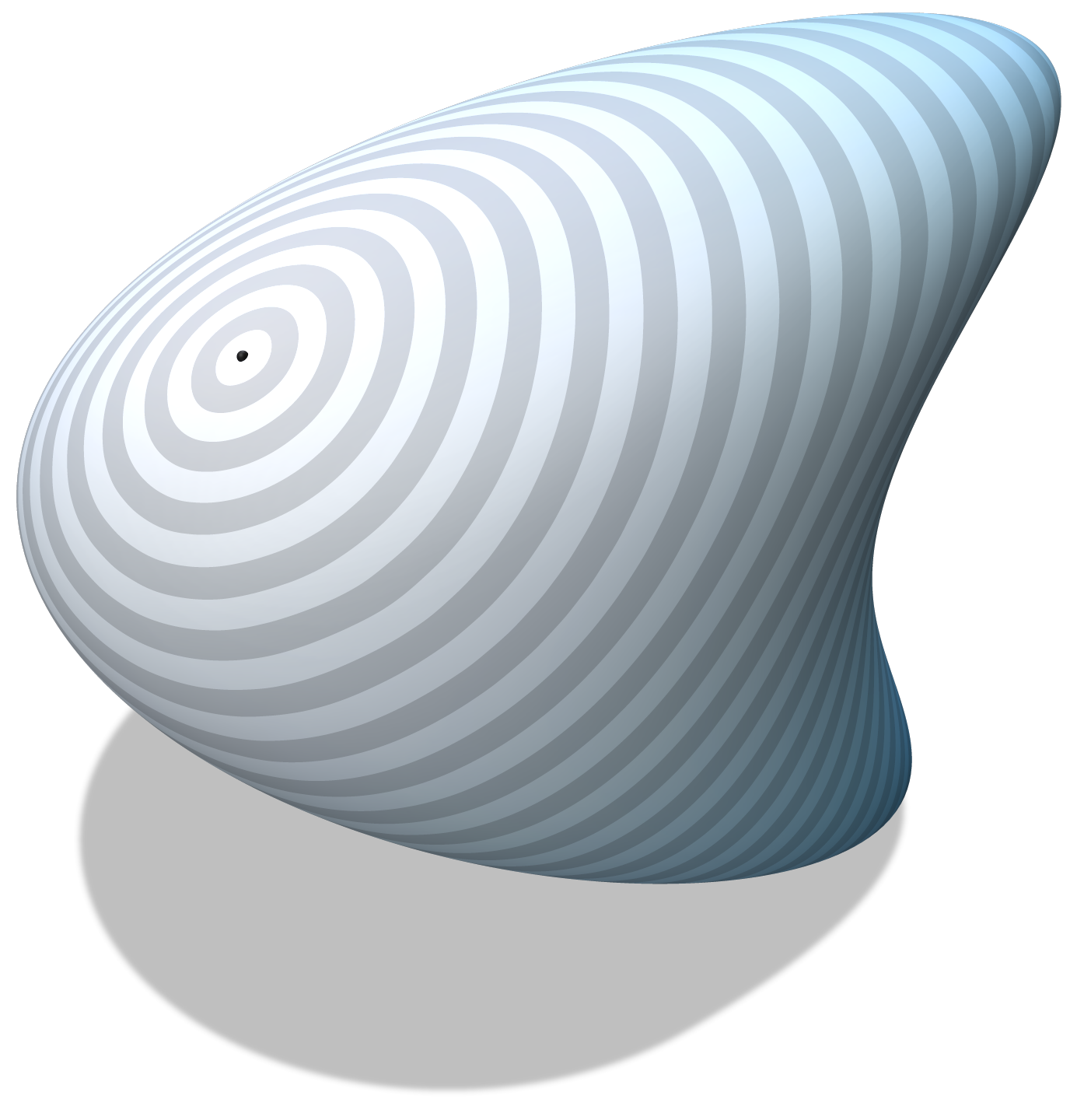}
         \end{subfigure}
         \hfill
         \begin{subfigure}[b]{0.49\textwidth}
             \centering
             \includegraphics[width=0.8\textwidth]{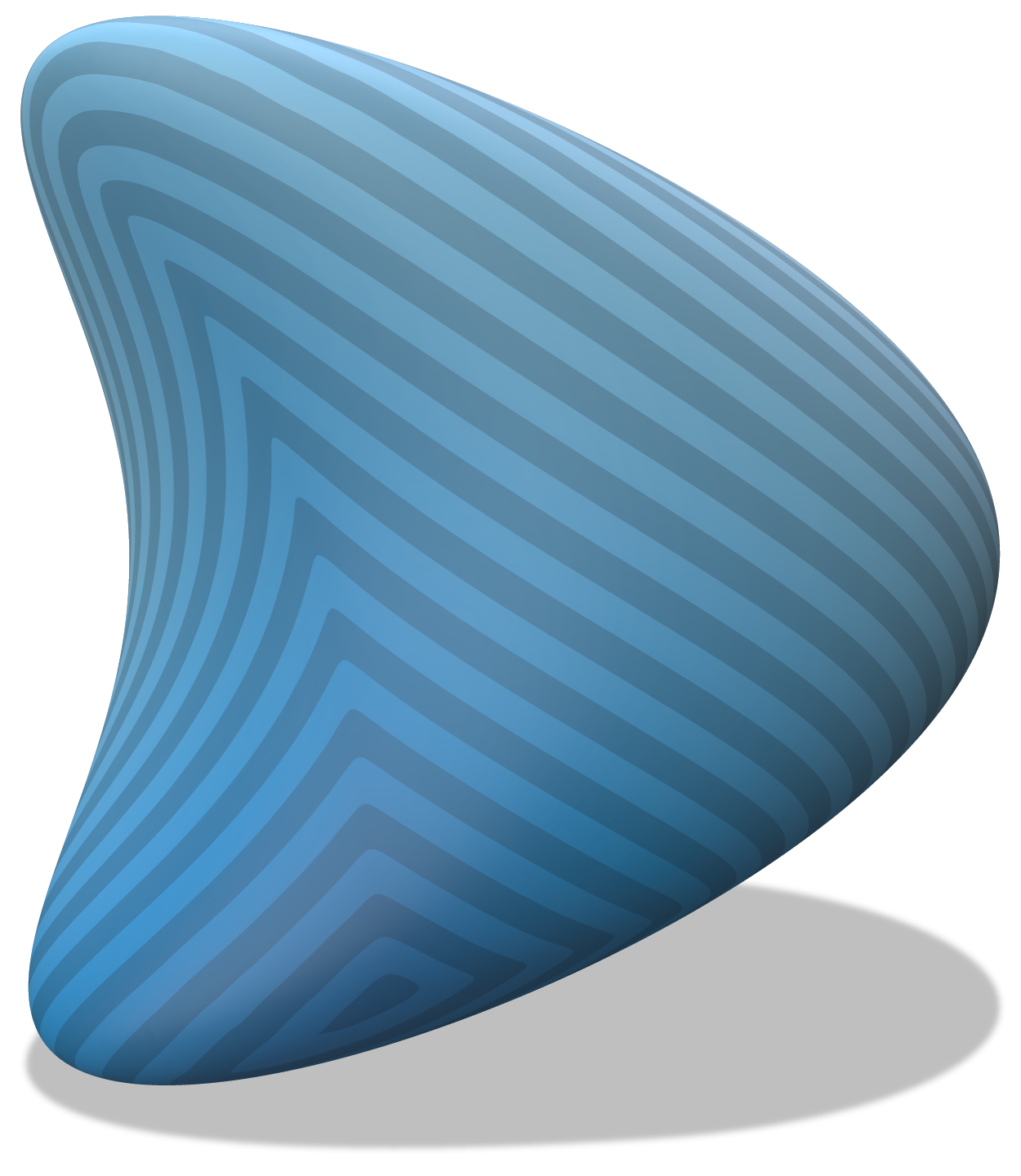}
         \end{subfigure}
     \end{subfigure}
     \hfill
     \begin{subfigure}[b]{0.33\textwidth}
     \caption*{Mesh Heat Method}
         \begin{subfigure}[b]{0.49\textwidth}
             \centering
             \includegraphics[width=0.9\textwidth]{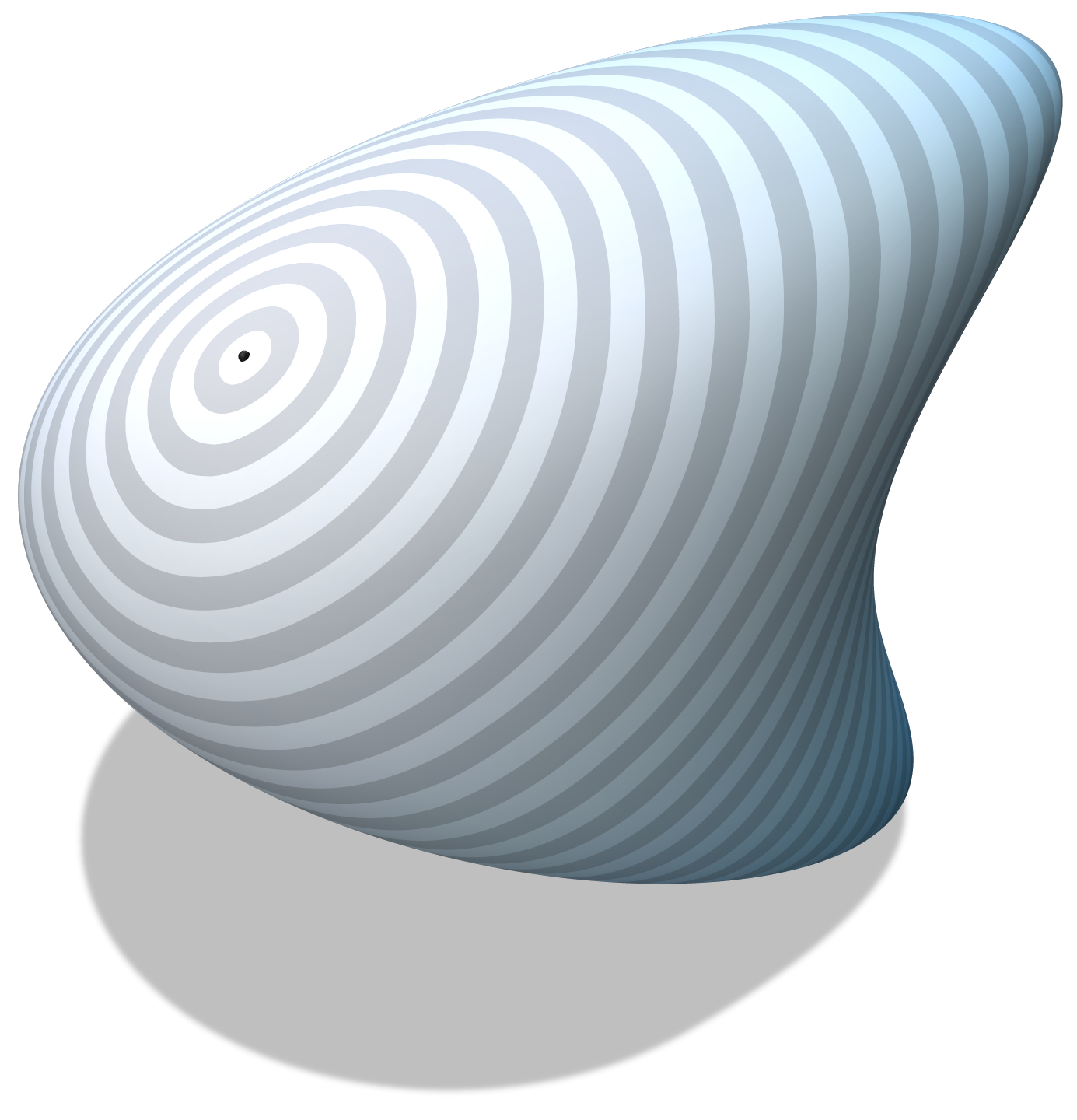}
         \end{subfigure}
         \hfill
         \begin{subfigure}[b]{0.49\textwidth}
             \centering
             \includegraphics[width=0.8\textwidth]{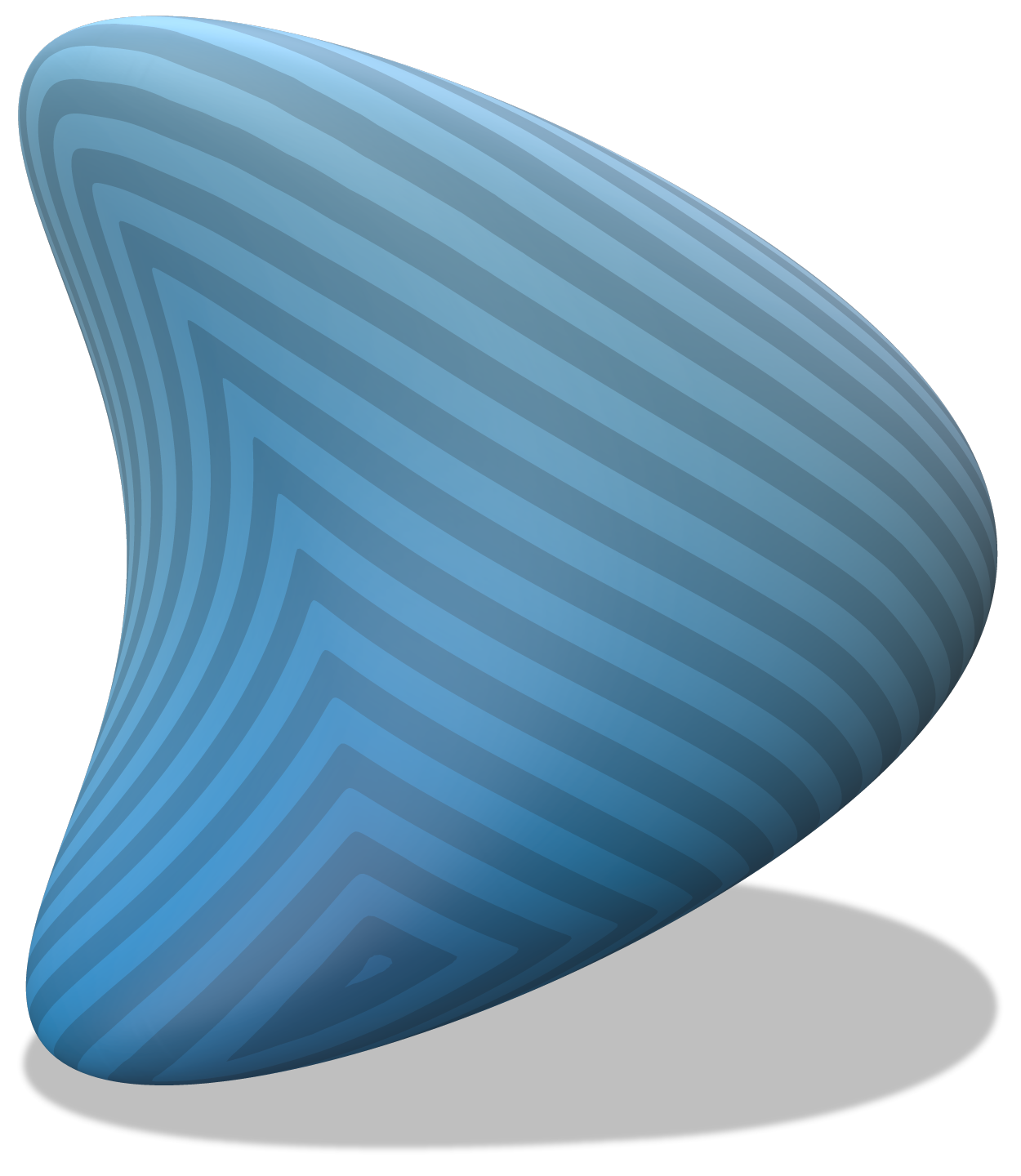}
         \end{subfigure}
     \end{subfigure}
        \caption{CPM vs.\ mesh-based methods for geodesic distances to a point on a triangulation of the Dziuk surface. Consistent results are observed.}
        \label{fig:geodesic-distance-comparison}
\end{figure*}

We use Eigen's SparseLU to solve (only) the linear systems arising from step (1) of the heat method. Using BiCGSTAB (either Eigen's or our custom solver) results in an incorrect solution despite the iterative solver successfully converging, even under a relative residual tolerance of $10^{-15}$. We observed that the small time-step of the heat method, $\Delta t = \Delta x^2,$ causes difficulties for BiCGSTAB. The reason is that values far from the heat sources are often extremely close to zero. Tiny errors in these values are tolerated by BiCGSTAB, but lead to disastrously inaccurate gradients in step (2), and thus incorrect distances in step (3). Another option is to calculate smoothed distances (see Section 3.3 of \cite{Crane2013}) using larger time-steps $\Delta t = m \Delta x^2$ with $m \geq 100$; in this scenario BiCGSTAB encounters no problems. Our partially matrix-free BiCGSTAB solver is nevertheless successfully used for step (3) of the heat method.

Figure~\ref{fig:geodesic-distance-comparison} shows the geodesic distance to a single source point on the Dziuk surface, where our CPM-based approach (with $\Delta x = 0.0125$) is compared to exact polyhedral geodesics \cite{mitchell1987discrete} and the mesh-based heat method. Implementations of the latter two methods are drawn from \texttt{geometry-central} \cite{geometrycentral}. All three approaches yield similar results.

For the example in Figure~\ref{fig:geodesic-distance-comparison}, closest points are computed from the same triangulation used in the exact polyhedral and mesh-based heat method. However, closest points can also be directly computed from the level-set Dziuk surface (as in Section~\ref{sec:conv-shifted-poisson}). To our knowledge, the heat method has not been applied on level-set surfaces before.  

\begin{figure}
     \centering
    \begin{subfigure}[b]{0.475\textwidth}
        \begin{subfigure}[b]{0.475\textwidth}
             \centering
             \includegraphics[width=\textwidth]{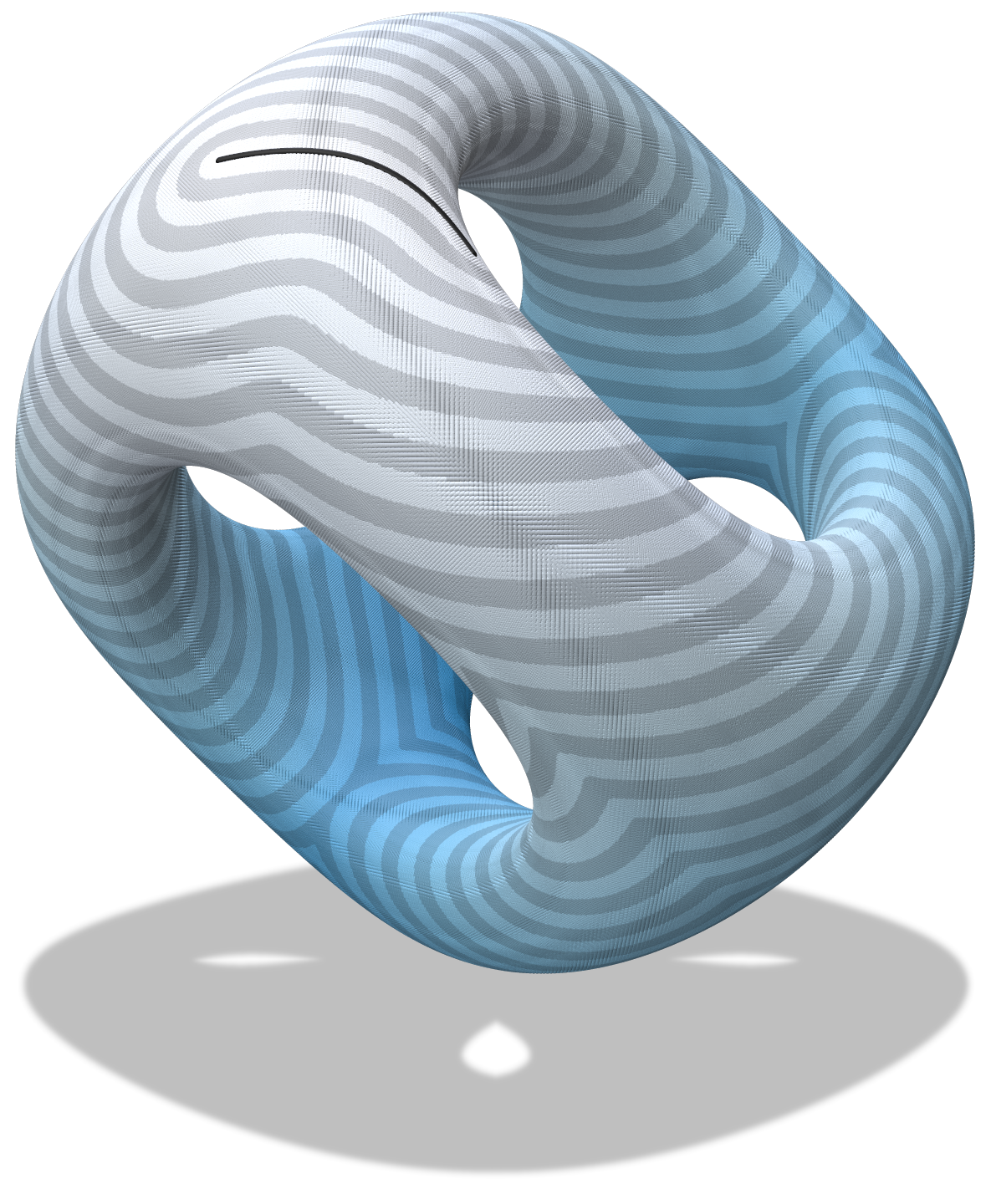}
         \end{subfigure}
             \hfill
        \begin{subfigure}[b]{0.475\textwidth}
             \centering
             \includegraphics[width=\textwidth]{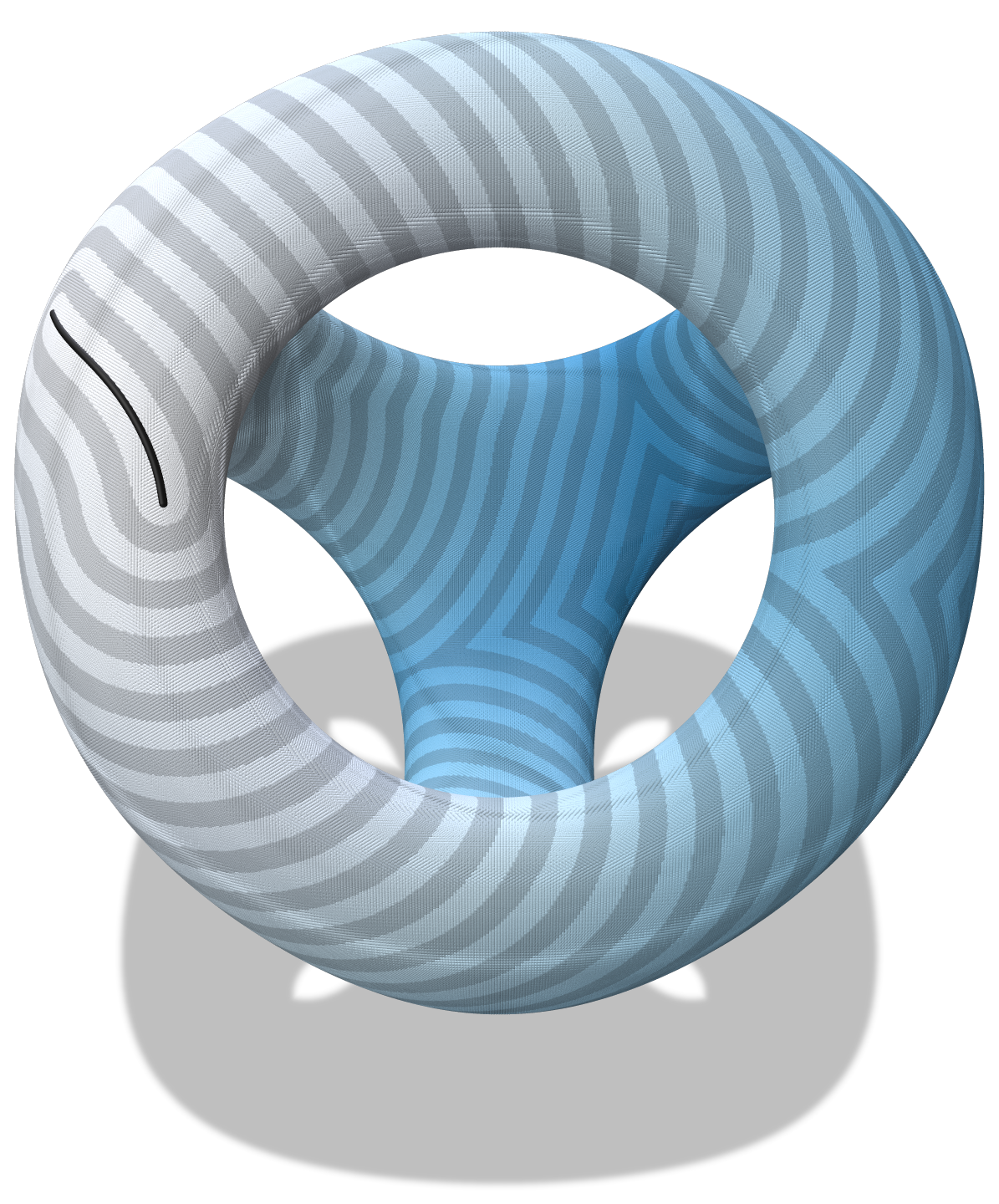}
         \end{subfigure}
     \end{subfigure}
\caption{Geodesic distance to a polyline curve (black) visualized on the ``DecoTetrahedron'' level-set surface computed using CPM with $\Delta x = 0.025$. The closest points themselves are directly rendered.}
\label{fig:geodesic-distance}
\end{figure}

We showcase the ability of our CPM to compute geodesic distance on general manifold representations. Figure~\ref{fig:geodesic-distance} visualizes the geodesic distance to an open curve on the ``DecoTetrahedron'' \cite{3DXM} level-set surface, 
\begin{align*}
\S = \bigg\{\x \in \mathbb{R}^3 \;\bigg|\; &\sum_{i = 1}^3 \left((x_i-2)^2 (x_i+2)^2 - 10 x_i^2\right) \\ 
&+ 3 \left(x_1^2 x_2^2 + x_1^2 x_3^2 + x_2^2 x_3^2\right) + 6 x_1 x_2 x_3 = -22 \bigg\}.
\end{align*}
$\S$ and $\C$ can also have mixed representations. For example, Figure~\ref{fig:teaser} (b) shows the geodesic distance (using $\Delta x = 0.00625$) to the trefoil knot (a.k.a. torus knot with $a = 2$ and $b = 3$, see~\eqref{eqn:torus-knot}) on a torus with minor and major radii 1 and 2, respectively. The trefoil knot uses a parametric representation while the torus uses an analytical closest point representation.

\subsection{Vector Field Design}
Designing tangent vector fields on surfaces is useful in many applications including texture synthesis, non-photorealistic rendering, quad mesh generation, and fluid animation \cite{de2016vector, zhang2006vector}. One approach for vector field design involves the user specifying desired directions at a sparse set of surface locations, which are then used to construct the field over the entire surface. Adapting ideas from \citet{turk2001texture} and \citet{wei2001texture}, we interpret the user-specified directions as Dirichlet IBCs and use diffusion to obtain the vector field over the whole surface.

We iterate between heat flow of the vector field and projections onto the tangent space to obtain the tangent vector field over all of $\S$. Specifically, each iteration involves the following steps:
\begin{enumerate}
    \item Perform heat flow independently for each component of $\u = [u^1, u^2, u^3]^T$ according to
\begin{equation*}
 \frac{\partial u^i}{\partial t} = \Delta_{\S} u^i, \quad {\rm with} \quad
\begin{cases}
    u^i = g^i, \;{\rm or}\\
    \nabla_{\S} u^i \cdot \b_{\C} = 0,
\end{cases}
{\rm on}\; \C,
\end{equation*}
starting from the vector field after the previous iteration.
\item Project $\u(\x_j)$ onto the tangent space of $\S$ using $\n_{\S}$ at $\cp_{\S}(\x_j)$
\begin{equation*}
    \u(\x_j) = \left(\mathbf{I} - \n_{\S} \n_{\S}^T\right) \u(\x_j).
\end{equation*}
\end{enumerate}
One time-step of heat flow is performed on each iteration using implicit Euler with $\Delta t = 0.1 \Delta x.$ A total of 10 iterations are used for all examples. The vector field for the first iteration consists of zero vectors unless the direction is specified by an IBC.

Dirichlet IBCs $\mathbf{g} = [g^1, g^2, g^3]^T$ can be specified at points or curves. For point Dirichlet IBCs the direction of $\mathbf{g}$ is any direction in the tangent space of $\S$. Dirichlet IBCs on curves could also specify any direction in the tangent space of $\S$, but designing vector fields is more intuitive when $\mathbf{g}$ is the unit tangent direction $\t_{\C}$ along $\C$. Zero-Neumann IBCs are also used within our framework to block the vector field from diffusing across $\C$.

Figure~\ref{fig:teaser} (c) shows an example of a vector field designed on the M\"obius strip using $\Delta x = 0.0064$. The M\"obius strip is actually a triangulated surface in this example, although its parametric form could be used instead (see \cite{Macdonald2011}). Zero-Neumann exterior BCs are imposed automatically by CPM with first-order accuracy on the geometric boundary. This example shows the ability of our approach to handle open and nonorientable surfaces. There are four points and two curves specifying the IBCs in Figure~\ref{fig:teaser} (c). A circular closed curve demonstrates that vortices can be created. The other curve on the M\"obius strip enforces a zero-Neumann IBC that blocks direction changes in the vector field (see Figure~\ref{fig:teaser} (c) zoom).

\begin{figure}
     \centering
    \includegraphics{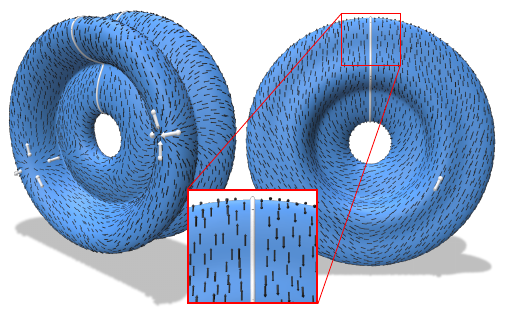}
    \caption{Vector field design on a parametric surface of revolution, with Dirichlet IBCs on a parametric curve and points shown in white.}
    \label{fig:vector-field-design}
\end{figure}

Figure~\ref{fig:vector-field-design} shows another example on a parametric surface of revolution (with $\Delta x = 0.025$), which is constructed by revolving the planar parametric curve~\eqref{eqn:planar-param-curve} with $c = \frac{1}{2}$ around the $z$-axis. All IBCs in this example are Dirichlet IBCs. Sinks and sources in the vector field are created with four Dirichlet point IBCs. The curve IBC is a two-sided Dirichlet IBC that flips the direction of the vector field across $\C$ (see Figure~\ref{fig:vector-field-design} zoom).

\begin{figure}
     \centering
     \begin{subfigure}[b]{0.49\textwidth}
         \begin{subfigure}[b]{0.49\textwidth}
                 \centering
                 \includegraphics[trim={24.5cm 0cm 22cm 0cm},clip,width=0.9\textwidth]{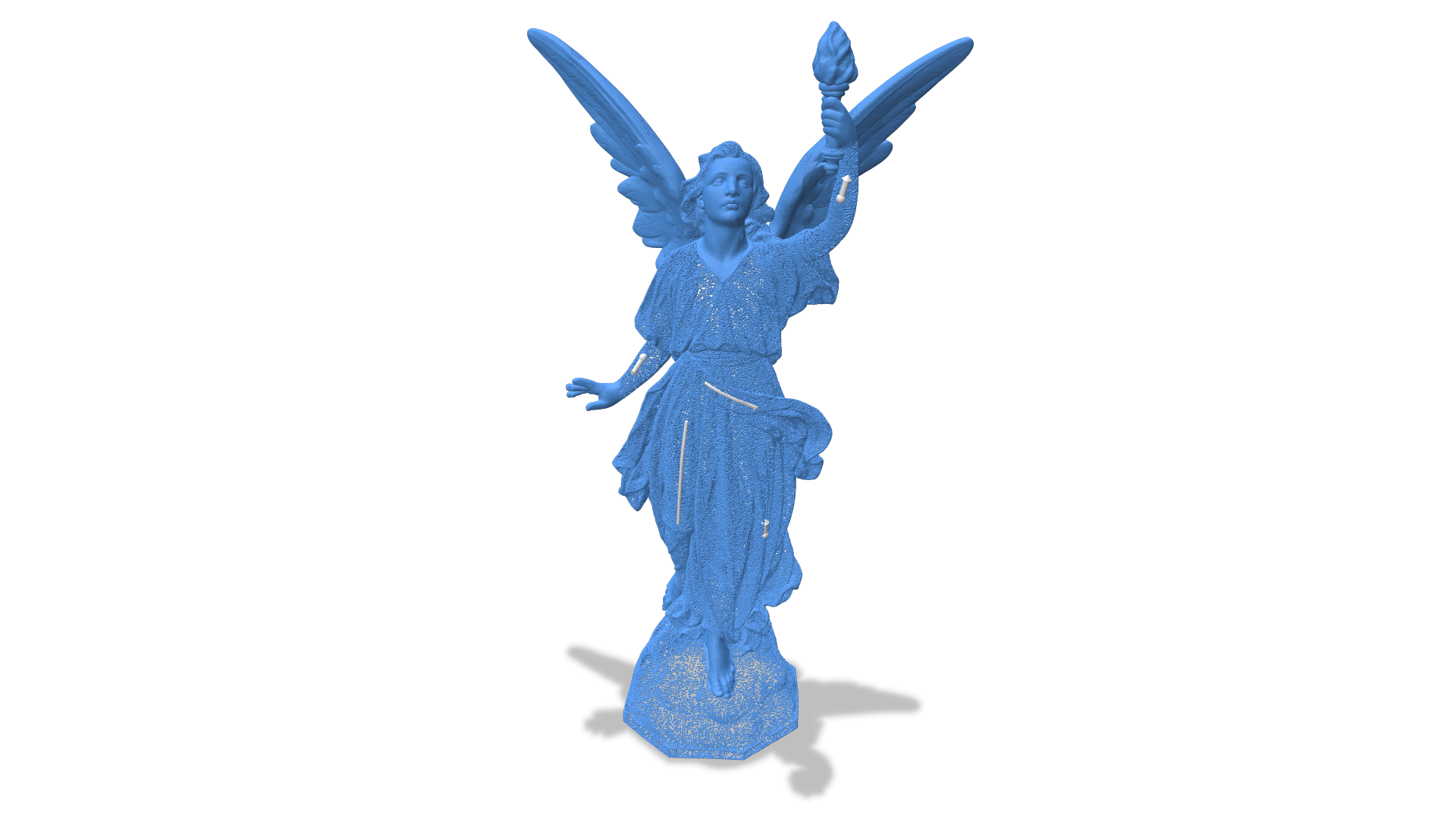}
         \end{subfigure}
         \hspace{-1.6cm}
         \begin{subfigure}[b]{0.49\textwidth}
                 \centering
           \includegraphics{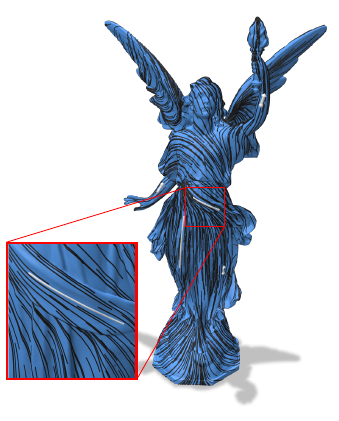}
        \end{subfigure}
    \end{subfigure}
\caption{Vector field design on a point cloud surface (left), with
Dirichlet IBCs on polyline curves and points shown in white. The resulting vector field is visualized with flow lines on a triangulation of the point cloud (right).}
\label{fig:vfd-lucy-point-cloud}
\end{figure}

A final vector field design example, on the Lucy surface, is given in Figure~\ref{fig:vfd-lucy-point-cloud}. A point cloud representation of the Lucy surface (vertices of a mesh~\cite{Stanford} with \textasciitilde1 million vertices) is used and the closest point function is defined to return the nearest neighbour; for dense enough point clouds this suffices. For less dense point clouds a smoother closest point function is required, for example using a moving-least-squares based projection method~\cite{Liu2006, Yingjie2011}. Nevertheless, the variable point density (i.e., higher density on head, wings, hands, and feet) of the Lucy point cloud in Figure~\ref{fig:vfd-lucy-point-cloud} (left) does not present any issue in this example. 

\subsection{Harmonic Maps}
A map between two manifolds, $\S_1$ and $\S_2$, matches locations on $\S_1$ with locations on $\S_2$. The map can be used to analyze differences between $\S_1$ and $\S_2$ or to transfer data from one manifold to the other. Harmonic maps are a specific type of map that appears in numerous domains, e.g., mathematical physics \cite{Bartels2005} and medical imaging \cite{Shi2007, Shi2009}. In computer graphics, harmonic maps can be used for many applications such as texture transfer, quad mesh transfer, and interpolating intermediate poses from key-frames of a character \cite{Ezuz2019}.


\citet{King2017} considered applying CPM to compute harmonic maps $\u(\y):\S_1 \rightarrow \S_2$.
Adapting their approach, we compute the harmonic map using the gradient descent flow
\begin{equation}
\begin{aligned}
\frac{\partial \u}{\partial t} &= \Pi_{T_{\u} \S_2} (\Delta_{\S_1} \u),\\
\u (\y,0) &= \mathbf{f} (\y),\\
\u(\y, t) &= \mathbf{g}(\y), \; {\rm for} \; \y\in\C_1,
\end{aligned} 
\label{eqn:harmGradDescent}
\end{equation}
where $\Pi_{T_{\u} \S_2}$ is the projection operator at the point $\u$ onto the tangent space of $\S_2$. The vector $\Delta_{\S_1} \u$ is defined componentwise, i.e., $\Delta_{\S_1} \u = [\Delta_{\S_1} u^1, \Delta_{\S_1} u^2, \Delta_{\S_1} u^3]^T.$ The $\mathbf{f}(\y)$ and $\mathbf{g}(\y)$ are the initial map (from $\S_1$ to $\S_2$) and the landmark map (from $\C_1 \subset \S_1$ to $\C_2 \subset \S_2$), respectively. The subsets $\C_1$ and $\C_2$ can be landmark points or curves on $\S_1$ and $\S_2$ that are enforced to match using our new Dirichlet IBC treatment; such IBCs were \emph{not} considered by \citet{King2017}.  

An operator splitting approach was used by \citet{King2017}, which allows~\eqref{eqn:harmGradDescent} to be solved with a PDE on $\S_1$ alone. Specifically, one time-step consists of the following:
\begin{enumerate}
\item Solve~\eqref{eqn:harmGradDescent} \emph{without the $\Pi_{T_{\u} \S_2}$ term} using CPM on $\Omega(\S_1)$ with $\Omega(\C_1)$ to enforce the IBC.
\item Project the solution from (1) onto $\S_2$.
\end{enumerate}
Denote the solution from step (1) at $\x_i\in \Omega(\S_1)$ and time-step $k$ by $\bfv^k_i$. The projection in step (2) simply moves $\bfv^k_i$ to its closest point on $\S_2$ by setting $\u^k_i = \cp_{\S_2}(\bfv^k_i)$. One time-step of explicit Euler is used for step (1) with $\Delta t = 0.1 \Delta x^2$ starting from $\u^{k-1}$.

To perform the above gradient descent flow a valid initial map $\u^0$ is needed to start from. Generating such initial maps in the general case has not yet been addressed for CPM \cite{King2017}. Approaches based on geodesic distance to landmark curves/points $\C_1,$ $\C_2$ could potentially be adapted \cite{Ezuz2019, Shi2007}. However, for our illustrative example of incorporating IBCs while computing harmonic maps, we opt for a simple (but restrictive) initial map construction. The surface $\S_1$ is given by a triangulation and deformed to create $\S_2$ while maintaining the same vertex connectivity. Therefore, the barycentric coordinates of each triangle can be used to initially map any point on $\S_1$ to a point on $\S_2$.

\begin{figure}
     \centering
     \begin{subfigure}[b]{0.47\textwidth}
         \begin{subfigure}[b]{0.49\textwidth}
                 \centering
                 \caption{$\S_1$ with Texture}
                 \vspace{5pt}
                 \includegraphics[trim={15cm 0cm 16cm 4cm},clip,width=\textwidth]{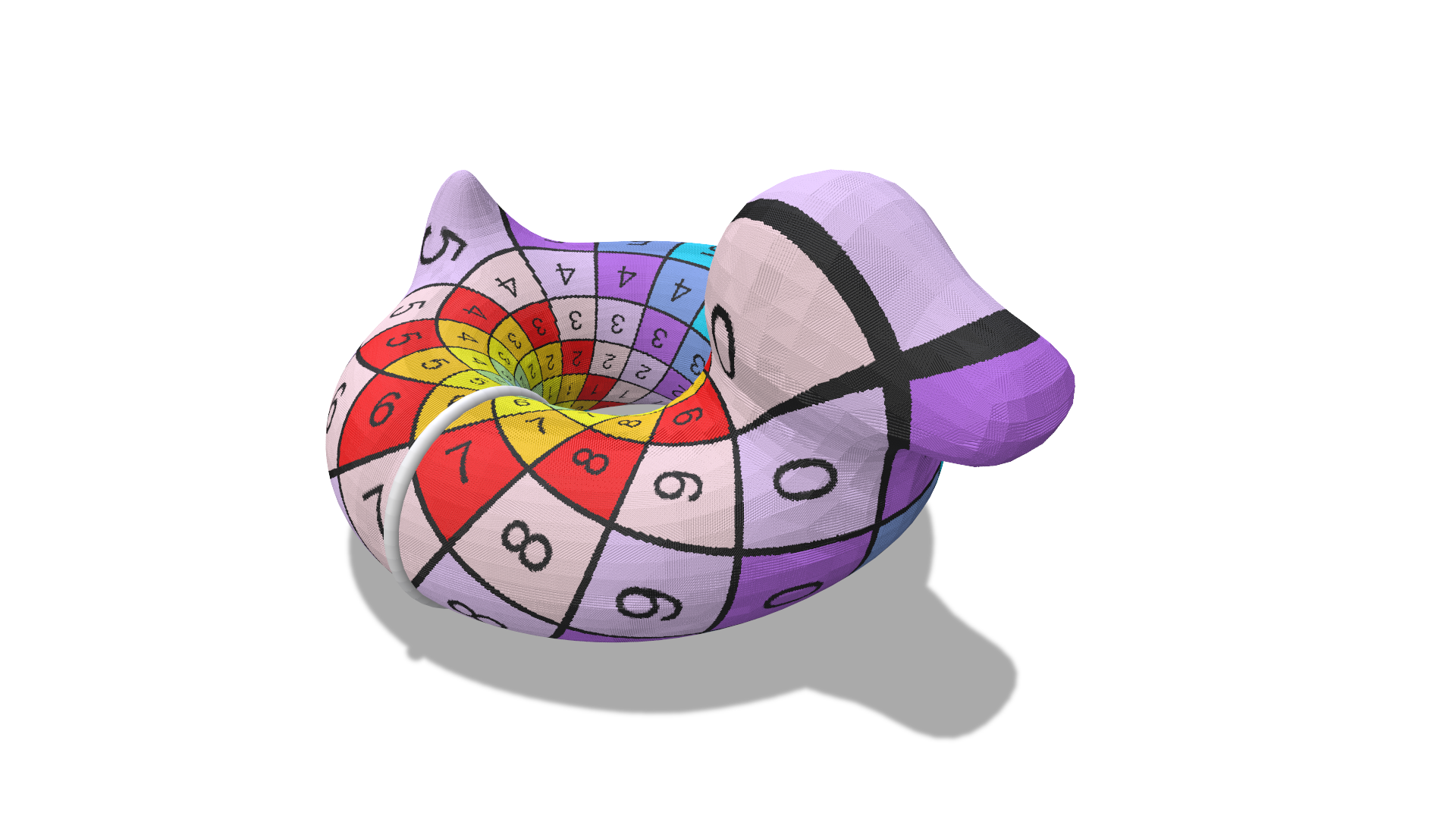}
             \label{fig:map-source}
         \end{subfigure}
         \hfill
         \vspace{-22pt}
         \begin{subfigure}[b]{0.49\textwidth}
                 \centering
                 \caption{Initial Map onto $\S_2$}
                 \vspace{5pt}
                 \includegraphics[trim={15.5cm 0cm 15.5cm 4cm},clip,width=\textwidth]{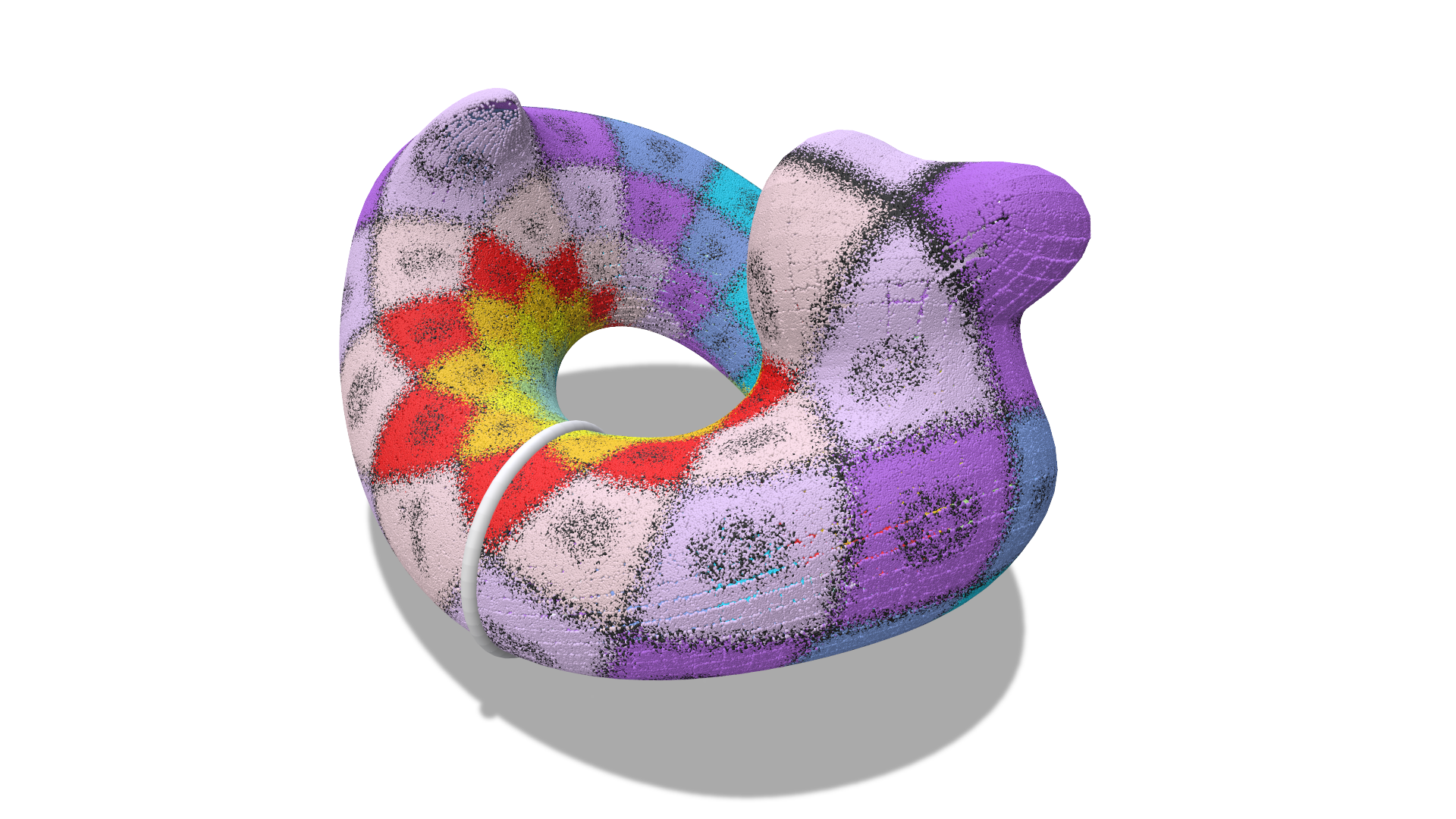}
             \label{fig:map-initial-noisy}
        \end{subfigure}
    \end{subfigure}
\hfill
    \begin{subfigure}[b]{0.47\textwidth}
        \begin{subfigure}[b]{0.49\textwidth}
             \centering
           \includegraphics{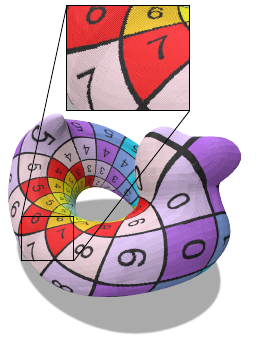}
             \caption{Harmonic Map onto $\S_2$ w/o IBCs}
             \label{fig:map-without-constraint}
         \end{subfigure}
         \hfill
        \begin{subfigure}[b]{0.49\textwidth}
             \centering
           \includegraphics{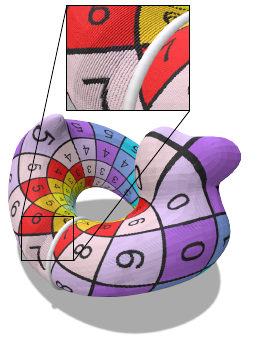}
             \caption{Harmonic Map onto $\S_2$ w/ IBCs}  
             \label{fig:map-with-constraint}
         \end{subfigure}
    \end{subfigure}
\caption{Maps from $\S_1$ to $\S_2$ with a texture for visualizing the mapping. Landmark curves (Dirichlet IBCs) $\C_1$ and $\C_2$ are shown in white. (a) $\S_1$ with texture. (b) $\S_2$ with texture from a noisy initial map. (c) $\S_2$ with a CPM harmonic mapped texture without IBCs. (d) $\S_2$ with a harmonic mapped texture using our CPM approach satisfying the IBCs. The surfaces are displayed as point clouds. The $\cp_{\S_1}$ and $\cp_{\S_2}$ are computed from triangulations, while $\cp_{\C_1}$ and $\cp_{\C_2}$ are computed from polylines.}
\label{fig:harmonic-map}
\end{figure}

Figure~\ref{fig:harmonic-map} shows an example of computing harmonic maps from the Bob~\cite{Bob-mesh} surface $\S_1$ to its deformed version $\S_2$. Grid spacing $\Delta x = 0.00663$ is used for $\Omega(\S_1)$.  The surfaces are visualized as point clouds. $\S_1$ is visualized with the set of closest points of grid points in $\Omega(\S_1)$. Each point in the point cloud for $\S_1$ has a corresponding point location on $\S_2$ given through the mapping $\u$. 
A texture is added to the surface of $\S_1$ and transferred to $\S_2$ through the mapping $\u$.

To emphasize the effect of computing the harmonic map, noise is added to the initial map (see Figure~\ref{fig:harmonic-map} (b)) before performing the gradient descent flow. The gradient descent flow is evolved to steady state using 1500 and 200 time steps with and without the IBC in Figure~\ref{fig:harmonic-map} (d) and (c), respectively. The harmonic map with a Dirichlet IBC stretches on one side of $\C_2$ and compresses on the other side to satisfy both the PDE and IBC. Comparing the zoom of Figure~\ref{fig:harmonic-map} (c) and (d), the point cloud density in (d) is more sparse on one side of $\C_2$ than in (c) due to the stretching of the map, leaving visual gaps between points in the cloud. The distortion is expected unless the IBC map $\mathbf{g}$ is a harmonic map itself.

\subsection{Reaction-Diffusion Textures}
\label{sec:RD-textures}
Much research in geometry processing has focused on Poisson and diffusion problems. There are however applications that require solving more general PDEs, e.g., reaction-diffusion textures~\cite{turk1991generating}. Reaction-diffusion textures involve solving coupled equations on surfaces. These PDEs can form patterns from random initial conditions and have been solved on meshes~\cite{turk1991generating}, level sets~\cite{Bertalmio2001}, and closest point surfaces~\cite{Macdonald2013}. Here we impose IBCs to control regions of the texture, emphasizing the generality of CPM, and our novel boundary condition treatment, with respect to PDE type.

\begin{figure}
     \centering
     \begin{subfigure}[b]{0.47\textwidth}
                 \centering
                 \includegraphics[width=\textwidth]{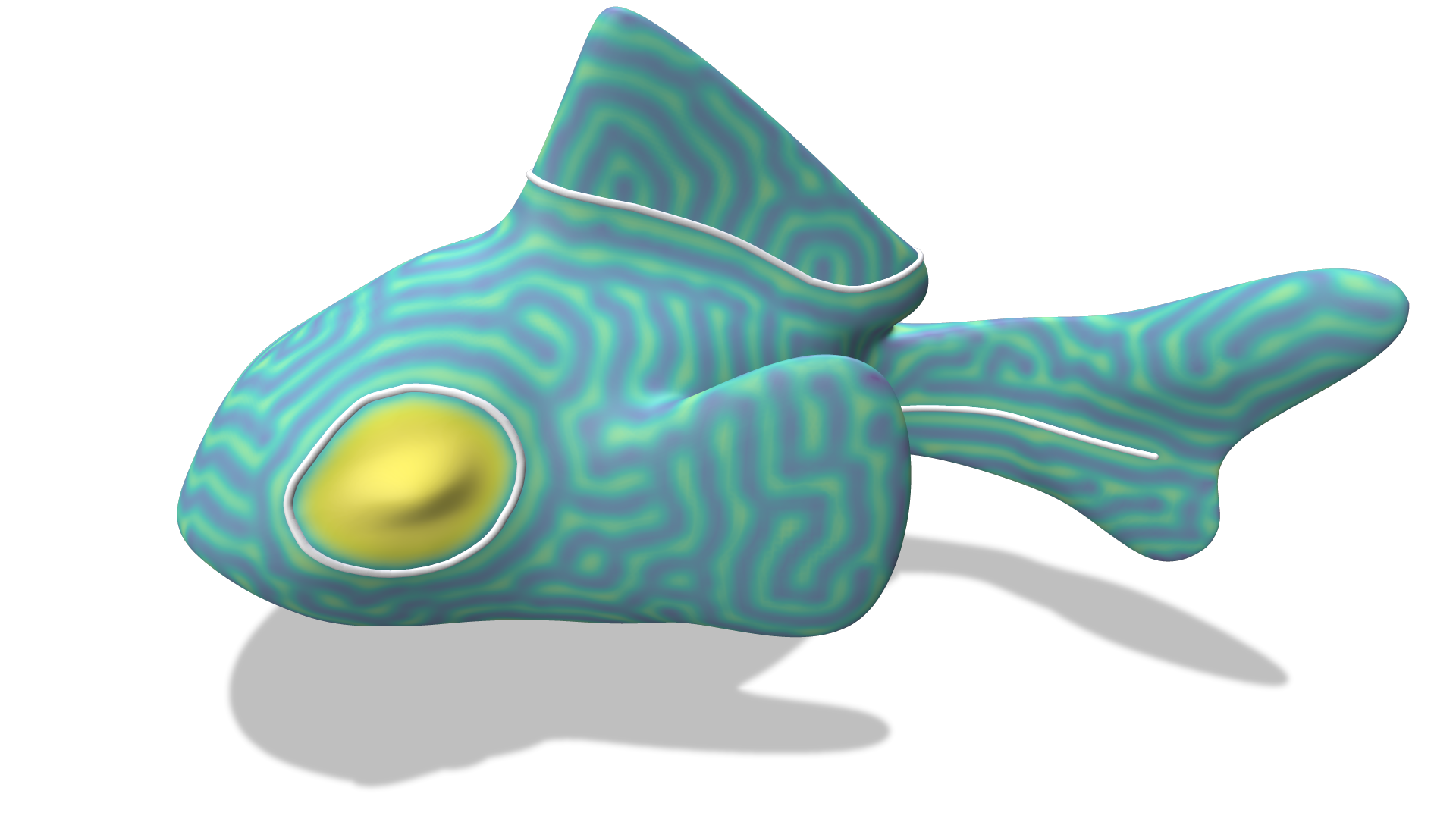}
    \end{subfigure}
\caption{Reaction-diffusion texture on a fish surface with zero Dirichlet IBCs around the eye and on the tail. A two-sided zero Dirichlet-Neumann IBC is imposed on the dorsal fin. The surface is coloured yellow for high concentrations of reactant $u$ and purple for low concentrations. The $\cp_{\S}$ are computed from a triangulation, while the $\cp_{\C}$ are computed from polylines.}
\label{fig:RD-fish}
\end{figure}

The Gray-Scott model~\cite{pearson1993complex}
\begin{equation}
\begin{cases}
 \frac{\partial u}{\partial t} = \mu_u \Delta_{\S} u -uv^2 + F(1-u), \\
 \frac{\partial v}{\partial t} = \mu_v \Delta_{\S} v + uv^2 -(F+k)v, 
 \end{cases}
 \end{equation}
 with
 \begin{equation}
\begin{cases}
    u = g \; {\rm or} \; \nabla_{\S} u \cdot \mathbf{b}_{\C} = 0,\\
    v = h \; {\rm or} \; \nabla_{\S} v \cdot \mathbf{b}_{\C} = 0,
\end{cases}
{\rm on}\; \C,
\end{equation}
is solved with CPM. Figure~\ref{fig:RD-fish} shows $u$ on a fish~\cite{fish-mesh} for a set of IBCs. The constants $\mu_u = 1.11\times 10^{-5}$, $\mu_v = \mu_u / 3$, $F = 0.054$, $k = 0.063$ are used with forward Euler time-stepping until $t = 10,000$ with $\Delta t = 0.9$ and $\Delta x = 0.01$. The initial condition is taken as $u = 1 - p$, $v = p/2$ where $p$ is given by small random perturbations around $$\frac{1}{2} e^{100(z - 0.1)^2} + \frac{1}{2}.$$  Zero Dirichlet IBCs allow stripes to be placed around the dorsal fin and tail. The upper side of the dorsal fin IBC is a zero Neumann IBC, which causes the pattern to intersect perpendicular to the IBC curve. A closed (zero Dirichlet) IBC curve allows for control of concentrations of the reactants $u$ and $v$ in the eye.

\section{Limitations and Future Work}
As we have discussed and demonstrated, CPM is a powerful tool for solving manifold PDEs since it provides a unified framework for general manifold characteristics, general manifold representations, and general PDEs. Our work extends CPM to solve manifold PDEs with interior boundary conditions (Dirichlet and zero-Neumann) while obtaining up to second-order accuracy. The ability to enforce IBCs enables CPM to be applied to many PDE-based geometry processing tasks and applications which were not previously possible. Additionally, we have developed a runtime and memory-efficient implementation allowing for the treatment of higher-detail surfaces without specialized hardware. To encourage wider adoption of CPM, we have made the code for our framework publicly available at \url{https://github.com/nathandking/cpm-ibc}. Below, we outline some of CPM's existing limitations and describe a few exciting directions for future work.



\subsubsection*{Grid Resolution in Practice}
Existing CPM theory assumes a unique closest point function $\cp_{\S}$ in the computational tube $\Omega(\S)$. 
For general $\S$, the closest point $\cp_{\S}(\x)$ is rarely unique for all $\x \in \mathbb{R}^d$. For smooth, compact manifolds, however, $\cp_{\S}(\x)$ is unique for $\x$ in a tubular neighbourhood $\N(\S) \subseteq \mathbb{R}^d$ surrounding $\S$ with sufficiently small tube radius $r_{\N(\S)}$ \cite{Marz2012}.

Uniqueness of $\cp_{\S}$ is equivalent to requiring $\N(\S) \cap {\rm med}(\S)  = \emptyset$, since by definition the medial axis of $\S$, denoted ${\rm med}(\S),$ is the subset of $\mathbb{R}^d$ that has at least two closest points on $\S$. The ${\rm reach}(\S)$ is the minimum distance from $\S$ to ${\rm med}(\S)$. Thus, for a uniform radius tube, to ensure uniqueness of $\cp_{\S}$ the tube radius must satisfy $r_{\N(\S)} < {\rm reach}(\S)$.  Hence, $\N(\S)$ depends on the geometry of $\S$ since ${\rm reach}(\S)$ depends on curvatures and bottlenecks (thin regions) of $\S$ (see Section 3 of \cite{Aamari2019}). 

In the discrete setting, the computational tube-radius $r_{\Omega(\S)}$ must be less than ${\rm reach}(\S)$. Rearranging~\eqref{eqn:bandwidth} means $\Delta x$ must satisfy
\begin{equation*}
    \Delta x < \frac{{\rm reach}(\S)}{\sqrt{(d-1) \left(\frac{p+1}{2} \right) ^2 + \left(q+\frac{p+1}{2}\right)^2}}
\end{equation*}
to ensure a unique $\cp_{\S}$ on $\Omega({\S})$. However, in practice CPM can often be used successfully with larger $\Delta x$, depending on the PDE to be solved and the accuracy requirements of the application.

\begin{figure}
     \centering
     \begin{subfigure}[b]{0.49\textwidth}
             \centering
             \includegraphics[width=0.873\textwidth]{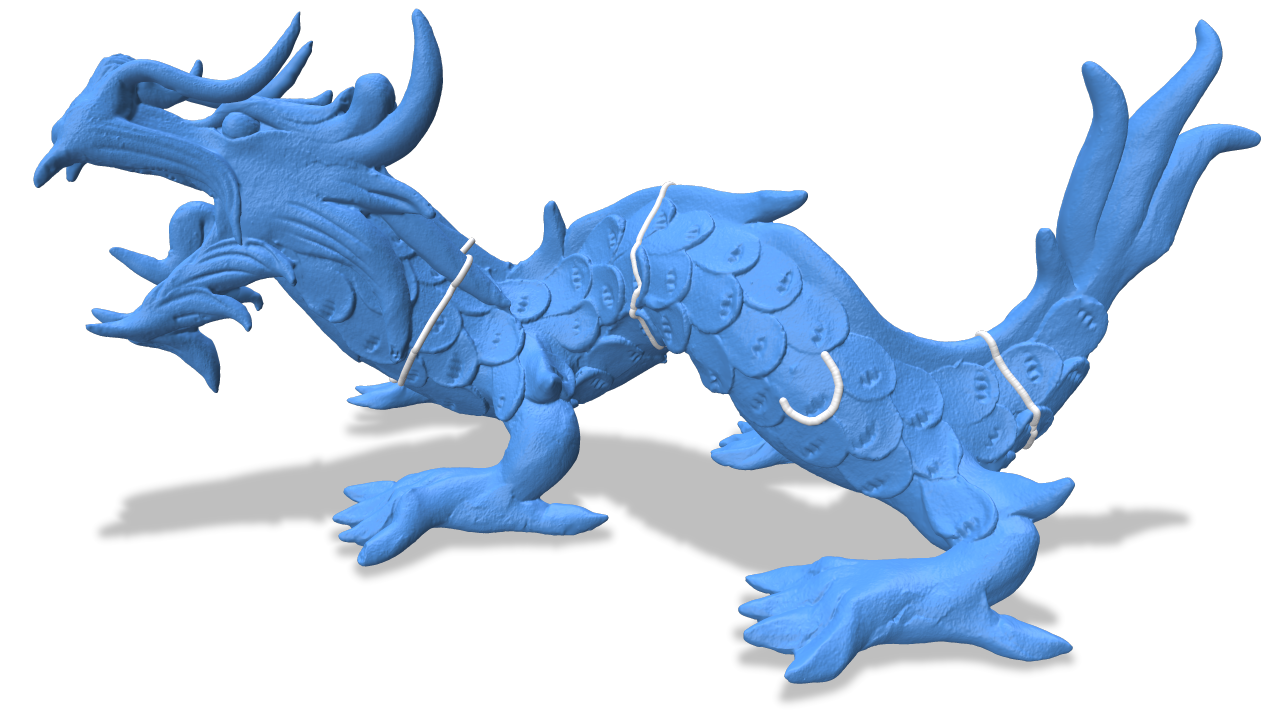}
     \end{subfigure}     
     \hfill
     \vspace{-0.45cm}
     \begin{subfigure}[b]{0.49\textwidth}
             \centering
            \includegraphics[scale=0.97]{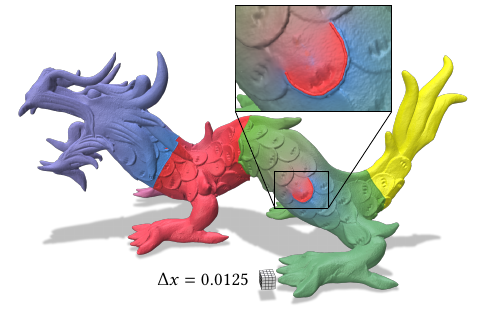}
     \end{subfigure}
     \hfill
     \vspace{-0.5cm}
     \begin{subfigure}[b]{0.49\textwidth}
             \centering
            \includegraphics[scale=0.97]{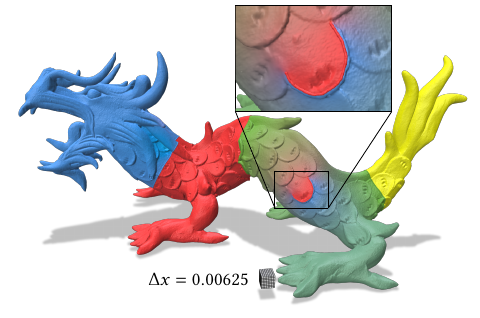}
     \end{subfigure}
     \hfill
     \vspace{-0.5cm}
     \begin{subfigure}[b]{0.49\textwidth}
             \centering
            \includegraphics[scale=0.97]{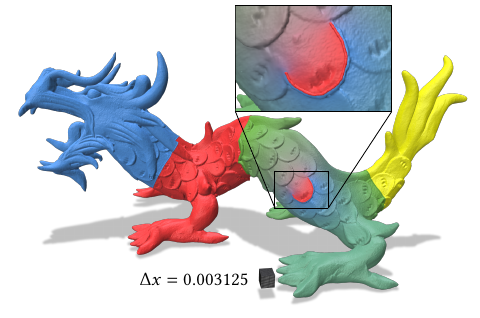}
     \end{subfigure}
    \caption{Results for three grid resolutions used to solve a diffusion curves problem to colour the surface of a dragon. The resolution is illustrated by a small block of grid cells (best viewed by zooming). The $\cp_{\S}$ are computed from a triangulation, while the $\cp_{\C}$ are from polylines.}
    \label{fig:DC-dragon}
\end{figure}

In many graphics applications the visual appearance is paramount. Consider a diffusion curves example on a dragon~\cite{Stanford}. Figure~\ref{fig:DC-dragon} shows the resultant surface colouring at different grid resolutions. 
Artifacts can be observed for $\Delta x = 0.0125$: unintended blending of blue and red on the head yields purple, while the zoomed-in dragon scale incorrectly shows hints of blue appearing in a red region. For $\Delta x = 0.003125$ (and arguably $\Delta x = 0.00625$) the result has converged to a visually acceptable, artifact-free result. However, the $\Delta x$ required to give a unique $\cp_{\S}$ for the dragon is $\Delta x < 1.28 \times 10^{-6}$. This assumes no thin bottlenecks exist, i.e., ${\rm reach(\S)}$ is computed based on only principal curvatures (computed directly on the mesh using \texttt{geometry-central} \cite{geometrycentral}). Therefore, $\Delta x$ has always been determined empirically for practical applications of CPM.
 
The need to choose $\Delta x$ experimentally is a limitation that costs the user time. A priori determination of a ``correct'' grid spacing $\Delta x$ is an open challenge: it will require knowledge about the specific PDE to be solved, the manifold it is to be solved on, and the accuracy requirements (perceptual, numerical, etc.) of the user. In general, a priori error estimation has been rare in computer graphics applications. A notable exception is the p-refinement FEM scheme of \citet{schneider2018decoupling}, which uses an a priori error estimate based on the geometry of the (volumetric) domain.

\subsubsection*{BC Types, Higher-Order Accuracy, and Other PDEs}
CPM work to date has only addressed Dirichlet and zero-Neumann (exterior) BCs. \citet{Macdonald2013} solved a surface-to-bulk coupled PDE with Robin BCs on the boundary of the bulk (but $\S$ was closed, i.e., $\partial \S = \emptyset$). Extending CPM to impose inhomogeneous-Neumann, Robin, and other types of BCs is an important area of future work. Fortunately, the interior BC framework developed here directly generalizes existing CPM approaches for exterior BCs; therefore, our work likely makes any future extensions of CPM for other exterior BC types immediately applicable as interior BCs as well. 

Third-order and higher (exterior and interior) BCs are also important for higher-order PDE discretizations. CPM itself extends naturally to higher order, but CPM with higher-order exterior BCs has not yet been explored. \citet{Macdonald2011} pointed out that a replacement for $\overline{\cp}_{\S}$ is required to incorporate the curvature of $\S$ near $\partial \S$. For higher-order interior BCs an improved $\S_{\perp}$ crossing test~\eqref{eqn:two-point-crossing-test}, involving curvatures of $\S$ near $\C,$ is likely also needed.

We primarily focused on Poisson and diffusion problems, but CPM has been applied to numerous other PDEs (see Section~\ref{sec:related-work}). In principle, our approach to IBC enforcement should also readily extend to those settings. This was confirmed for reaction-diffusion equations in Section~\ref{sec:RD-textures}. Extending CPM to approximate previously unexplored operators, such as the relative Dirac operator \cite{Liu2017} or the connection Laplacian \cite{Sharp2019}, would allow other geometry processing applications to benefit from CPM. 

\subsubsection*{Efficiency}
The discrete setup using a uniform grid near $\S$ was chosen for its simplicity and use of well-studied Cartesian numerical methods (i.e., Lagrange interpolation and finite differences). However, the ideal radius of CPM's computational tube is dictated by the curvature and/or bottlenecks of $\S$ and $\C$ (see Section~\ref{sec:cpm-continuous}). Higher curvatures or narrow bottlenecks force the uniform grid spacing to be small, leading to inefficiency due to a large number of DOFs. 

One way to improve the runtime and memory efficiency of CPM on uniform grids is to use parallelization on specialized hardware, e.g., GPU \cite{Auer2012} or distributed memory \cite{may2022closest}. However, the number of DOFs with a uniform grid can be higher than necessary, since the grid is allowed to be coarser in low curvature regions and away from tight bottlenecks of $\S$ and $\C$. Near bottlenecks with \emph{low} curvature, duplicate DOFs on either side of the medial axis could be introduced to avoid refining while ensuring data is extended from the correct part of $\S$ (similar to how the current work distinguishes different sides of an IBC). This would result in a nonmanifold grid similar to the work of \citet{Mitchell2015} and \citet{chuang2013grid}. Conversely, near high curvature regions, spatial adaptivity (e.g., octrees) could be used to provide locally higher resolution. Combining duplicate DOFs and adaptivity is, therefore, a promising direction to make CPM more efficient (both in runtime and memory) for complex surfaces, without recourse to specialized hardware.

Exploring other approximations of the CP extension and differential operators in the Cartesian embedding space could also improve efficiency. 
For example, combining Monte Carlo methods \cite{Sawhney2020, Sawhney2022, rioux2022monte, Sugimoto:2023:WoB} with CPM is one interesting avenue. Monte Carlo methods can avoid computing the global solution, so they may be more efficient when the solution is only desired on a local portion of $\S$.

\subsubsection*{Smoothness of $\S$ and $\C$}
Most CPM work and theory is based on smooth manifolds. 
However, WENO interpolation has been used to improve the grid-based CPM (i.e., the form used in this paper) for nonsmooth surfaces (e.g., surfaces with sharp features) \cite{Macdonald2008, Auer2012}. \citet{cheung2015localized} used duplicated DOFs (similar to the current work) near the sharp feature with a radial-basis function discretization of CPM. However, such discretizations can suffer from ill-conditioned linear systems. Therefore, it would be interesting to instead explore altering stencils (similar to our IBC approach) for the grid-based CPM near sharp features to use data from the ``best side'' of the sharp feature. In this context, the BC curve $\C$ would instead be the sharp feature and the PDE is still imposed on $\C$ instead of a BC.

The theoretical restriction of smoothness also applies to the curve $\C$. Therefore, our IBC approach is theoretically restricted to curves without kinks or intersections. In practice, we are still able to obtain the expected result when $\C$ has sharp features or intersections, e.g., Figure~\ref{fig:teaser} (a) involves many intersecting curves (in the band of the headdress) that also create sharp corners. Similarly CPM gives expected results in practice for mixed-codimensional objects as seen in Figure~\ref{fig:mixed-codim-diff-curves} where sharp features are present when differing codimensional pieces meet (one does however observe a decrease in the empirical convergence order). The development of a sound theoretical understanding of CPM's behaviour near sharp features and intersections is interesting future work.\\

\noindent CPM offers an exciting, unified framework for manifold PDEs on ``black box'' closest point representations, which we have extended with accurate interior BCs. Above, we have outlined a partial roadmap of CPM's significant untapped potential; we hope that others in the computer graphics community will join us in exploring it.

\begin{acks}
Nathan King was supported in part by the QEII-GSST and Ontario Graduate Scholarships.
Mridul Aanjaneya was supported in part by the National Science Foundation under awards CCF-2110861, IIS-2132972, IIS-2238955 and CCF-2312220 as well as a research gift from Red Hat, Inc. and Houdini licenses from SideFX Software. Any opinions, findings and conclusions, or recommendations expressed in this material are those of the authors and do not necessarily reflect the views of the National Science Foundation.
Steven Ruuth was supported in part by the NSERC Discovery grant program (RGPIN 2022-03302).
Christopher Batty was supported in part by the NSERC Discovery grant program (RGPIN-2021-02524) and the CFI-JELF program (Grant 40132).
\end{acks}

\bibliographystyle{ACM-Reference-Format}
\bibliography{InteriorBCs}

\appendix

\section{Closest Point Computation}
\label{sec:cp_computation}
Some manifolds allow closest points to be computed analytically, e.g., lines, circles, planes, spheres, cylinders, and tori. We use the analytical expressions for exact closest points in all examples for which they exist. For parameterized manifolds, closest points can be computed using standard numerical optimization techniques, e.g., \citet{Ruuth2008} used Newton's method for various manifolds, such as a helix. For examples in this paper, we solve 
$$\argmin_{\mathbf{t}} \frac{1}{2} \|\mathbf{p}(\mathbf{t}) - \x_i\|^2,$$ 
for the parameters $\mathbf{t}$ (e.g., $\mathbf{t} = t$ for a 1D curves and $\mathbf{t} = [u,v]^T$ for a 2D surface), where $\mathbf{p}(\mathbf{t})\in\S$ and $\x_i\in\Omega(\S)$. LBFGS++ \cite{LBFGSpp} is used to solve the optimization problem. An initial guess for $\cp_{\S}(\x_i)$ is taken as the nearest neighbour in a point cloud $\mathcal{P}_{\S}$ of the parametric manifold. The point cloud $\mathcal{P}_{\S}$ is constructed using $N$ equispaced points of the parameter $\mathbf{t}$. 

Computing closest points to triangulated surfaces is also well-studied \cite{Strain1999, Mauch2003, Auer2012}. Notably, the work of \citet{Auer2012} implements the closest point evaluation on a GPU. There also exist open source libraries that support computing closest points to triangle meshes, e.g., \texttt{libigl} \cite{libigl}. Here we use the library \texttt{fcpw} \cite{fcpw} to compute closest points to triangulated surfaces and polyline curves.

The simplest way to compute closest points to a point cloud is to take the nearest neighbour as the closest point. As discussed by \citet{Macdonald2013} this choice can be inaccurate if the point cloud is not dense enough. \citet{wang2020codimensional} (Figure 17) showed the inaccuracy of using nearest neighbours as closest points with CPM on a diffusion problem. Several more accurate approaches for closest points to point clouds have been developed \cite{Martin2020, Petras2022, Liu2006, Yingjie2011}. 

Closest points can also be computed from analytical signed-distance functions $d(\x)$ as 
\begin{equation}
\cp_{\S}(\x) = \x - d(\x) \nabla d(\x).
\label{eqn:cp_from_sd}
\end{equation}
Equation~\eqref{eqn:cp_from_sd}, however, is not valid for more general level-set functions $\phi$. High-order accuracy of closest points from level-set functions (sampled on a grid) can be obtained using the method of \citet{Saye2014}. For the examples in this paper, we use the ideas of \citet{Saye2014} but with analytical expressions for $\phi$. Specifically, an initial guess $\cp^{\star}$ of the closest point is obtained using a Newton-style procedure, starting with $\cp_0 = \x_i,$ and iterating
\begin{equation*}
    \cp_{k+1} = \cp_k - \frac{\phi(\cp_k) \nabla \phi(\cp_k)}{\|\nabla \phi(\cp_k)\|^2},
\end{equation*}
with stopping criterion $\|\cp_{k+1} - \cp_k\| < 10^{-10}.$ Then Newton's method
\begin{equation*}
    \y_{k+1} = \y_k - (D^2f(\y_k))^{-1} \nabla f(\y_k),
\end{equation*}
is used to optimize
\begin{equation*}
    f(\cp, \lambda) = \frac{1}{2} \|\cp - \x_i\|^2 + \lambda \phi(\cp),
\end{equation*}
where $\y = [\cp, \lambda]^T$ and $\|\y_{k+1} - \y_k\| < 10^{-10}$ is used as the stopping criterion. The initial Lagrange multiplier is $\lambda_0 = (\x_i - \cp^{\star}) \cdot \nabla \phi(\cp^{\star}) / \|\nabla \phi(\cp^{\star})\|^2$. Analytical expressions for $\nabla f(\y)$ and $D^2 f(\y)$ are computed using analytical expressions of $\nabla \phi$ and $D^2 \phi$.

Closest points for objects composed of multiple parts can be computed by obtaining the closest point to each independent manifold first. Then the closest point to the combined object is taken as the closest of the independent manifold closest points (e.g., the torus and sphere joined by line segments in Figure~\ref{fig:mixed-codim-diff-curves}).

Closest points can be computed for many other representations. For example, closest points to neural implicit surfaces can be computed using the work of \citet{Sharp2022}. Further references for closest point computation are given in Section 5.1 of \cite{Sawhney2020}.

\end{document}